\documentclass{aa}  

\usepackage{graphicx}
\usepackage{supertabular}
\usepackage{multirow}
\usepackage{url}
\DeclareGraphicsExtensions{.ps,.png,.jpg}
\usepackage{txfonts}
\newcommand{\teff}{$T_{\text{eff}}$\,}
\newcommand{\logg}{$\log \text{g}$\,}
\newcommand{\FeI}{Fe\,{\sc i}\,}
\newcommand{\FeII}{Fe\,{\sc ii}\,}
\newcommand{\kms}{km\,s$^{-1}$}
\newcommand{\isologg}{$\log \text{g}_{\text{iso}}$\,}

\begin{document}

   \title{SPECIES I: Spectroscopic Parameters and atmosphEric ChemIstriEs of Stars}

   \author{M. G. Soto
          \inst{1}
          \and
          J. S. Jenkins\inst{1,2}\fnmsep
          }

   \institute{$^1$Departamento de Astronom\'ia, Universidad de Chile.\\
   $^2$Centro de Astrof\'isica y Tecnolog\'ias Afines (CATA), Casilla 36-D, Santiago, Chile.\\
   \and
              \email{maritsoto@ug.uchile.cl}
             \email{jjenkins@das.uchile.cl}
             }

%   \date{Received September 15, 1996; accepted March 16, 1997}

% \abstract{}{}{}{}{} 
% 5 {} token are mandatory
 
  \abstract
   {The detection and subsequent characterisation of exoplanets are intimately linked to the characteristics of their host star.
   Therefore, it is necessary to study the star in detail in order to understand the formation history and characteristics of their companion(s).}
   {Our aims were to develop a community tool that allows the automated calculation of stellar parameters for a large number of stars, using high resolution echelle spectra and minimal photometric magnitudes, and introduce the first catalogue of these measurements in this work.}
   {We measured the equivalent widths of several iron lines and used them to solve the radiative transfer equation assuming local thermodynamic equilibrium in order to obtain the atmospheric parameters (\teff, [Fe/H], \logg and $\xi_t$). We then used these values to derive the abundance of 11 chemical elements in the stellar photosphere (Na, Mg, Al, Si, Ca, Ti, Cr, Mn, Ni, Cu and Zn). 
   Rotation and macroturbulent velocity were obtained using temperature calibrators and synthetic line profiles to match the observed spectra of five absorption lines.
   Finally, by interpolating in a grid of MIST isochrones, we are able to derive the mass, radius and age for each star using a Bayesian approach.}
   {Our SPECIES code obtains bulk parameters that are in good agreement with measured values from different existing catalogues, including when different methods are used to derive them. We find excellent agreement with previous works that used similar methodologies, in particular when the \teff is calculated using model fitting to the spectra themselves.
   We find discrepancies in the chemical abundances for some elements with respect to other works, which could be produced by differences in \teff, or in the line list or the atomic line data used to derive them. We also obtained analytic relations to describe the correlations between different parameters, and we implemented new methods to better handle these correlations, which provides a better description of the uncertainties associated with the measurements.}
   {}


\maketitle
%
%________________________________________________________________

\section{Introduction}\label{sec:introduction}

The characterisation of exoplanetary systems has become a booming field of study in astronomy over the last 20 years, thanks to the large amount of detections provided by different surveys from different telescopes and instruments (CORALIE, Keck, HARPS, AAT, WASP, Kepler, K2, etc). Unfortunately, the low surface brightness and size of planets compared to their host star, makes them extremely difficult to study directly, therefore it is necessary to study the behaviour and physical parameters of the host stars in order to better characterise their planetary companions. These parameters include the temperature, metallicity, surface gravity, mass, and age, which in turn gives us an estimate of their evolutionary stages. 

Calculation of the stellar bulk parameters, like temperature, metallicity and mass, is vital to derive the physical characteristics of the companions. The minimum mass of the planetary candidates can be obtained by the amplitude of the star's radial velocity, which in turn depends on the mass of the host star, among other parameters. Planetary sizes can be inferred by studying the decrease in brightness of the host star when the planet transits, which in turn depends on the diameter of the star. 
By knowing the mass and physical size of a planetary companion, it is possible to understand its chemical composition, since the planet bulk density can be calculated. This information, combined with the knowledge of the stellar effective temperature (\teff) and the orbital distance of the planet, allows a probability to be placed on the likelihood that the planet has liquid water in its atmosphere, and/or on its surface. Knowledge of the stellar parameters is also needed in order to study the formation of planetary companions \citep{Fischer2005,Buchhave2012,Jenkins2013}, how the system has evolved to its current stage \citep{Ida2004,Ida2005,Mordasini2012}, and how the subsequence evolution of the host star will affect the planetary system \citep{Villaver2009,Kunitomo2011,Jones2016}. 

The derivation of stellar parameters is not something new in astrophysics. Many works have dealt with this task, employing different methods in order to obtain them. The most common methods in the literature are using equivalent width (EW) measurements \citep[e.g.][]{Edvardsson1993,Feltzing1998,Santos2004,bond2006,Neves2009}, and the spectral synthesis approach \citep[e.g.][]{spocs2005,Jenkins2008,pavlenko2012}. The results produced by different methods show significant systematic differences \citep{torres2012,ivanyuk2017}, which then can affect the physical characteristics of any detected companions. When the values for the stellar parameters are retrieved from different sources, it can lead to problems when studying populations of stars. This is often necessary because not all catalogues of stellar parameters have all the quantities needed, or uncertainties in the values are not listed, making it difficult to implement them in other studies \citep[e.g. see the analysis presented in ][]{Jenkins2017}. Another barrier one finds when studying stellar parameters is that most works are limited to the stars included in their resulting catalogues, making it difficult to compute parameters for new stars in a homogeneous way. All of these issues were behind the development of the SPECIES code, an open source method that can compute stellar parameters for large numbers of stars in a homogeneous and self-consistent fashion, and crucially, that is publicly available to the scientific community\footnote{\url{https://github.com/msotov/SPECIES/}}.

The SPECIES code is written mostly in the python programming language, making use of some previously developed software (e.g. \texttt{MOOG}, \citealt{sneden_moog}; \texttt{ARES}, \citealt{sousa_ares}) that allows automatic calculations of specific jobs to be performed, with the goal of increasing the speed of the process whilst subsequently decreasing the user input for the derivation of the stellar parameters. The code is automated in the computation of all the parameters, and the only input from the user is a high resolution spectrum of the desired star. It can be used with only one star at a time, or several stars at once running in parallel. This makes it possible to derive the parameters for large samples of stars, a necessity in this new era of exoplanet surveys (e.g. NGTS, ESPRESSO, TESS, etc), with the number of planetary candidates increasing each month.

%Structure of the paper.

The paper structure is as follows. Section \ref{sec:stellar_params} explains the inputs needed to run the code, and its final output. Here we also list the atomic lines used (subsection \ref{sec:atomic_lines}), we explain in detail the derivation of the atmospheric parameters and their corresponding uncertainties (subsections \ref{sec:atmospheric_parameters} and \ref{sec:uncertainty}, respectively), the stellar mass, radii and age (subsection \ref{sec:mass_age_plogg}), the chemical abundances of different elements (subsection \ref{sec:chemical_abundances}), and the computation of the macroturbulence and rotational velocity (subsection \ref{sec:broadening}). Section \ref{sec:results} shows the results obtained using our code for a sample of 522 stars, how those values compare to others in the literature (subsection \ref{sec:comparison}), and the difference obtained when using spectra from different instruments (subsection \ref{sec:compare_inst}). In Section \ref{sec:correlations} we show the correlations between the parameters that we find in our results. Finally, in Section \ref{sec:conclusions} we give a summary of the characteristics and use of SPECIES and how we plan to continue to develop the code in the future.

\section{Stellar parameter computation}\label{sec:stellar_params}

\begin{figure*}
\centering
\includegraphics[width=17cm]{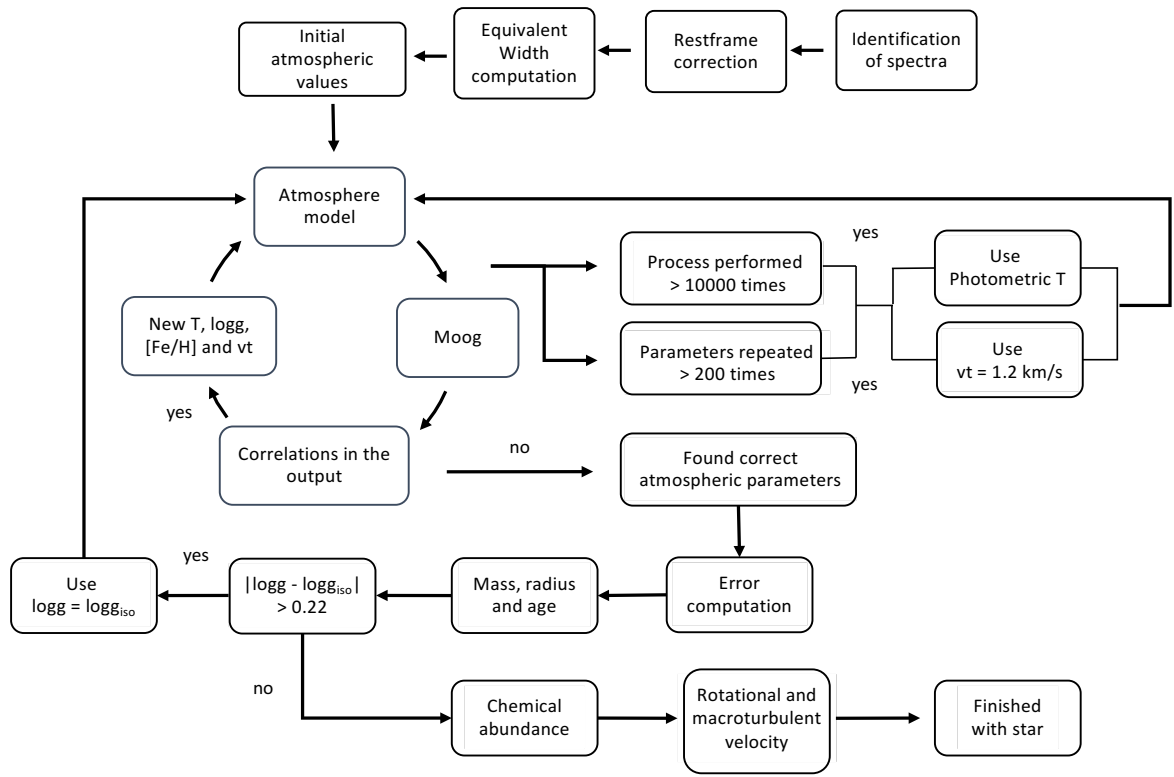}
\caption{Flow diagram showing the process sequence of SPECIES. Each step is explained in Section \ref{sec:stellar_params}.}
\label{fig:diagrama}
\end{figure*}

SPECIES is an automatic code that computes stellar atmospheric parameters in a self-consistent and homogeneous way: effective temperature, surface gravity, metallicity with respect to the Sun, and microturbulent velocity (\teff, \logg, [Fe/H] and $\xi_t$ respectively). Our code also derives chemical abundances for 11 additional atomic elements, rotational and macroturbulence velocity, along with stellar mass, age, radius, and photometric \logg. The uncertainties of each parameter are also computed in a consistent way, dealing with parameter correlations and propagation of uncertainties, all of which will be discussed in the following sections.  

The inputs needed for our code to perform all computations are:

\begin{itemize}
\item A high resolution (R > 40000) spectrum. It handles spectra acquired with HARPS \citep[High Accuracy Radial velocity Planet Searcher;][]{mayor2003}, FEROS \citep[Fiber-fed Extended Range Optical Spectrograph;][]{kaufer1999}, UVES \citep[Ultraviolet and Visual Echelle Spectrograph;][]{dekker2000}, HIRES \citep[High Resolution Echelle Spectrometer;][]{vogt1994}, AAT \citep[Anglo-Australian Telescope;][]{tinney2001} and Coralie instruments \citep{queloz2000}. The spectra do not need to be normalized, because then will be locally normalized when measuring the equivalent widths (section \ref{sec:EW}). The optimal wavelength range should go from 5500 to 6500 \r{A}, or cover most of this range. It is not necessary for the spectra to have continuous wavelength coverage, except for the regions where included iron lines are located (Table \ref{tab:linelist}). A minimum of 15 \FeI lines and 5 \FeII lines should be present in the input spectrum.\newline

\item Coordinates. They can be input by the user, retrieved from the fits header of the spectra, or retrieved from the following catalogues: 2MASS \citep{cutri2003}, GAIA DR1 \citep{GaiaDR1}, the Hipparcos catalogue \citep{vanleeuwen2007}, or the Tycho-2 Catalogue \citep{hog2000}. \newline

\item Parallax data. It can be input by the user, but otherwise it will be automatically retrieved from the GAIA DR1 \citep{GaiaDR1}, or from the Hipparcos catalogue \citep{vanleeuwen2007}.\newline 

\item Apparent magnitudes for each star. These can be either given by the user, or they can be retrieved from the following catalogues: 2MASS \citep{cutri2003} for the $JHKs$ bands, Tycho-2 Catalogue \citep{hog2000} for the Tycho-2 $(BV)t$ magnitudes, \cite{hauck1998} or \cite{holmberg2009} for Str\"{o}mgren $b-y$, $m_1$ and $c_1$, and \cite{Koen2010}, \cite{casagrande2006}, \cite{beers2007}, or \cite{ducati2002} for the Johnson $BV(RI)c$ magnitudes.

\end{itemize}

All the data from these catalogues were obtained using Vizier\footnote{\url{http://vizier.u-strasbg.fr}}. 
Other files used by SPECIES, like the atomic line list and binary masks are included in the SPECIES package.

A diagram showing a representation of the process followed by SPECIES to derive all the stellar parameters is shown in Figure \ref{fig:diagrama}. Each step of the computation will be explained in the next sections.

\subsection{Atomic Line Selection}\label{sec:atomic_lines}

We selected all of the lines used in our analysis from the Vienna Atomic Line Database version 3 \citep[VALD3,][]{Piskunov1995,Kupka1999,vald3}.  The lines were selected based on comparing the line database to a HARPS solar spectrum to ensure they appeared strong and clearly detectable at the resolution offered by HARPS (note that the macroturbulence of the solar envelope ensures that the spectra have an effective resolution $R$ of around 70'000).  We also note that the lines were cross-validated by literature searches, since each of the lines has previously been employed in atomic abundance calculations by other teams using different methods.  The parameters drawn from the VALD3 catalogue for each of these lines are the excitation potential ($\chi_l$), central rest wavelengths, and oscillator strengths ($\log$gf).  The final line lists contain 149 \FeI lines and 21 \FeII lines, along with 6 (Na\,{\sc i}\,), 4 (Mg\,{\sc i}\,), 3 (Al\,{\sc i}\,), 22 (Si\,{\sc i}\,), 14 (Ca\,{\sc i}\,), 22 (Ti\,{\sc i}\,), 3 (Ti\,{\sc ii}\,), 37 (Cr\,{\sc i}\,), 8 (Mn\,{\sc i}\,), 52 (\FeI), 15 (\FeII), 24 (Ni\,{\sc i}\,), 4 (Cu\,{\sc i}\,) and 1 (Zn\,{\sc i}\,) lines.  We show all data for the iron lines used in the computation of the atmospheric stellar parameters in Table~\ref{tab:linelist}, and for the lines used in the computation of the chemical abundances in Table~\ref{tab:linelist_ab}.

\subsection{Equivalent Width Computation}\label{sec:EW}

\begin{figure}
   \centering
   \includegraphics[width=8cm]{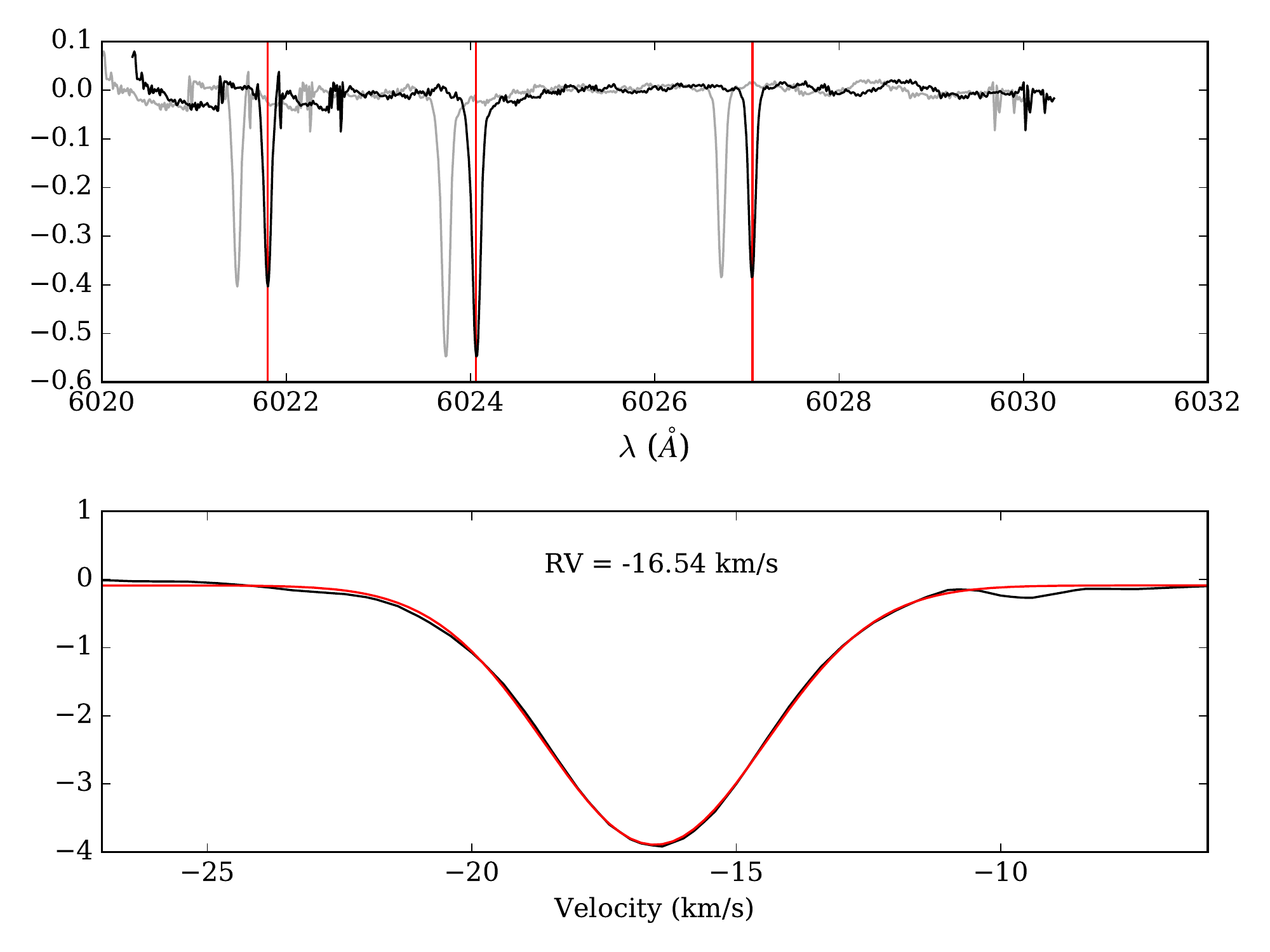}
      \caption{Restframe correction applied to HD10700. Top panel: Original spectra (grey), corrected spectra (black), and three reference lines at 6021.8, 6024.06, and 6027.06 \r{A} (red). Bottom panel: Cross-correlation function (CCF) between the binary mask and the spectra. The red line corresponds to the Gaussian fit to the CCF, with a mean equal to -16.54 km s$^{-1}$.}
         \label{fig:restframe_correction}
   \end{figure}

The EWs were measured using the ARES code \citep{sousa_ares,sousa2015}. 
The input required by ARES is a one-dimensional spectrum and a line list. For each line, the code performs a local normalisation, over a window of 4\r{A} across the line centre. The normalisation is done by adjusting a third-order polynomial to that portion of the spectrum, and selecting only the points laying above $rejt$ times the obtained fit. This process is repeated three times, and the final local continuum is subtracted from the data. $rejt$ is computed per spectrum, and depends on the signal-to-noise of the data, where large ($\sim 1$) values would correspond to high S/N. We use the S/N given in the image header, and when that is not provided, we compute it as the median of the S/N for different portions of the spectra, free of absorption lines. More information about this parameter can be found in \citet{sousa_ares}.
For the computation of the atmospheric parameters, we only consider lines with EWs in the range 10 $\leq$ EW $\leq$ 150 \textbf{m\r{A}}, in order to avoid lines too weak that could be affected by the continuum fitting, and lines too strong for which the LTE approximation might no longer be valid. We also discarded the lines for which $\sigma_{\text{EW}}/\text{EW}>1$, where $\sigma_{\text{EW}}$ is the error in the EW measured with ARES. We note that for these rejected lines, the Gaussian fit performed by ARES is not accurate enough, leading to an incorrect computation of the stellar parameters.
We also perform a restframe correction in all our spectra before computing the EW of the lines. This was done by cross-correlating \citep{Tonry1979} a portion of the spectrum, between 5500 \r{A} and 6050 \r{A}, with a G2 binary mask within the same wavelength range. The wavelength was then scaled as $\lambda = \lambda_o/(1 + v/c),$ where $\lambda_o$ is the observed wavelength, $v$ is the derived velocity of the star that SPECIES computes using a cross-correlation method, and $c$ is the speed of light. This spectrum will be used for the rest of the calculations.
One example of this cross-correlation function (CCF) and subsequent velocity correction is shown in Figure \ref{fig:restframe_correction}. 
We have made sure that the instruments accepted by our code have a wavelength coverage that is wide enough so that the region of the spectra used for the restframe correction is included.

\subsection{Initial conditions}\label{sec:inital_conditions}

Initial values for the atmospheric parameters (\teff, \logg, [Fe/H]) are needed for their subsequent derivation through SPECIES, in a manner which will be explained in the next section. The initial conditions can be input by the user, or can be derived from the photometric information for each star.

In the case that the photometric magnitudes are retrieved from existing catalogues, they first should be corrected for interstellar extinction. We use the coordinates and parallax, along with the \citet{Cardelli1989} extinction law, and $R_V = 3.1$, to obtain the extinction in the V-band, $A_V$ for each star using the \citet{Arenou1992} interstellar maps. Extinction in the rest of the photometric bands, $A_{\lambda}$, was derived from the \citet{Cardelli1989} relations. The corrected magnitudes for each band is then obtained as $m_{\lambda, C} = m_{\lambda, O} - A_{\lambda}$, where $m_{\lambda, O}$ is the magnitude retrieved from the catalogue, and $m_{\lambda, C}$ is the extinction corrected magnitude.

\subsubsection{Luminosity class}\label{sec:luminosity_class}

\begin{figure}
   \centering
   \includegraphics[width=8cm]{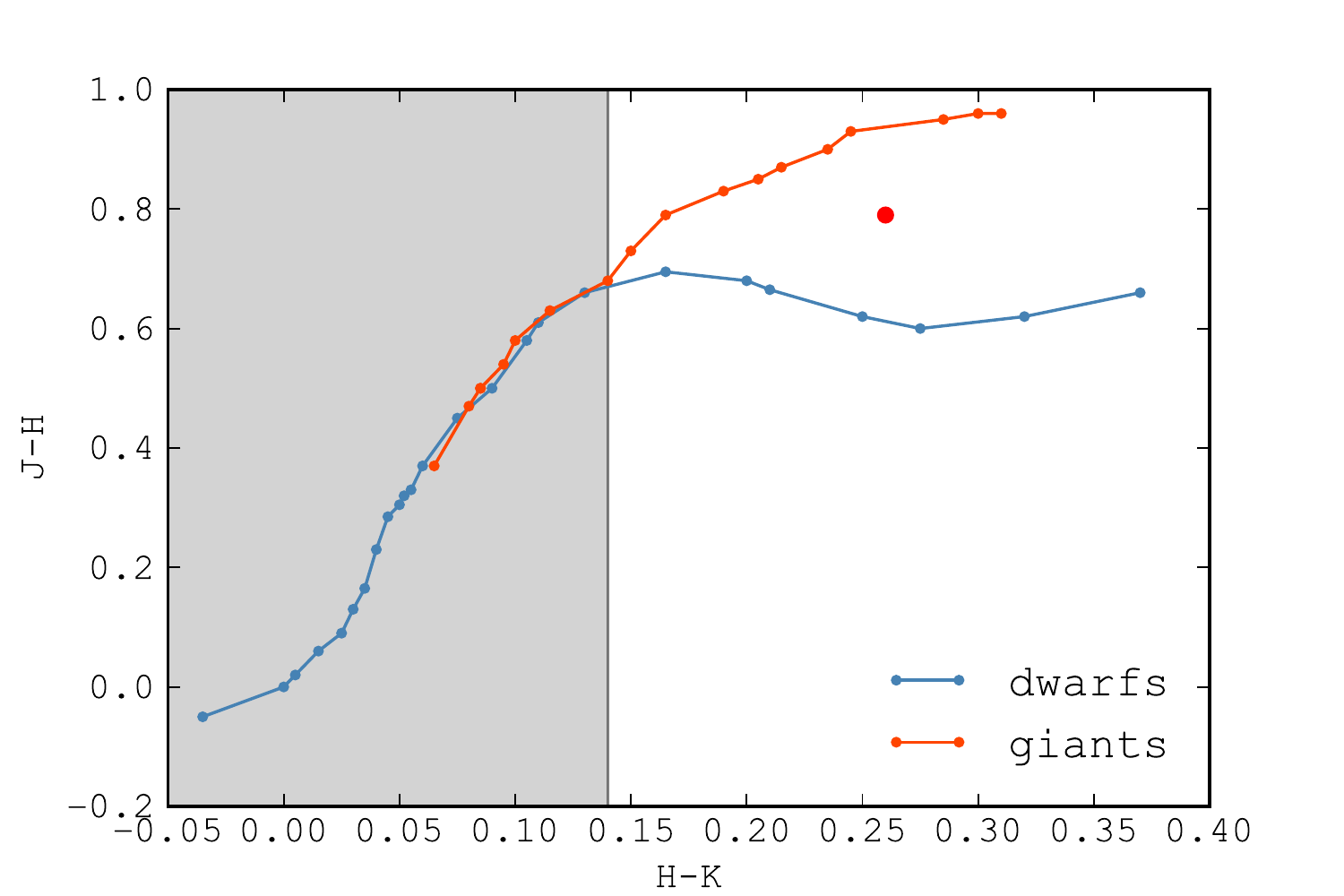}
      \caption{Intrinsic JHK colors for dwarf (blue line) and giant (red line) stars, from \citet{Bessell1988}. The red point represents a star with $H-K = 0.26$ and $J-H = 0.79$ The distance between the point to the giant curve is 0.08 dex, and to the dwarf curve is 0.13 dex, therefore the star is classified as a giant. The shaded area represents the $H-K$ range for which the curves overlap, so no classification can be correctly performed. In those cases, the star is classified as dwarf.}
         \label{fig:luminosity_classification}
   \end{figure}

Before deriving the initial conditions, it is necessary to classify the star as a dwarf or giant. That is done by using the JHK magnitudes, and the intrinsic colors of dwarfs and giants, for different spectral types, from \citet{Bessell1988}. We first converted the JHK magnitudes to the Bessel and Brett system using the relations from \citet{Carpenter2001}\footnote{\url{http://www.astro.caltech.edu/~jmc/2mass/v3/transformations/}}. We then computed the distance to the dwarf and giant evolutionary models, and classify the star from the curve for which the distance is the shortest. This procedure, as well as the dwarf and giant curves from Bessel and Brett, are shown in Figure \ref{fig:luminosity_classification}. If the JHK magnitudes are missing, the star is classified as a dwarf. This procedure is also performed only when $H-K > 0.14$, which is when the dwarf and giant curves no longer overlap. For stars with $H-K < 0.14$, they are classified as dwarfs.

\subsubsection{Metallicity}\label{sec:initial_met}

The metallicity is derived from Equation 1 of \citet{Martell2002}, using the Str\"{o}mgren coefficients $b-y$, $m_1$ and $c_1$. This relation is valid for $-2.0 < $ [Fe/H] $< 0.5$. In the case the Str\"{o}mgren coefficients are missing, or the derived metallicity is outside of the permitted ranges, then [Fe/H] is set to zero.

\subsubsection{Temperature}\label{sec:ini_temperature}

\begin{figure}
   \centering
   \includegraphics[width=8cm]{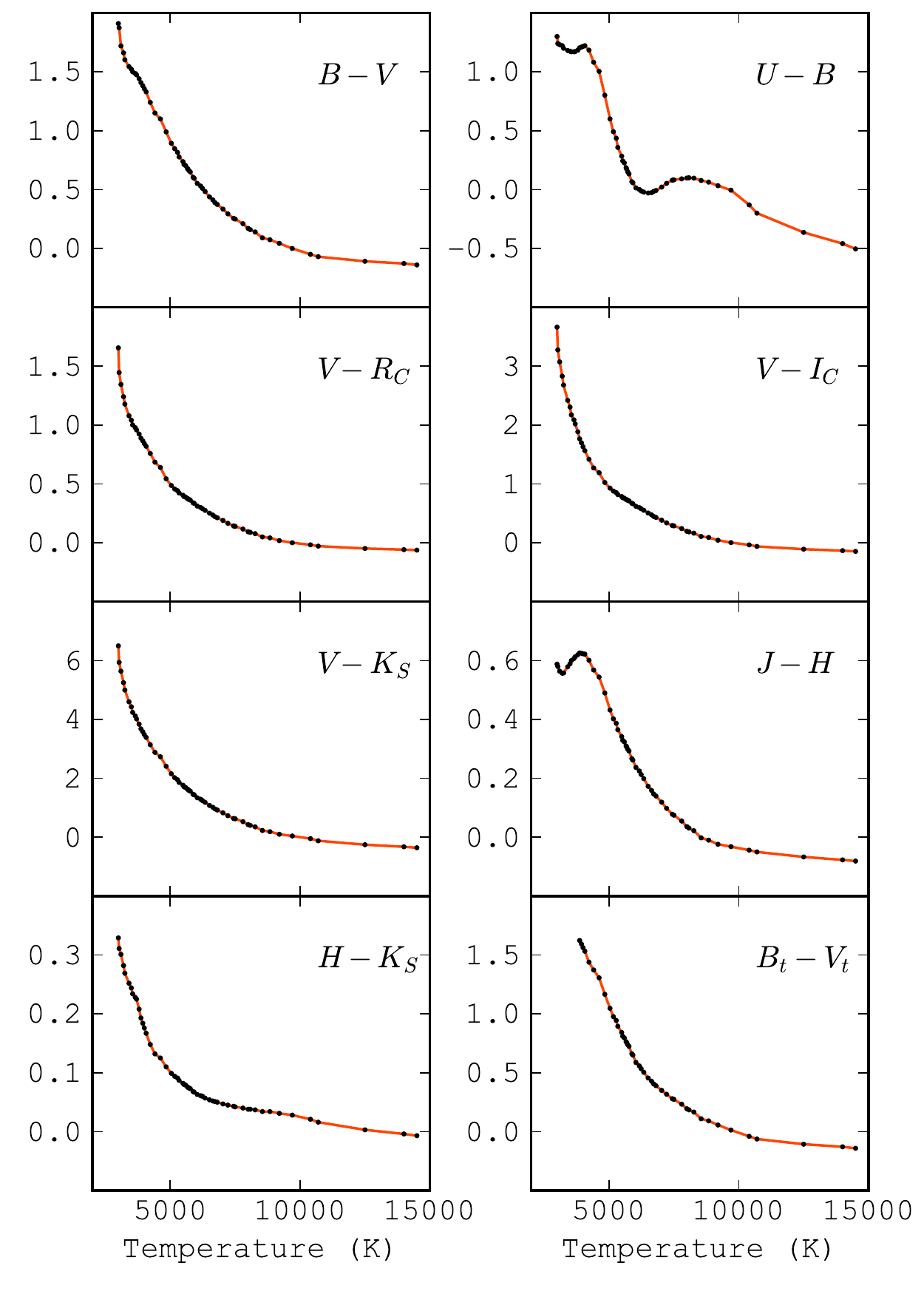}
      \caption{Photometric colors of main sequence stars, for different effective temperatures, from \citet{Pecaut2013}. Red lines represent the spline approximation for each curve. The temperature for a given color is computed as the root of the ($m_1$ - $m_2$) - ($m_{1,\star}$ - $m_{2, \star}$) curve, where $m_1 - m_2$ is the chosen color, and $m_{1,\star}$, $m_{2,\star}$ are the star's colors.}
         \label{fig:mamajek}
   \end{figure}

The derivation of the initial effective temperature ($T_{ini}$) will depend of the luminosity class. In the case of dwarf stars, the photometric relations from \citet[][hereafter C10]{casagrande2010} or \citet[][hereafter M15]{mann2015} are used. The difference between both relations is that the one from M15 is optimized for M dwarf stars, and the relations from C10 are applicable for FGK stars. In order to infer which type of star we are dealing with, we use its apparent magnitudes and the intrinsic colors for each spectral type derived in \citet[][hereafter P13]{Pecaut2013}. If the colors are in agreement with a K07 star or later, we use the M15 relations. If the photometric colors are outside of the permitted ranges for the C10 or M15 relations, we then infer the initial temperature by interpolating from the photometric colors and temperatures from P13. The color in different bands and temperatures, along with the spline representation of each curve, are shown in Figure \ref{fig:mamajek}.

In C10, M15 and P13, several relations are available for different photometric colors. For each star, we compute the average of the temperature derived for each photometric color, weighted by their uncertainty.

If the star is classified as a giant, then the relations from \citet{gonzalez-hernandez2009} are used. Just like for C10 and M15, different temperatures are derived for each photometric color, and the final value corresponds to the average of the individual temperatures for each color, weighted by their uncertainty.

The initial temperature will also set the boundaries of the parameter space through which SPECIES will search for the final temperature. These boundaries are set to be 200\,K from $T_{ini}$. In section \ref{sec:compare_photometry} we will show the reasons for choosing 200\,K as the window over $T_{ini}$. This can be disabled by the user at any time.

\subsubsection{Surface gravity}\label{sec:initial_logg}

The initial surface gravity is derived by comparing \logg and \teff obtained for stars in the literature with different luminosity classes. We used the sample of stars from the NASA Exoplanet Archive\footnote{\url{https://exoplanetarchive.ipac.caltech.edu}}, and separated the points into two classes: dwarf (\logg $\gtrsim$ 4.0) or giants (\logg $\lesssim$ 4.0). Then, we adjusted a second order polynomial to each group, obtaining the following relations, depending on the luminosity class: 

\begin{equation}\label{eq:ini_logg}
\text{logg} = \begin{cases}
	4.68\times 10^{-8}\, T^2 - 8.33\times 10^{-4}\, T - 7.547 & \text{dwarf}\\
    -2.8\times 10^{-7}\, T^2 + 3.79\times 10^{-3}\, T - 9.335 & \text{giant}
\end{cases}
\end{equation}

\begin{figure}
   \centering
   \includegraphics[width=8cm]{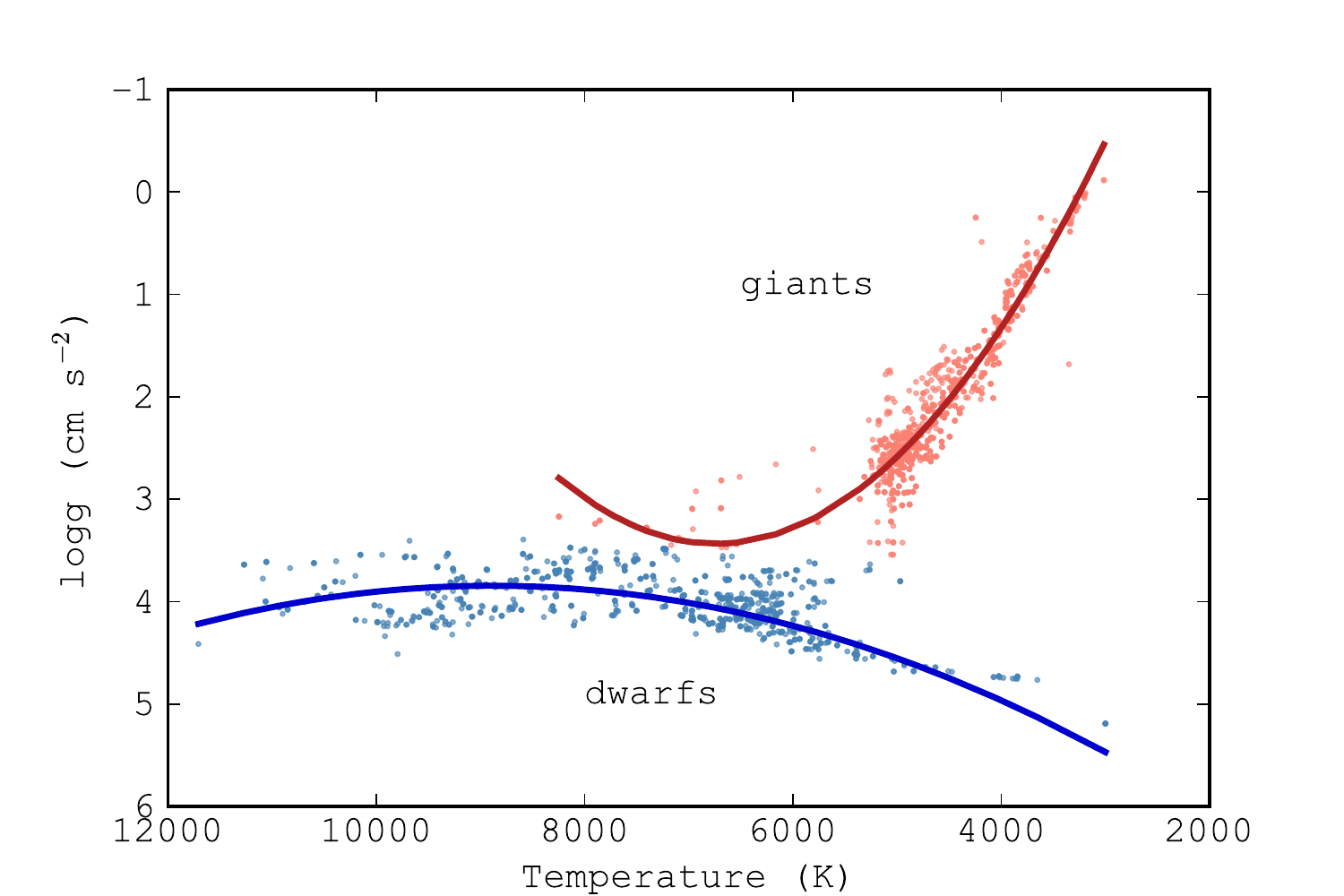}
      \caption{logg vs temperature for several stars in the literature. Red points represent giant stars, and blue points dwarf stars. The solid lines are the polynomials adjusted to each group, showed in Equation \ref{eq:ini_logg}.}
         \label{fig:logg_computation}
   \end{figure}

\subsection{Atmospheric parameters}\label{sec:atmospheric_parameters}

The atmospheric parameters (\teff, \logg, [Fe/H] and $\xi_t$) are derived using equivalent widths (EWs) for the set of \FeI and \FeII lines discussed above.
These EW values, along with an appropriate atmosphere model obtained by interpolating through a grid of ATLAS9 atmosphere models \citep{ATLAS9}, are given to the 2017 version of the MOOG code \citep{sneden_moog}, using the driver \textit{abfind}, which solves the radiative transfer equation assuming local thermodynamic equilibrium (LTE) conditions. 
The atmospheric parameters are then derived through an iterative process that stops when no correlation
is found to a tolerance level of 0.02 dex between the abundance of each individual \FeI line and both the excitation potential and the 
reduced equivalent width ($\log EW/\lambda$), and also when the average abundances for \FeI and \FeII are equal to the iron abundance given to the atmosphere model at the level of 0.02\%.
   
The ranges in parameters accepted by the code are [3500\,$\--$\,15000 K] for the temperature, [-3.0\,$\--$\,+1.0 dex] for the metallicity, [0.0\,$\--$\,5.0] for the logarithm of the surface gravity in cm\,s$^{-2}$, and [0.0\,$\--$\,2.0 \kms] for the microturbulent velocity. If, during the iterative process, all four parameters are outside of those ranges, or the same values are repeated more than 200 times, the computation stops. This last case would mean that SPECIES is stuck in one section of the parameter space, prolonging the time the code runs, without reaching final convergence. From our experience using SPECIES, we find that after the same parameters are repeated over 200 times, the code is not able to search the rest of the parameters space. For those cases, it is recommended that the user specifies the initial conditions, or fixes one of the parameters to a certain value. If no convergence was reached, we perform the derivation another time but now setting the temperature to be equal to $T_{ini}$, and keeping it fixed. SPECIES also has an option to set the microturbulence to a fixed value through the computation. We found that, for some stars, all the atmospheric parameters would reach convergence except for the microturbulence, therefore for those stars we set $\xi_t = 1.2$\,\kms. These options to fix the temperature or/and the microturbulence are only used when there was no convergence on the atmospheric parameters, and can be disabled by the user.

\subsection{Uncertainty estimation}\label{sec:uncertainty}

An important facet of the SPECIES code is the handling of uncertainties for each of the calculated parameters.  A number of the parameters derived by SPECIES are heavily correlated, such as temperature and iron abundance, due to them being derived simultaneously with MOOG via the curve of growth analysis. The code tries to take into consideration these correlations to return a more representative uncertainty estimate for each of the elements, and to consider the uncertainty in the EW (derived with ARES) in the equation. In order to do so, we took as a reference the uncertainty estimation method used in \citet{gonzalez1998} and \citet{santos2000}. In Table \ref{tab:uncertainties} we show the typical uncertainties obtained for each atmospheric parameter, separated in ranges of stellar temperature.

\begin{table}
\caption{Estimate of the uncertainties for the stellar parameters, separated in ranges of stellar temperature, for dwarf stars.}
\centering
\begin{tabular}{l|c|c|c|c}
& 4500 - & 5125 - & 5750 - & 6375 - \\
& 5125 K & 5750 K & 6375 K & 7000 K \\
\hline
$\sigma_T$ & 52.835 & 30.706 & 32.36 & 74.668 \\
$\sigma_{\log \text{g}}$ & 0.314 & 0.267 & 0.238 & 0.752 \\
$\sigma_{\text{[Fe/H]}}$ & 0.118 &  0.0705 & 0.064 & 0.117 \\
$\sigma_{\xi_t}$ & 0.046 &  0.0235 & 0.026 &  0.057
\end{tabular}
\label{tab:uncertainties}
\end{table}

\subsubsection{Microturbulence}\label{sec:error_microturbulence}

The microturbulence is computed as the value for which the slope of the linear fit performed between the individual FeI abundances and the reduced equivalent width reaches zero. This value will be referred to as $S_{RW}$. Therefore, the resulting uncertainty will depend on this slope, resulting in:
\begin{equation}
\sigma_{\xi_t}^2 = \left( \left.\frac{\partial \xi_t}{\partial S_{RW}}\right|_{S_{RW} = 0} \right)^2 \sigma_{S_{RW}}^2,
\end{equation}
where $\sigma_{S_{RW}}$ corresponds to the uncertainty in $S_{RW}$.  

\begin{figure}
   \centering
   \includegraphics[width=6cm]{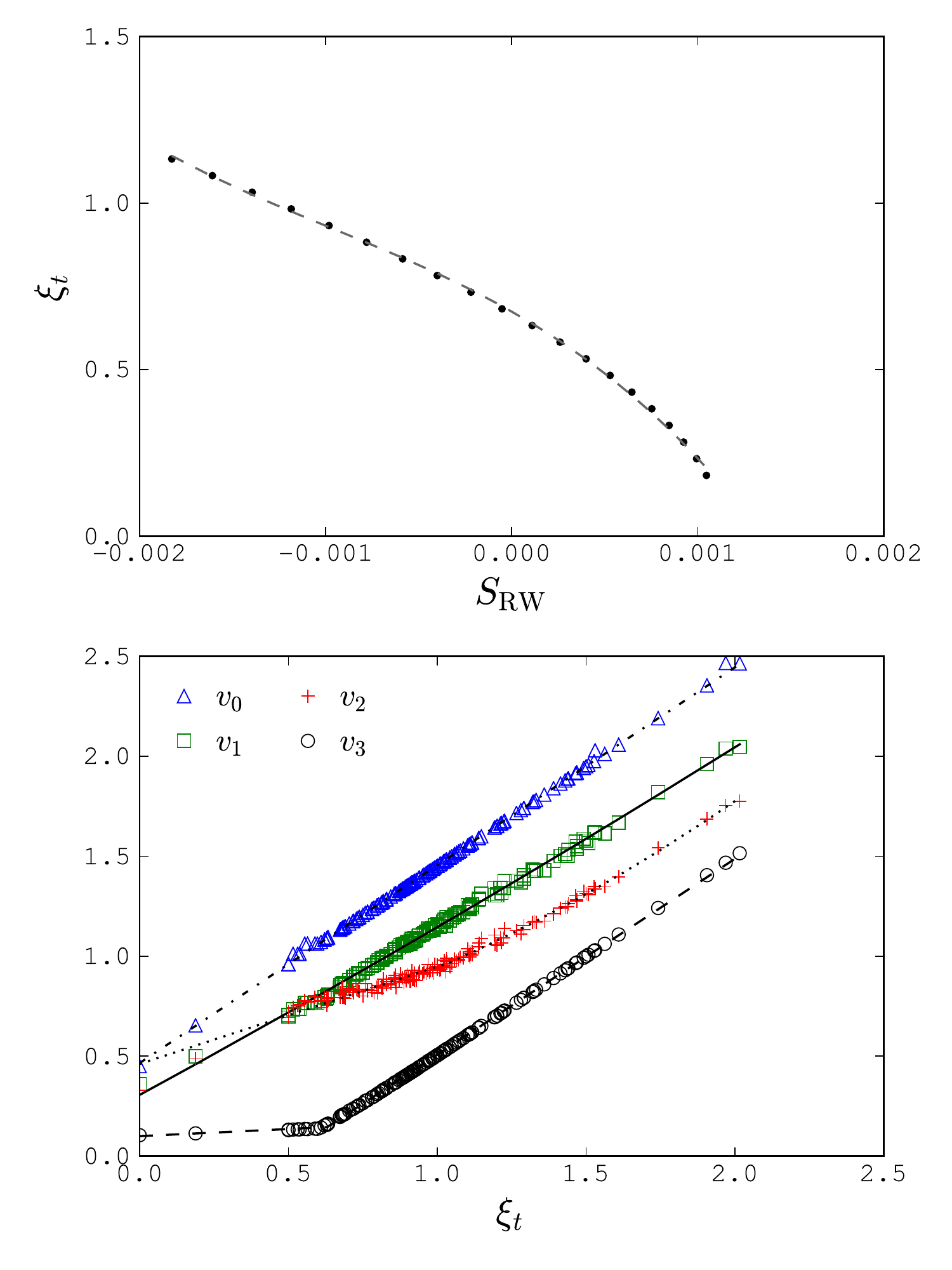}
      \caption{Top panel: $\xi_t$ vs $S_{RW}$ for a Solar spectrum. 
      Bottom panel: Fit of the spline coefficients as a function of microturbulence.}
         \label{fig:coefs_err_vt}
   \end{figure}

We computed this uncertainty for 160 stars from the sample studied in \cite{sousa2008}, and for 10 solar spectra, all taken using HARPS (more details about this sample of stars will be given in Section \ref{sec:comparison}). We found that, for each case, there was a dependency of $\xi_t$ with $S_{RW}$, which can be adjusted with a cubic spline,
\begin{equation}\label{eq:vt_cubic_spline}
\xi_t = Y(v_0, v_1, v_2, v_3, S_{RW}),
\end{equation}
where $Y$ represents a cubic spline, with coefficients $v_i$. The coefficients are a function of microturbulence velocity, shown in Figure \ref{fig:coefs_err_vt}, and have the following dependence: 

\begin{align}\label{eq:coefs_vt}
v_0& = 0.99\cdot \xi_t \: + \: 0.466 \nonumber\\
v_1& = 0.03\cdot \xi_t^2 \: +\: 0.81\cdot \xi_t \: +\: 0.306 \nonumber\\
v_2& = \begin{cases}
    0.49\cdot \xi_t \: +\: 0.46 & \xi_t < 1.04\\
    0.19\cdot \xi_t^2 \: +\: 0.26\cdot \xi_t \: +\: 0.49 & \xi_t \geq 1.04 
\end{cases} \\
v_3& = \begin{cases}
    0.07\cdot \xi_t \: +\: 0.1 & \xi_t < 0.63\\
    0.98\cdot \xi_t \: -\: 0.47 & \xi_t \geq 0.63 
\end{cases} \nonumber
\end{align}

Another way SPECIES computes the uncertainty in the microturbulence is explained in the appendix, and although we performed this method on all our stars, it is not the preferred final value. We note however that it does appear in the SPECIES catalogue as \texttt{err\_vt2}.     
   
\subsubsection{Temperature}\label{sec:error_temperature}

The temperature is obtained when the slope of the dependence between the individual \FeI abundances, and the excitation potential, is zero. We will call this slope as $S_{\text{EP}}$. Since all the atmospheric parameters are derived simultaneously, the microturbulence will have an effect on the final temperature, and its uncertainty. The final expression for the error in the temperature is then:

\begin{equation}
\sigma_T^2\: =\: \left(\left.\frac{\partial T}{\partial \xi_t}\right|_{\xi_t}\right)^2\, \sigma_{\xi_t}^2 \: +\: \left(\left.\frac{\partial T}{\partial S_{\text{EP}}}\right|_{S_{\text{EP}} = 0}\right)^2\,\sigma_{S_{\text{EP}}}^2,
\end{equation}
where $\partial T/\partial \xi_t$ is evaluated at the microturbulence derived by our code, $\sigma_{\xi_t}$ is its uncertainty, and $\sigma_{S_{\text{EP}}}$ is the uncertainty in $S_{\text{EP}}$, when the temperature reaches convergence.

\begin{figure}
   \centering
   \includegraphics[width=9cm]{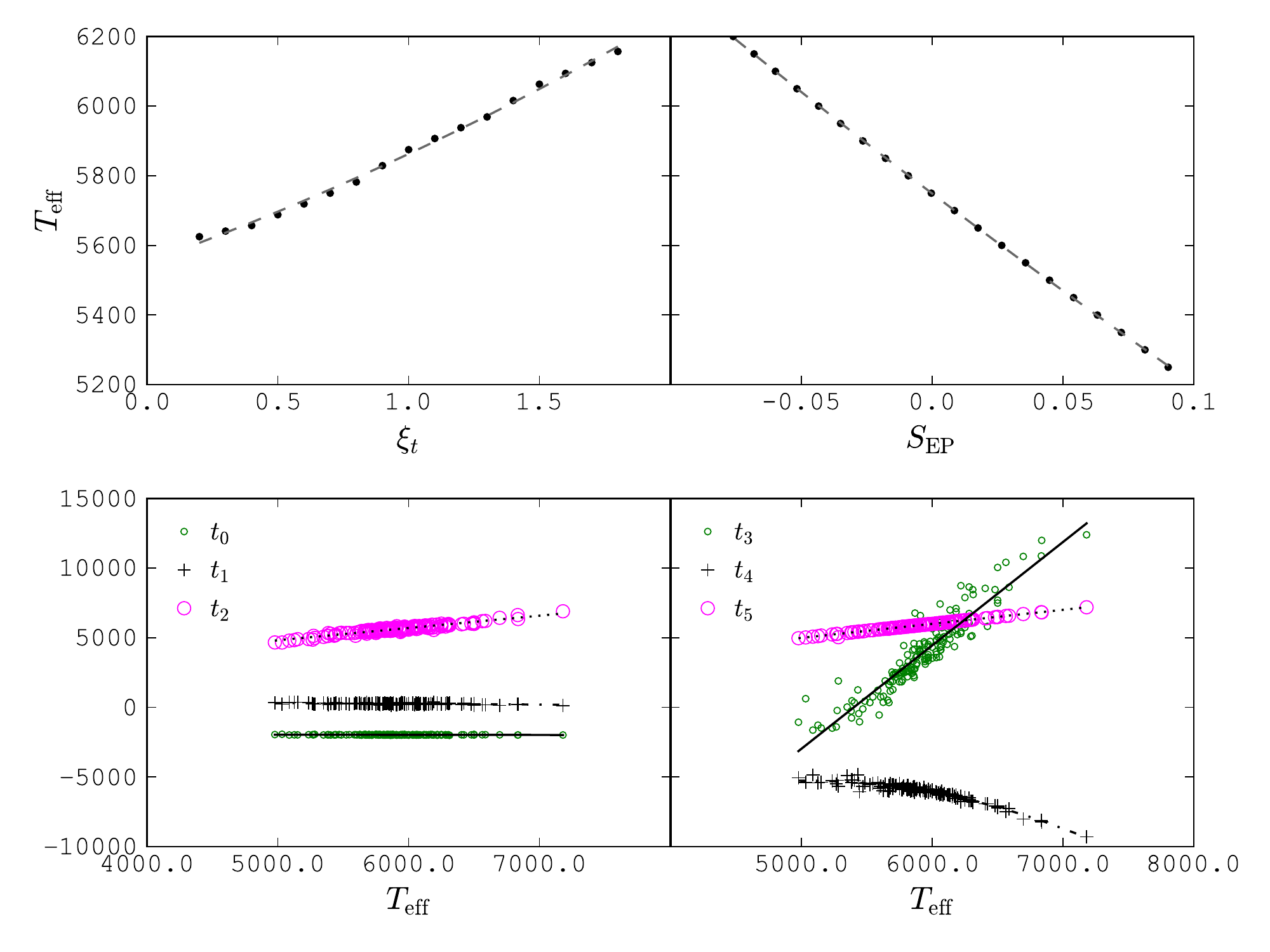}
      \caption{Top panel: Dependence of temperature with microturbulence (left) and $S_{\text{EP}}$ (right), for a Solar spectrum. 
      Bottom panel: Coefficients of the fit between $T$ and $\xi_t$ (left), and between $T$ and $\chi_I$ (right).}
         \label{fig:coefs_err_T}
   \end{figure}

In order to find the first term, we computed the value of the temperature after fixing $\xi_t$ to a specific value. We obtained a quadratic fit, which is shown in the top-left panel of Figure \ref{fig:coefs_err_T}. For the second term, we followed a similar procedure to the one discussed for the microturbulence, and computed the temperature we would obtain as a function of the $\chi_IS_{\text{EP}}$. In this case, the dependence could be fitted by a quadratic curve, and is shown in the top-right panel of Figure \ref{fig:coefs_err_T}. 

As we did before, to find the values for the coefficients of the dependence between $T$ and $\xi_t$, and between $T$ and $S_{\text{EP}}$, we computed the uncertainties for the same sample of stars than in the previous section. The coefficients corresponding to the relation between $T$ and $\xi_t$ are plotted in the bottom-left panel of Figure \ref{fig:coefs_err_T}, and depend on $T$ in the following way:

\begin{align}\label{eq:coefs_T1}
T& = t_0\cdot \xi_t^2 \: +\: t_1\cdot \xi_t \: +\: t_2 \nonumber\\
t_0& = -8.7\times 10^{-3}\cdot T \: +\: 81.74\\
t_1& = -5.4\times 10^{-2}\cdot T \: +\: 580.5 \nonumber\\
t_2& = 0.88\cdot T \: +\: 405.8 \nonumber
\end{align}

For the relation between $T$ and $\chi_I$, the coefficients are plotted in the bottom-right panel of Figure \ref{fig:coefs_err_T}, and have the following dependency with $T$:

\begin{align}\label{eq:coefs_T2}
T& = t_3\cdot S_{\text{EP}}^2 \: +\: t_4\cdot S_{\text{EP}} \: +\: t_5 \nonumber \\
t_3& = 7.4\cdot T \: -\: 4\times 10^4\\
t_4& = -8.77\times 10^{-4}\cdot T^2 \: +\: 8.8\cdot T \: -\: 2.74\times 10^4 \nonumber\\
t_5& = T \: -\: 41 \nonumber
\end{align}

Again we highlight a second way to compute the uncertainty in temperature that is explained in the appendix. We performed it for all our stars but it is used mainly as a cross-check and is again not the preferred value. It appears in the catalogue as \texttt{err\_T2}.

\subsubsection{Metallicity}\label{sec:error_metallicity}

The final value for the metallicity is reached when the average of the individual \FeI abundances matches the one from the input model atmosphere, and will depend on the scatter found in the \FeI abundances. As was mentioned previously, the final metallicity will also depend on the rest of the atmospheric parameters, and in this case, it will depend on the temperature and the surface gravity.
The final expression for the uncertainty in the metallicity will be:

\begin{equation}
\sigma_{\text{[Fe/H]}}^2 \: =\: \left(\left.\frac{\partial \text{[Fe/H]}}{\partial \xi_t}\right|_{\xi_t}\right)^2\, \sigma_{\xi_t}^2 \: +\: \left(\left.\frac{\partial \text{[Fe/H]}}{\partial T}\right|_{T}\right)^2\, \sigma_{T}^2 \: +\: \sigma_{\text{\FeI}}^2,
\end{equation}
where $\partial \text{[Fe/H]}/\partial \xi_t$ and $\partial \text{[Fe/H]}/\partial T$ are evaluated at $\xi_t$ and $T$ derived by our code, respectively, and $\sigma_{\text{\FeI}}$ is the scatter over the abundances of each \FeI lines.

\begin{figure}
   \centering
   \includegraphics[width=8cm]{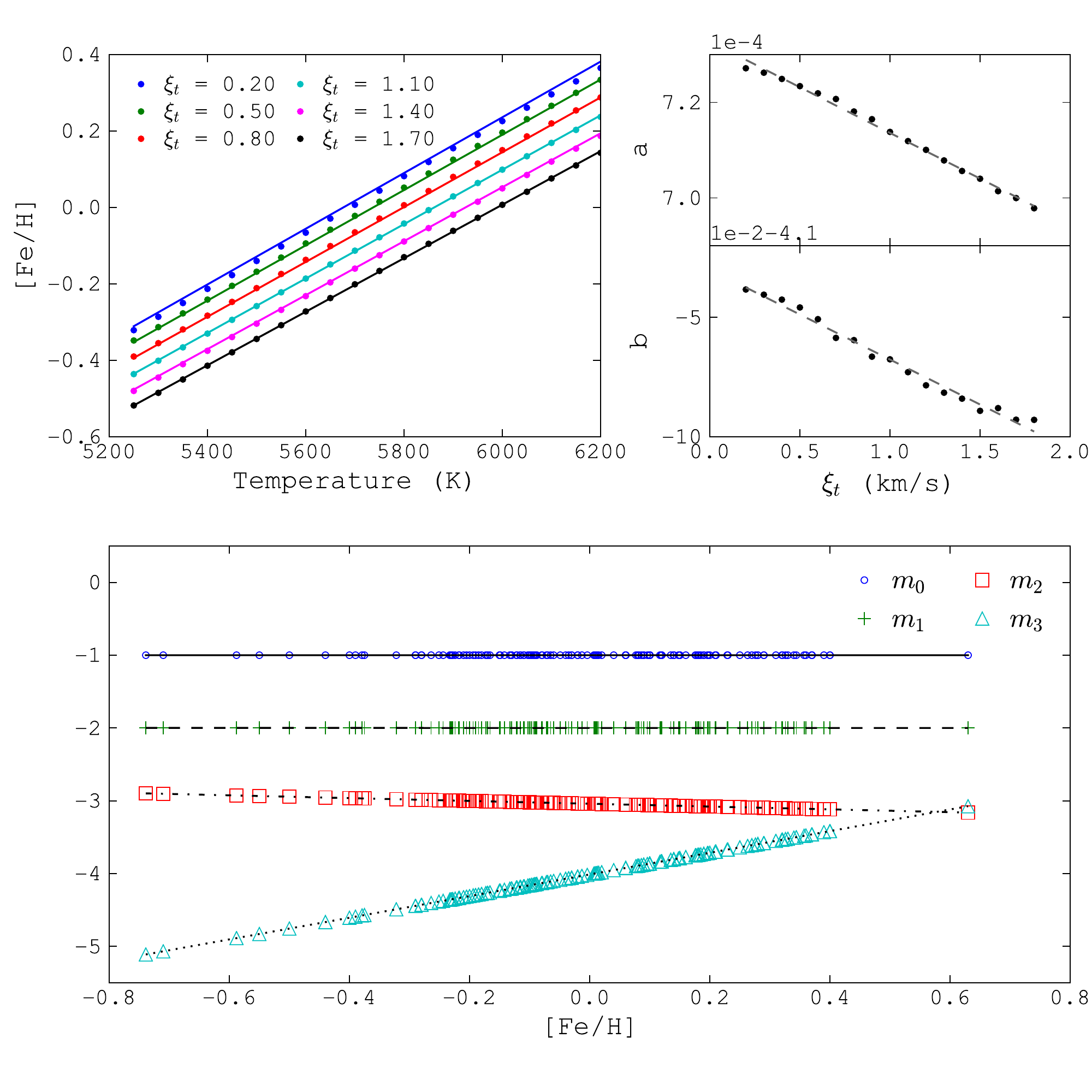}
      \caption{Top panel:  (left) Dependence of metallicity with temperature for several values of microturbulence, (right) dependence of the coefficients  $a$ and $b$ with $\xi_t$, so that [Fe/H] = $a\cdot T + b$.
      Bottom panel: Coefficients of the fit of $[Fe/H]$ with $T$ and $\xi_t$, so that $a = m_0\cdot \xi_t + m_1$ and $b = m_2\cdot \xi_t + m_3$.}
         \label{fig:coefs_err_met}
   \end{figure}

We computed the metallicity for different values of the temperature and microturbulence, and obtained the following relation: 
\begin{equation}
\text{[Fe/H]}\: =\: (m_0\cdot \xi_t + m_1)\cdot T \: +\: (m_2\cdot \xi_t + m_3),
\end{equation}
which can also be seen in the top panels of Figure \ref{fig:coefs_err_met}.

We followed the same procedure than in the previous sections to find the dependence of the fit coefficients with the metallicity. The fits we obtained are plotted in the bottom panel of Figure \ref{fig:coefs_err_met}, and correspond to

\begin{align}\label{eq:coefs_met}
m_0& = 2.02\times 10^{-5}\cdot \text{[Fe/H]} \: -\: 1.8\times 10^{-5} \nonumber \\
m_1& = -9.57\times 10^{-5}\cdot \text{[Fe/H]} \: +\: 7.03\times 10^{-4}\\
m_2& = -0.2\cdot \text{[Fe/H]} \: -\: 0.04 \nonumber \\
m_3& = 1.49\cdot \text{[Fe/H]} \: -\: 4.01 \nonumber
\end{align}

\subsubsection{Surface gravity}\label{sec:error_logg}

The surface gravity depends on the average abundance obtained for the \FeII lines, as well as on the final temperature from the iterative process. Therefore, the uncertainty in \logg will be given by:

\begin{equation}
\sigma_{\log \text{g}}^2 \: =\: \left(\left.\frac{\partial \log \text{g}}{\partial T}\right|_{T}\right)^2\, \sigma_{T}^2 \: +\: \left(\left.\frac{\partial \log \text{g}}{\partial\, \text{\FeII}}\right|_{\,\text{\FeII}}\right)^2\, \sigma_{\text{\FeII}}^2,
\end{equation}
where $\partial \log \text{g}/\partial T$ and $\partial \log \text{g}/\partial\, \text{\FeII}$ are evaluated for $T$ and [Fe/H] found by our code, respectively.

\begin{figure}
   \centering
   \includegraphics[width=8cm]{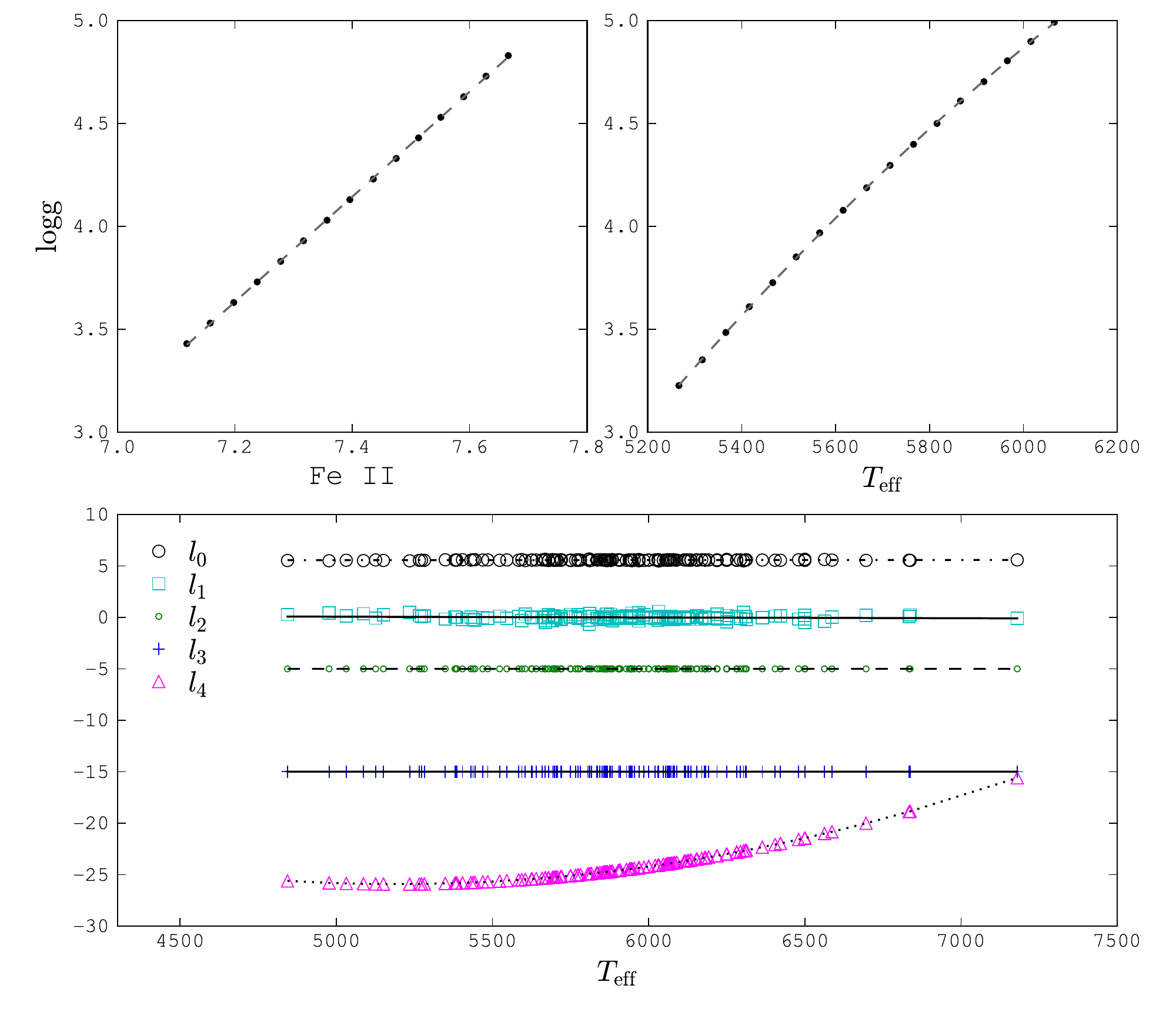}
      \caption{Top panel:  (left) Dependence of surface gravity (\logg) with individual \FeII abundance, (right) dependence of \logg with temperature.
      Bottom panel: Coefficients of the fits between \logg with \FeII, and between \logg with \teff, as shown in Equation \ref{eq:coefs_logg}. A vertical offset was applied to each coefficient for plotting purposes.}
         \label{fig:coefs_err_logg}
   \end{figure}

As for the previous parameters, we found that $\; \log \text{g} \: =\: l_0\cdot \text{\FeII} + l_1,\;$ and $\:\log g \: =\: l_2\cdot T^2 + l_3\cdot T + l_4.$ Both of these relations are plotted in the top panel of Figure \ref{fig:coefs_err_logg}. The coefficients $l_0, l_1, l_2, l_3$ and $l_4$ all depend on the temperature and are shown in the bottom panel of Figure \ref{fig:coefs_err_logg}.  The relations and coefficient values are as follows:

\begin{align}\label{eq:coefs_logg}
l_0& = 8.3\times 10^{-5}\cdot T \: +\: 2.1 \nonumber \\
l_1& = -5.36\times 10^{-4}\cdot T \: - \: 11.8 \nonumber \\
l_2& = 9.2\times 10^{-10}\cdot T \: -\: 5.83\times 10^{-6}\\
l_3& = -2.23 \times 10^{-6}\cdot T \: +\: 2.1\times 10^{-2} \nonumber \\
l_4& = 2.61\times 10^{-6}\cdot T^2 \: -\: 2.71\times 10^{-2}\cdot T \: +\: 44.4 \nonumber
\end{align}

\subsection{Mass, age, and radius}\label{sec:mass_age_plogg}

\begin{figure}
   \centering
   \includegraphics[width=8.5cm]{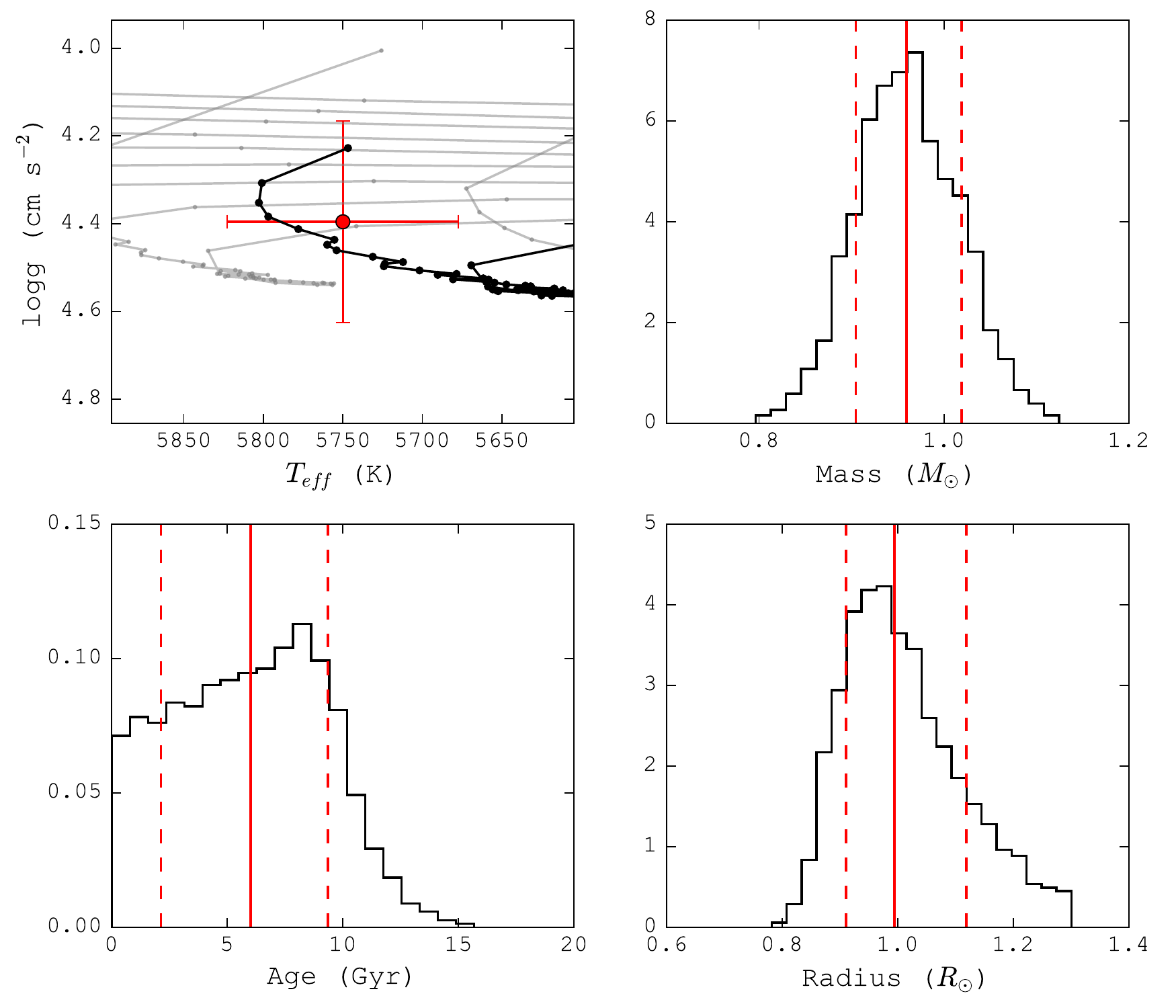}
      \caption{Results obtained for one of the solar spectrum used in SPECIES.
      Top panel: \logg-\teff space diagram. The red point represents the position of the Sun, with the final values obtained using our code (T = 5776 $\pm$ 73 K, [Fe/H] = 0.0$\pm$ 0.1 dex, \logg = 4.5 $\pm$ 0.2 cm\,s$^{-2}$). The dotted lines represent the evolutionary tracks for stars with masses from 0.5 to 1.5 $M_{\odot}$, and [Fe/H] = 0.0. The black line is the track for a 1.0 $M_{\odot}$ star. Each point in the lines represent a different age.
       Bottom panels: Distribution of the mass, age, and radius, for one of our HARPS solar spectra. The red dashed-solid-dashed lines represent the (16, 50, 84) quantiles, respectively. The final values obtained are M = 0.97 $\pm$ 0.06 $M_{\odot}$, Age = 4.5 $\pm$ 3.5 Gyr, and R = 0.99 $\pm$ 0.14 $R_{\odot}$.
       The results for this spectrum are also listed in Table~\ref{tab:solar_values} as \texttt{sun03}.}
         \label{fig:sun_mass}
   \end{figure}
   
\begin{figure}
   \centering
   \includegraphics[width=8.5cm]{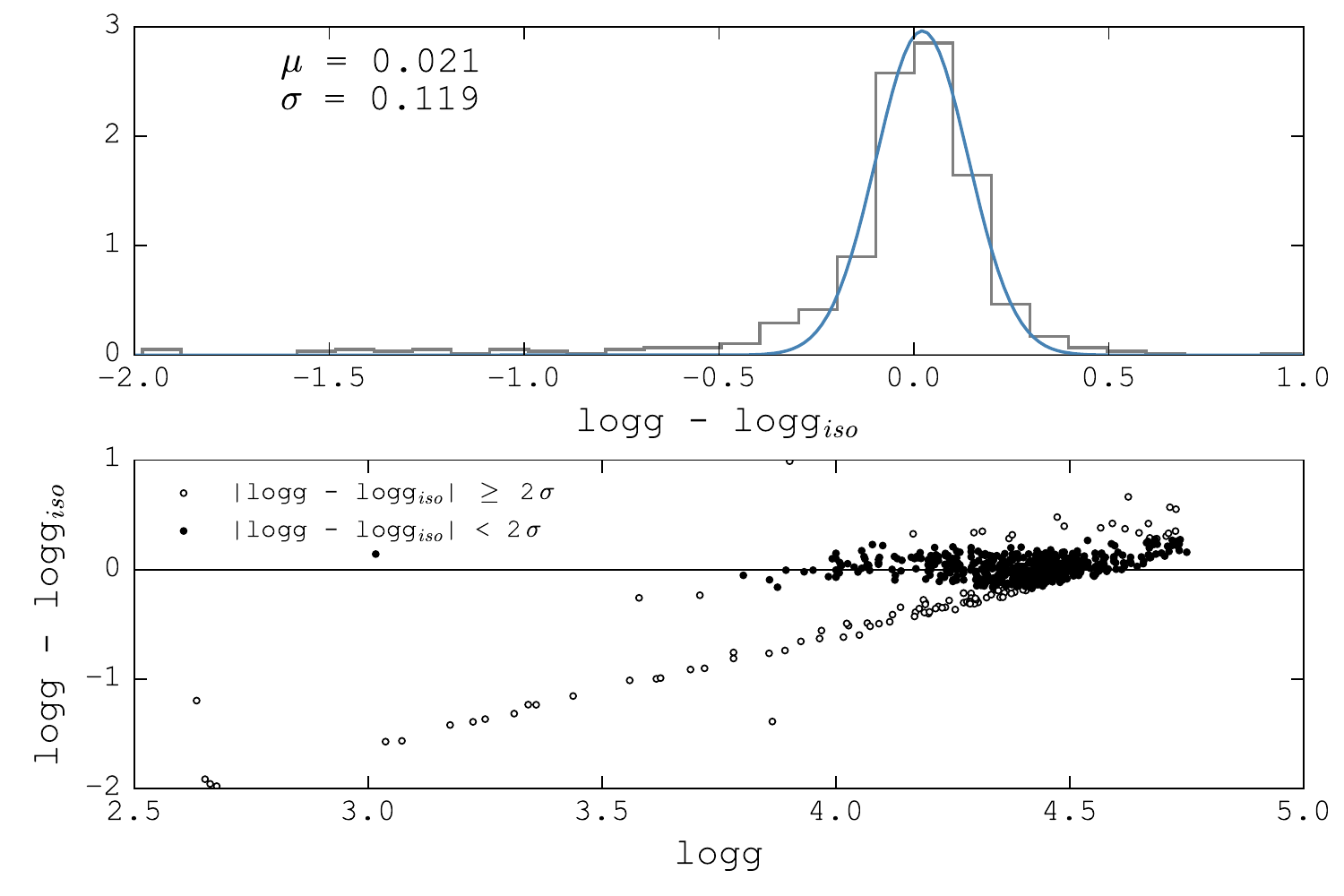}
      \caption{Top panel: histogram of the difference between \logg and \isologg. The blue line corresponds to the Gaussian distribution fit performed, with mean and sigma equal to $\mu = -0.004$ and $\sigma = 0.112$, respectively.
      Bottom panel: Difference between \logg and \isologg vs \logg. Filled circles are the stars laying within $2\sigma$ of the Gaussian fit performed to the histogram. Empty circles are the points laying beyond the $2\sigma$ level, for which the computation was performed a second time, but setting \logg = \isologg.}
         \label{fig:compare_logg}
   \end{figure}

SPECIES uses the python package \texttt{isochrones}\footnote{\url{https://github.com/timothydmorton/isochrones}} \citep{python_isochrones} in order to derive the mass, age, and radius for each star. It uses the previously derived [Fe/H], \logg and \teff, and the MESA Isochrones and Stellar Tracks \citep[MIST,][]{Dotter2016}. The package performs a MCMC fit, with priors given by the [Fe/H], \logg and \teff input values, plus their uncertainties. The samples generated correspond to the mass, age, and radius, evaluated at each chain link. The resulting values will be given by the median and standard deviation of the posterior distributions. It is also possible to input photometric data as priors. This data corresponds to apparent magnitude in several bands, as well as parallax in mas, and can either be given by the user, or retrieved from catalogues. The list of catalogues used, as well as the allowed magnitudes, are given in section \ref{sec:stellar_params}. Figure \ref{fig:sun_mass} shows an example posterior distribution obtained for one of our solar spectra.

Another value measured from the isochrones interpolation is the surface gravity a star would have for the mass, age, and radius derived previously. This quantity, which we will referred to as \isologg, should match the input \logg (referred to as the spectroscopic \logg within the text), and it does so in most cases, but we do find some exceptions. When using SPECIES on a sample of dwarf stars (which will be further explained in section \ref{sec:comparison}), we find that for some cases the value of \logg is $<4.0$. We also find better agreement between \isologg and the surface gravity from the literature (for the description of the catalogues used for the comparison, see section \ref{sec:comparison}), than when using the spectroscopic \logg (section~\ref{sec:comparison_atmospheric_parameters}, Figure \ref{fig:compare_logg_and_literature}). This leads us to conclude that \isologg is a better tracker of the true surface gravity than the \logg obtained from the iterative process explained in section \ref{sec:atmospheric_parameters}. In order to incorporate this result into the computation, we studied the distribution of \logg-\isologg (Figure \ref{fig:compare_logg}), and we found that it follows a Gaussian distribution centred around zero, and with a standard devitation equal to 0.11. Most of the stars are found contained within 2$\sigma$ of this distribution. For the cases when the discrepancy between both \logg measurements is larger than $2\sigma$, which translates into 0.22 dex, we perform a second iteration to derive the atmospheric parameters, following the same procedure than in section \ref{sec:atmospheric_parameters}, but setting \logg $=$ \isologg as the correct value. This option can be disabled when running SPECIES (section \ref{sec:results}).

It is important to mention that \isologg not only seems to provide a better estimate of the true surface gravity of a dwarf star, but it also agrees for evolved stars. This is shown in section \ref{sec:Gaia}, where we use SPECIES to derive the parameters for a sample of dwarf and evolved stars, and we find agreement between our values and those from the literature.

\subsection{Chemical abundances}\label{sec:chemical_abundances}

SPECIES allows the computation of chemical abundances for 11 elements (Na, Mg, Al, Si, Ca, Ti, Cr, Mn, Ni, Cu, Zn) and to test the level to which the code performs these measurements, we compared the SPECIES values with the solar values already studied in the literature, where we used the values for \teff, \logg, [Fe/H] and $\xi_t$ computed previously. We used the line list used in \citet{ivanyuk2017} (we refer the reader to that work for a detailed description of the line selection), and the solar abundances from \citet{asplund2009}. The EW were measured using the ARES code, and for the analysis only lines with 10 < EW $\leq$ 150 were used, as explained before.

For each element, we considered only the lines for which the individual abundance was within 1.5 $\sigma$ of the mean value. This was done in order to avoid the lines that deviate too much (more than 2 dex in some cases) from the abundance given by the rest of the lines for that element.
The final abundance for each element was computed as the average abundance from each individual line, after the sigma-clipping, and its uncertainty was taken as the standard deviation over the average. We weighted the abundance of each line as 1/$\sigma_{\text{EW}}$. When only one line per element is available, the uncertainty is taken to be the average error for the other elements used, and no sigma-clipping was performed.

For all the elements, except for Ti, only lines from neutral species were used. In the case of Ti, we list the abundances obtained for both Ti\,{\sc i}\, and Ti\,{\sc ii}. We also include in the output the abundances for \FeI and \FeII.
%, for which we use Ti\,{\sc i}\, and Ti\,{\sc ii}\, lines, we add a flag value for when the average abundance from each stage differs by more than 1.0 dex, and the abundance given by Ti\,{\sc i}\, is used as the final Ti value. This flag is explained in Section \ref{sec:results}. In the case of Fe, we computed its abundance using the lines shown in Table \ref{tab:linelist_ab} and compared its value to the metallicity found previously. In cases where these do not match, a flag is used (also explained in Section \ref{sec:results}). 

Currently, it is not possible to quickly modify the line list, nor add new species to the computation.

\subsection{Macroturbulence and rotational velocity}\label{sec:broadening}

In order to compute the macroturbulence ($v_{mac}$) and rotational ($v\sin i$) velocities, we followed the procedure described in \citet{dosSantos2016}. It consist of measuring both quantities individually for five different absorption lines, and then compares the results to those from the Sun. The lines used, as well as their atomic characteristics, are mentioned in Table \ref{tab:lines_vsini}. 

\begin{table}
\caption{Line list used to measure the rotational velocity for each star. $v_{macro, \odot}$ is the macroturbulence velocity found for the Sun in \citet{dosSantos2016}}
\centering
\begin{tabular}{cccccc}
Wavelength & Z & Ion & Exc. pot. & $\log(gf)$ &$v_{macro, \odot}$ \\
(\r{A}) & & & (eV) & & (km s$^{-1}$)\\
\hline
6027.050 & 26 & \FeII & 4.076 & -1.09 & 3.0\\
6151.618 & 26 & \FeI & 2.176 & -3.30 & 3.2\\
6165.360 & 26 & \FeI & 4.143 & -1.46 & 3.1\\
6705.102 & 26 & \FeI & 4.607 & -0.98 & 3.6\\
6767.772 & 28 & Ni\,{\sc i}\, & 1.826 & -2.17 & 2.9
\end{tabular}
\label{tab:lines_vsini}
\end{table}

\begin{figure*}
\centering
\includegraphics[width=17cm]{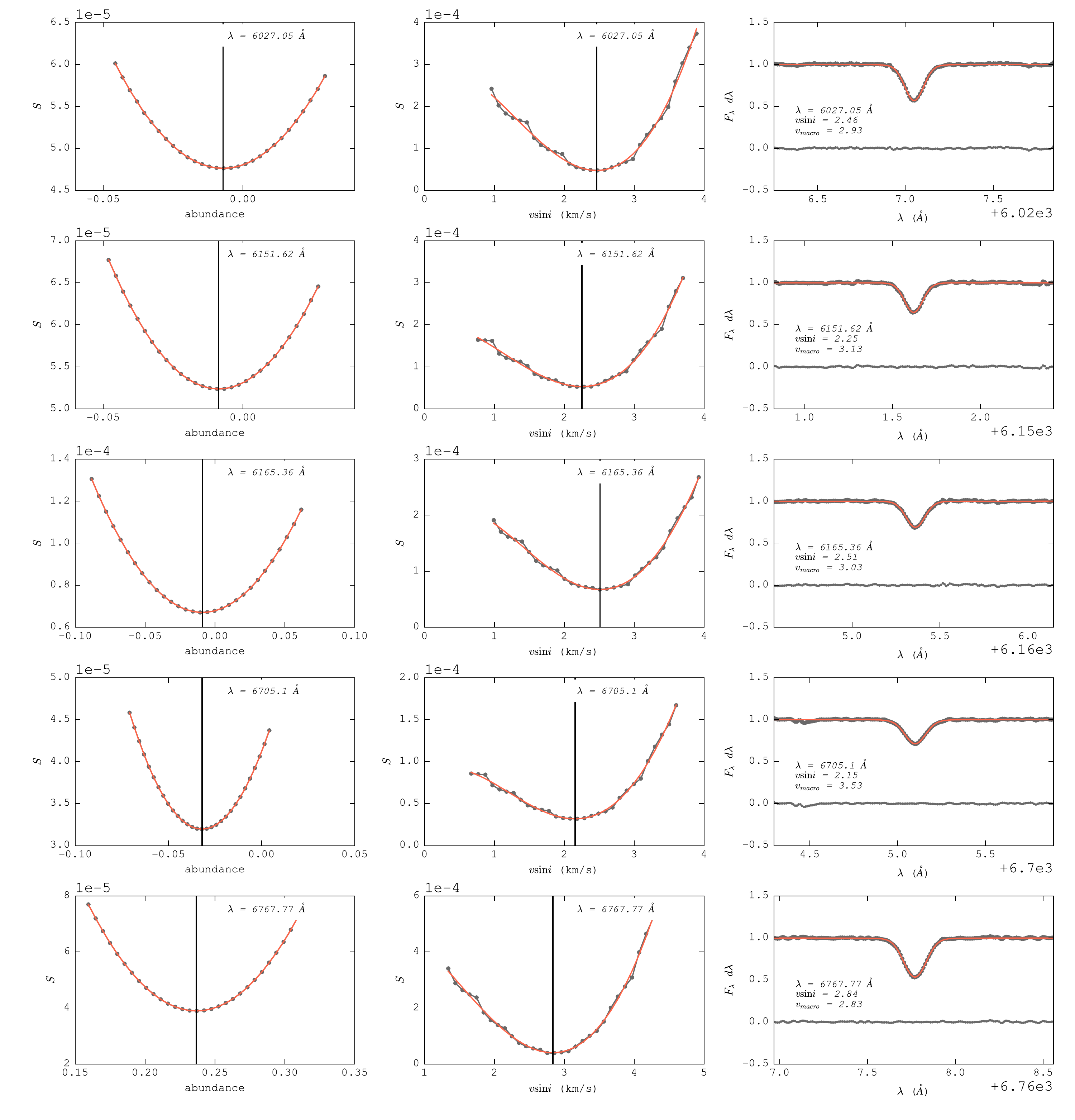}
\caption{Change of $S$ versus abundance (left panels) and rotational velocity (middle panels), for each line of the solar spectra. The red lines represent the cubic spline fit performed over the data, and the vertical line shows the values where the minimum of $S$ is reached. The right panels show the line profiles, along with the final fit (red line) combining the instrumental profile, macroturbulence and rotational velocity. The horizontal black line around zero represents the residuals of the fit.}
\label{fig:vsini_vmac}
\end{figure*}

The macroturbulent velocity was obtained from \citet[Eq. 1]{dosSantos2016}:

\begin{equation}\label{eq:vmac}
\begin{aligned}
v_{macro,\lambda} = {} v^{\odot}_{macro,\lambda} & - 0.00707\, T + 9.2422\times 10^{-7} T^2 \\
&+10.0 + k_1(\log \text{g} - 4.44) + k_2
\end{aligned}
\end{equation}

where $v^{\odot}_{macro, \lambda}$ are the velocities obtained for each line in the solar spectra, shown in Table \ref{tab:lines_vsini}. $k_1$ and $k_2$ are constants equal to -1.81 $\pm$ 0.26 and -0.05 $\pm$ 0.03, respectively. All the quantities mentioned were computed by \citet{dosSantos2016}. This relation is also very similar to the one used in \cite{spocs2005}. The uncertainty in $v_{mac,\lambda}$ for each line is given by:

\begin{equation}\label{eq:err_vmac}
\begin{aligned}
\sigma^2_{v_{mac, \lambda}} = \sigma^2_{v^{\odot}_{mac, \lambda}} + & (1.848 \times 10^{-6}\, T - 0.00707)^2 \sigma_{T}^2 \\
& + k_1^2 \sigma_{\log \text{g}}^2 + (\log g - 4.44)^2\sigma_{k_1}^2 + \sigma_{k_2}^2,
\end{aligned}
\end{equation}

where the error in $v^{\odot}_{mac, \lambda}$ is reported to be $\pm$ 0.1 \kms. The temperature and surface gravity, along with their uncertainties, are the ones produced by SPECIES. The final $v_{mac}$ corresponds to the average of the individual results, weighted by their uncertainties.

The rotational velocity for each line was obtained by comparing the line profiles with synthetic ones produced by the MOOG driver \textit{synth}. The driver receives a model atmosphere, obtained from the ATLAS 9 grids and the atmospheric values found by SPECIES, and the line abundance, found by measuring the EW of the line with ARES (following the same settings described in section \ref{sec:EW}), and using the MOOG driver \textit{abfind}. It also receives the macroturbulent velocity found previously, and the width of the line produced by the instrument resolution. The synthetic profile is then convolved with a rotational profile \citep{gray2005} for a certain $v\sin i$ value. This was performed using the PoWeRS\footnote{\url{https://github.com/RogueAstro/PoWeRS}} code, which was modified and optimized in order to fit into SPECIES. The code creates grids of different values for $v\sin i$ and line abundance, and finds the values (abundance, $v\sin i$) for which the synthetic profile best matches the original line profile. This is measured by the quantity $S$, which measures the goodness of the fit, and is given by
$$ S = \frac{1}{N}\sum_{i = 0}^N (y_{o,i} - y_{s,i})^2,$$
with $i = \{0,...,N\}$ the number of points in the line profile, which is considered from $\lambda - 0.5$ to $\lambda + 0.5$. $y_o$ represents the measured line profile, and $y_s$ the synthetic one. The code performs a cubic spline fit to the value of $S$ vs. the abundance and rotational velocity, and finds the values for which the minimum of $S$ is reached.
A minimum of four iterations are performed, refining the abundance and velocity grids by shifting the grid centre to match the values with the best goodness of fit, and making the delta between grid points smaller. This is done in order to obtain the most precise results (minimum of $S$). In Figure \ref{fig:vsini_vmac} we show the changes in $S$ for a grid of line abundances and rotational velocities, and the final fits obtained for each line, for a solar spectrum. The final $v\sin i$ corresponds to the average of the individual values found for each line, and its uncertainty is estimated as $\sigma^2_{v\sin i} = \sum (S^2_{\lambda} + \sigma^2_{v_{mac},\lambda})$, where the sum is performed over all the lines.

The stellar parameters obtained for the Sun, using 10 different solar spectra taken with the HARPS instrument\footnote{\url{https://www.eso.org/sci/facilities/lasilla/instruments/harps/inst/monitoring/sun.html}}, are listed in Table \ref{tab:solar_values}. The final values found, after performing a weighted average with the S/N of each solar spectrum, are T = 5754.2 $\pm$ 23.3 K, [Fe/H] = -0.02 $\pm$ 0.02 dex, \logg = 4.38 $\pm$ 0.05 cm s$^{-2}$, $\xi_t$ = 0.68 $\pm$ 0.11 \kms, v$_{mac}$ = 3.15 $\pm$ 0.06 \kms, v$\sin$i = 2.35 $\pm$ 0.27 \kms, mass = 0.96 $\pm$ 0.01 $M_{\odot}$, radii = 1.00 $\pm$ 0.01, and age = 6.17 $\pm$ 0.7 Gyr. 
%These values agree with what is found in the literature \citep[e.g.]{andreasen2016, bouvier2010, pavlenko2012} for the Sun.

\begin{table*}
\caption{Stellar parameters found for a sample of Solar spectra, taken using HARPS.}
%The last column in the signal-to-noise of the spectra, computed as the average of the S/N for each echelle order.}
\centering
\begin{tabular}{cccccccccc}
Name & [Fe/H] & Temperature & $\log$g & $\xi_t$ & Mass & Radius & Age & v$\sin$i & $v_{mac}$ \\
\hline
ceres01 & -0.0 $\pm$ 0.1 & 5766 $\pm$ 37 & 4.4 $\pm$ 0.2 & 0.67 $\pm$ 0.03 & 0.97 $\pm$ 0.04 & 1.02 $\pm$ 0.14 & 6.0 $\pm$ 3.7 & 2.2 $\pm$ 0.2 & 3.2 $\pm$ 0.2 \\
ceres02 & -0.0 $\pm$ 0.1 & 5778 $\pm$ 32 & 4.4 $\pm$ 0.2 & 0.85 $\pm$ 0.03 & 0.97 $\pm$ 0.04 & 1.00 $\pm$ 0.12 & 5.8 $\pm$ 3.7 & 2.0 $\pm$ 0.2 & 3.1 $\pm$ 0.2 \\
ceres03 & -0.0 $\pm$ 0.1 & 5707 $\pm$ 63 & 4.3 $\pm$ 0.2 & 0.78 $\pm$ 0.05 & 0.94 $\pm$ 0.05 & 1.01 $\pm$ 0.16 & 7.7 $\pm$ 4.7 & 2.1 $\pm$ 0.2 & 3.2 $\pm$ 0.2 \\
moon01 & 0.0 $\pm$ 0.1 & 5782 $\pm$ 48 & 4.4 $\pm$ 0.2 & 0.66 $\pm$ 0.04 & 0.98 $\pm$ 0.05 & 1.01 $\pm$ 0.14 & 5.4 $\pm$ 3.6 & 2.9 $\pm$ 0.3 & 3.1 $\pm$ 0.3 \\
ganymede01 & -0.0 $\pm$ 0.1 & 5782 $\pm$ 57 & 4.5 $\pm$ 0.2 & 0.84 $\pm$ 0.05 & 0.97 $\pm$ 0.05 & 1.00 $\pm$ 0.13 & 5.4 $\pm$ 3.6 & 2.0 $\pm$ 0.3 & 3.2 $\pm$ 0.3 \\
sun01 & -0.1 $\pm$ 0.1 & 5735 $\pm$ 37 & 4.3 $\pm$ 0.2 & 0.70 $\pm$ 0.03 & 0.94 $\pm$ 0.04 & 1.00 $\pm$ 0.13 & 7.0 $\pm$ 4.3 & 2.3 $\pm$ 0.2 & 3.2 $\pm$ 0.2 \\
sun02 & 0.0 $\pm$ 0.1 & 5766 $\pm$ 62 & 4.4 $\pm$ 0.3 & 0.44 $\pm$ 0.04 & 0.97 $\pm$ 0.05 & 1.01 $\pm$ 0.16 & 6.0 $\pm$ 3.8 & 2.4 $\pm$ 0.3 & 3.1 $\pm$ 0.3 \\
sun03 & -0.0 $\pm$ 0.1 & 5750 $\pm$ 73 & 4.4 $\pm$ 0.2 & 0.66 $\pm$ 0.06 & 0.96 $\pm$ 0.06 & 1.00 $\pm$ 0.14 & 6.0 $\pm$ 3.9 & 2.4 $\pm$ 0.2 & 3.1 $\pm$ 0.2 \\
sun04 & -0.0 $\pm$ 0.1 & 5735 $\pm$ 63 & 4.3 $\pm$ 0.3 & 0.52 $\pm$ 0.04 & 0.96 $\pm$ 0.05 & 1.00 $\pm$ 0.16 & 6.5 $\pm$ 4.2 & 2.3 $\pm$ 0.3 & 3.2 $\pm$ 0.3 \\
sun05 & -0.0 $\pm$ 0.1 & 5735 $\pm$ 39 & 4.4 $\pm$ 0.3 & 0.58 $\pm$ 0.03 & 0.95 $\pm$ 0.04 & 0.99 $\pm$ 0.14 & 6.5 $\pm$ 4.2 & 2.4 $\pm$ 0.2 & 3.1 $\pm$ 0.2 \\

\end{tabular}
\label{tab:solar_values}
\end{table*}

\section{Results obtained with SPECIES}\label{sec:results}

Currently, the SPECIES catalogue has 72 columns with the stellar parameters, described in section \ref{sec:column_description}. In order to test the accuracy of the results obtained with SPECIES, we derived the parameters for a sample of 584 dwarf stars, targeted by the HARPS GTO projects \citep[parameters derived in ][]{sousa2008} and the Calan-Hertfordshire Extrasolar Planet Search (CHEPS) program \citep[stellar parameters derived in \citealt{ivanyuk2017}]{jenkins2009}. They cover a wide range in temperature and metallicity, from 4300 to 6500 K, and -0.9 to 0.6 dex, respectively. We selected the highest S/N spectra taken with HARPS (given that it is the highest resolution instrument currently accepted by SPECIES), and compare them with what is obtained with photometric relations and other catalogues in the literature. Unless stated otherwise, the results presented here were computed using the temperature from photometry and/or fixing $\xi_t = 1.2$ km\,s$^{-2}$ when convergence is not reached in the atmospheric parameters, and setting \isologg as the correct value for the surface gravity when their differences are larger than 0.22 dex. If different options were used in the computation, it will be specified within the text and in the captions of the figures and/or tables.

A sample of the catalogue is shown in Table \ref{tab:sample_table}.

\subsection{Comparison with photometric relations}\label{sec:compare_photometry}

\begin{figure}
   \centering
   \includegraphics[width=9cm]{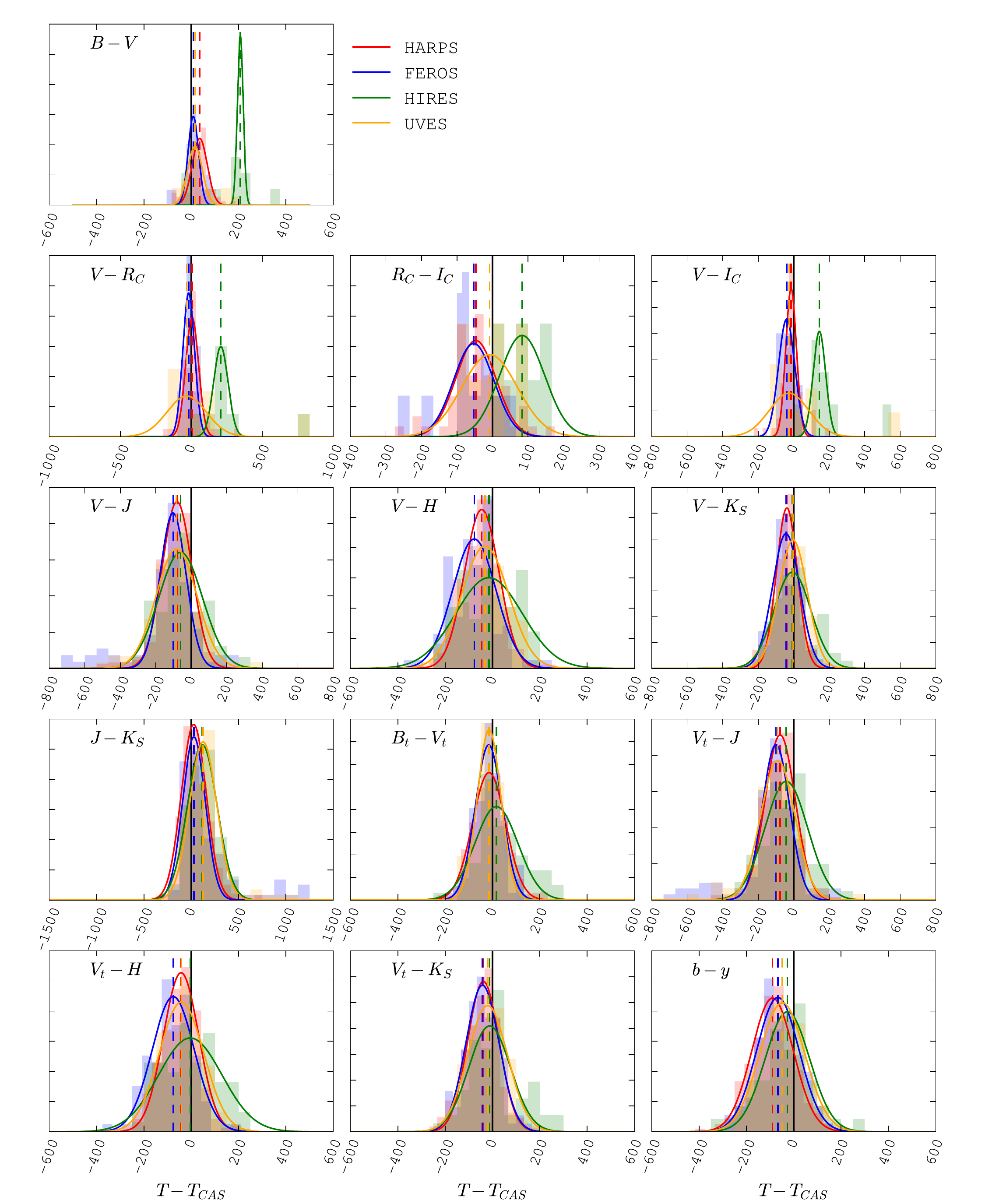}
      \caption{Histograms of the difference between the temperature computed by SPECIES, and those using the color relations from C10. The lines correspond to Gaussian distributions adjusted to the histograms, with mean values listed in Table \ref{tab:comparison_casagrande}. Each color represents a different instrument: red for HARPS, blue for FEROS, green for HIRES, and orange for UVES. The different panels show the temperatures computed using different photometric colors, mentioned in Section \ref{sec:stellar_params}.}
         \label{fig:comparison_casagrande}
   \end{figure}

\begin{table}
\caption{Differences between the temperatures (in K) derived by SPECIES, and from using the color relations from C10. Values inside the parenthesis are the number of points used for each color and instrument.}
\label{tab:comparison_casagrande}
\centering
\begin{tabular}{lllll}
Color & HARPS & FEROS & HIRES & UVES\\
\hline
$B-V$ & 35.5 (71) & 8.3 (19) & 206.9 (9) & 15.8 (17) \\
$V-R_C$ & 5.6 (59) & -18.4 (10) & 207.8 (8) & -33.0 (8) \\
$R_C-I_C$ & -47.1 (59) & -53.1 (11) & 83.3 (8) & -8.2 (8) \\
$V-I_C$ & -14.6 (58) & -39.7 (10) & 144.6 (8) & -26.4 (8) \\
$V-J$ & -80.8 (458) & -102.9 (82) & -60.2 (48) & -83.3 (88)  \\
$V-H$ & -45.7 (466) & -77.0 (84) & -14.7 (49) & -31.2 (88) \\
$V-K_S$ & -38.4 (468) & -43.3 (83) & -7.1 (50) & -3.8 (89) \\
$J-K_S$ & 25.2 (562) & 27.3 (94) & 115.4 (61) & 124.7 (91) \\
$B_t-V_t$ & -14.7 (573) & -16.1 (96) & 16.3 (61) & -14.4 (92) \\
$V_t-J$ & -76.3 (561) & -99.8 (94) & -41.7 (60) & -89.8 (91)\\
$V_t-H$ & -43.2 (567) & -76.4 (95) & -4.3 (61) & -41.7 (91) \\
$V_t-K_S$ & -38.6 (570) & -42.4 (95) & -12.1 (62) & -20.7 (92) \\
$b-y$ & -89.2 (458) & -66.8 (85) & -26.7 (47) & -48.2 (89)\\

\end{tabular}
\end{table}

\begin{figure}
   \centering
   \includegraphics[width=9cm]{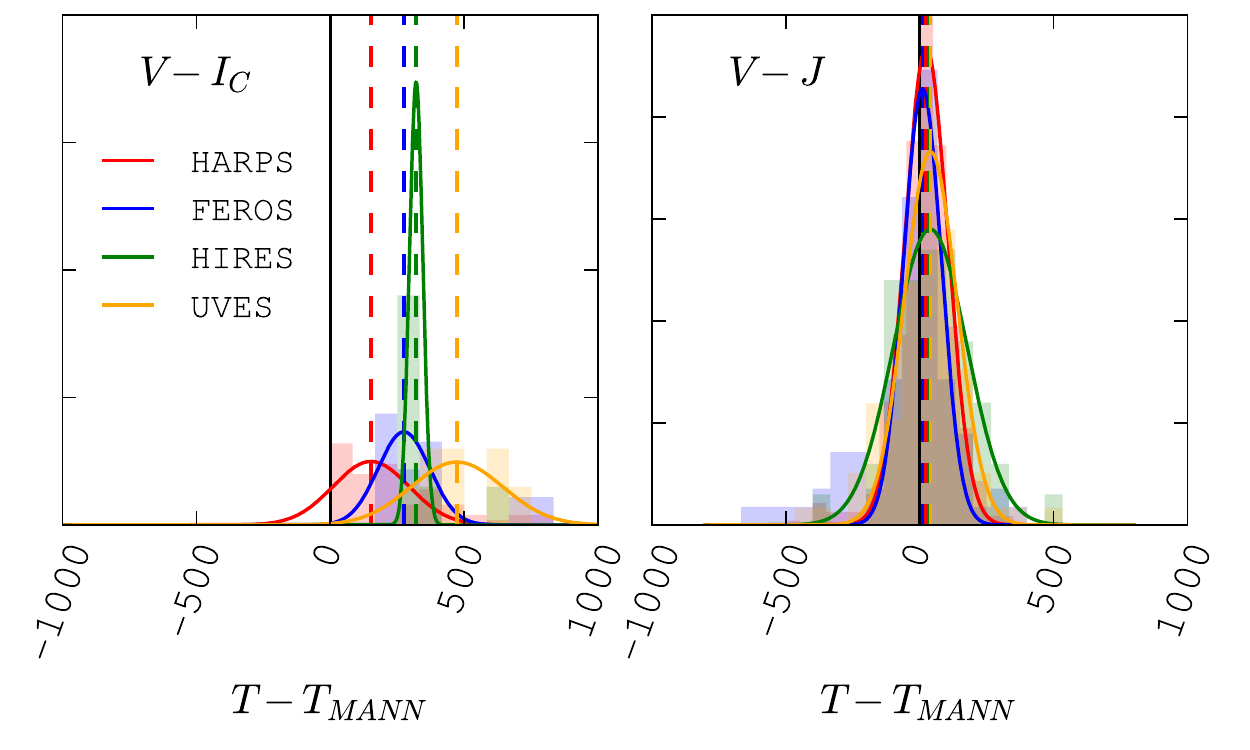}
      \caption{Histograms of the difference between the temperature computed by SPECIES, and by using the color relations from M15. The lines correspond to the Gaussian fits for each distribution, and the colors represent the same instruments as in Figure \ref{fig:comparison_casagrande}. The mean of the Gaussian fits are shown in Table \ref{tab:comparison_mann}.}
         \label{fig:comparison_mann}
   \end{figure}
   
\begin{table}
\caption{Mean of the Gaussian fits (in K) performed on the distribution of differences between the temperatures derived by SPECIES, and from using the color relations from M15. Values inside the parenthesis are the number of points used for each color and instrument. Inside the parenthesis are the number of points used for each distribution.}
\label{tab:comparison_mann}
\centering
\begin{tabular}{lllll}
Color & HARPS & FEROS & HIRES & UVES\\
\hline
$V-I_C$ & 151.8 (60) & 274.5 (11) & 320.1 (8) & 472.2 (8) \\
$V-J$ & 24.5 (462) & 10.6 (84) & 39.4 (50) & 40.7 (88) \\
\end{tabular}
\end{table}

\begin{figure}
   \centering
   \includegraphics[width=9cm]{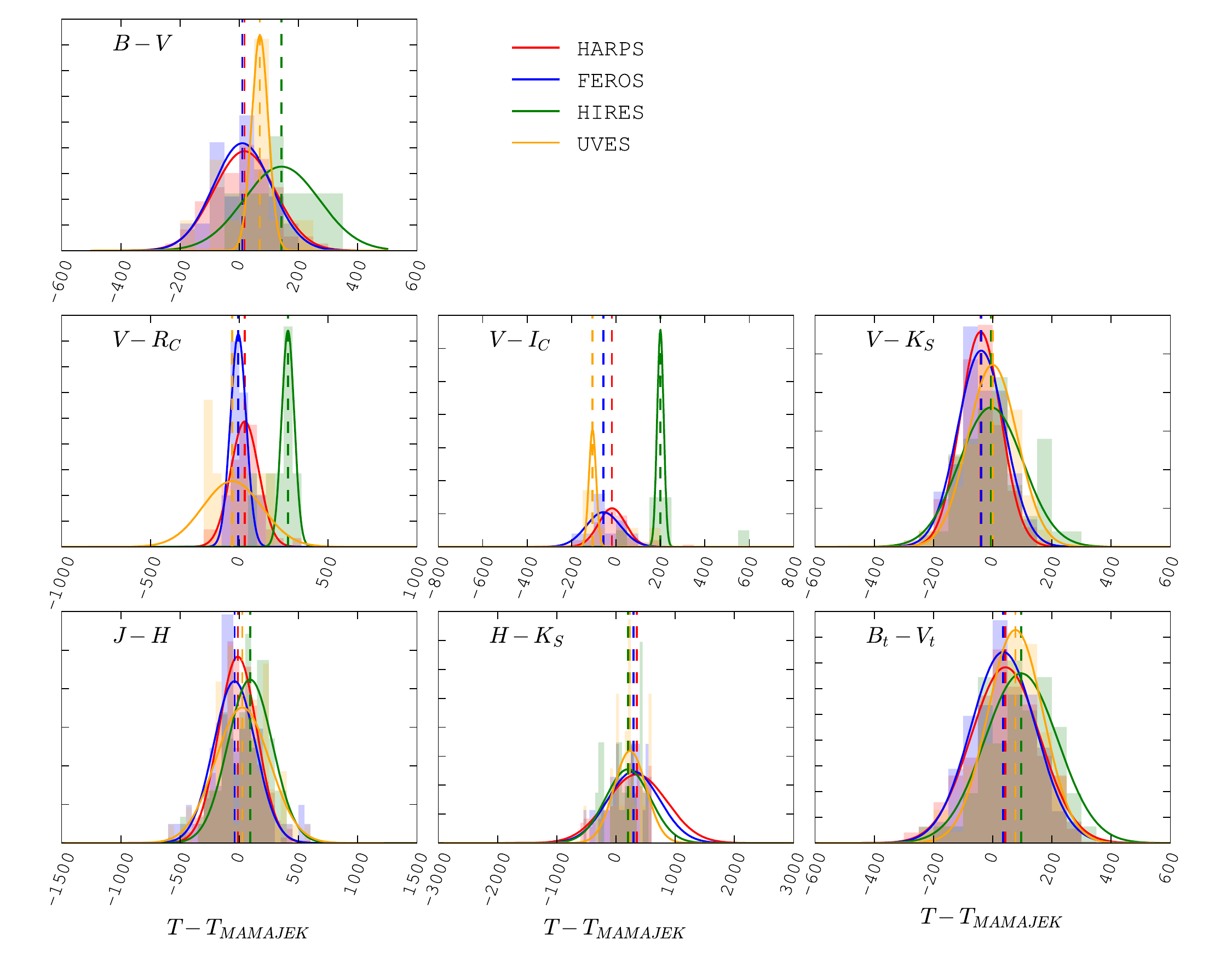}
      \caption{Histograms of the difference between the temperature computed by SPECIES, and those obtained by interpolating through the models of P13. The lines correspond to Gaussian distributions adjusted to the histograms, with mean values listed in Table \ref{tab:comparison_casagrande}. The mean of each Gaussian distribution is listed in Table \ref{tab:comparison_mamajek}. The colors represent the same instruments as in Figure \ref{fig:comparison_casagrande}.}
         \label{fig:comparison_mamajek}
   \end{figure}

\begin{table}
\caption{Mean of the Gaussian distribution (in K) adjusted to the histograms of the different between the temperature from SPECIES, and from interpolating between the models of P13. Values inside the parenthesis are the number of points used for each distribution.}
\label{tab:comparison_mamajek}
\centering
\begin{tabular}{lllll}
Color & HARPS & FEROS & HIRES & UVES\\
\hline
$B-V$ & 17.1 (73) & 10.5 (19) & 142.6 (9) & 69.6 (17)\\
$V-R_C$ & 30.5 (59) & -5.9 (10) & 273.8 (8) & -39.7 (8)\\
$V-I_C$ & -18.1 (59) & -56.9 (10) & 200.0 (8) & -105.6 (8)\\
$V-K_S$ & -41.0 (469) & -38.8 (84) & -5.2 (50) & -0.3 (90) \\
$J-H$ & -12.8 (562) & -40.5 (94) & 91.3 (61) & 25.9 (93) \\
$H-K_S$ & 349.4 (538) & 297.9 (89) & 202.0 (57) & 237.1 (87) \\
$B_t-V_t$ & 43.0 (579) & 35.6 (97) & 96.3 (62) & 76.7 (93) \\
\end{tabular}
\end{table}

The first comparison we performed was to analyse the differences between the temperatures derived from our code and from the photometric relations explained in Section \ref{sec:ini_temperature}. As was mentioned in Section \ref{sec:ini_temperature} and \ref{sec:atmospheric_parameters}, we used the temperatures derived from photometry (for the cases when that information was available) as an initial value for SPECIES, and in the cases when our code could not converge to a valid result for the atmospheric parameters.

In order to check that the temperature from photometry is in agreement with that from SPECIES, we compared both results for our sample of FGK dwarfs stars, observed with HARPS, FEROS, HIRES and UVES. We considered only the cases when it was not necessary to set the temperature from photometry as the correct value to reach converge in the derivation of the atmospheric parameters. For each star we retrieved the photometric information from Vizier, using the catalogues mentioned in Section \ref{sec:stellar_params}, and computed the temperature using the relations from Section \ref{sec:ini_temperature}.

We computed the difference between the temperature from SPECIES, and from using each photometric relation, for spectra taken with different instruments. We then adjusted Gaussian models to the distributions obtained, and define the mean of the model as the offset between each temperature measurement. This was done for every relation described in Section \ref{sec:ini_temperature}, except when using \citet{gonzalez-hernandez2009}, due to the small number of stars in our sample which met the requirement of being classified as giants. Instead of adjusting Gaussian models to the distribution, we just computed the mean of the difference, setting that value as the offset between both temperature measurements.
The results from the comparisons are shown in Figures \ref{fig:comparison_casagrande}, \ref{fig:comparison_mann} and \ref{fig:comparison_mamajek}, and Tables \ref{tab:comparison_casagrande}, \ref{tab:comparison_mann} and \ref{tab:comparison_mamajek}, for the relations from \citet{casagrande2010}, \citet{mann2015} and \citet{Pecaut2013}.

We use these offsets to correct the temperatures obtained using the photometric relations (section \ref{sec:ini_temperature}), to match the values with those obtained by SPECIES. For the case of the relations from \citet{gonzalez-hernandez2009}, we did not perform this comparison because we had no giants in our test.

\begin{figure}
   \centering
   \includegraphics[width=9cm]{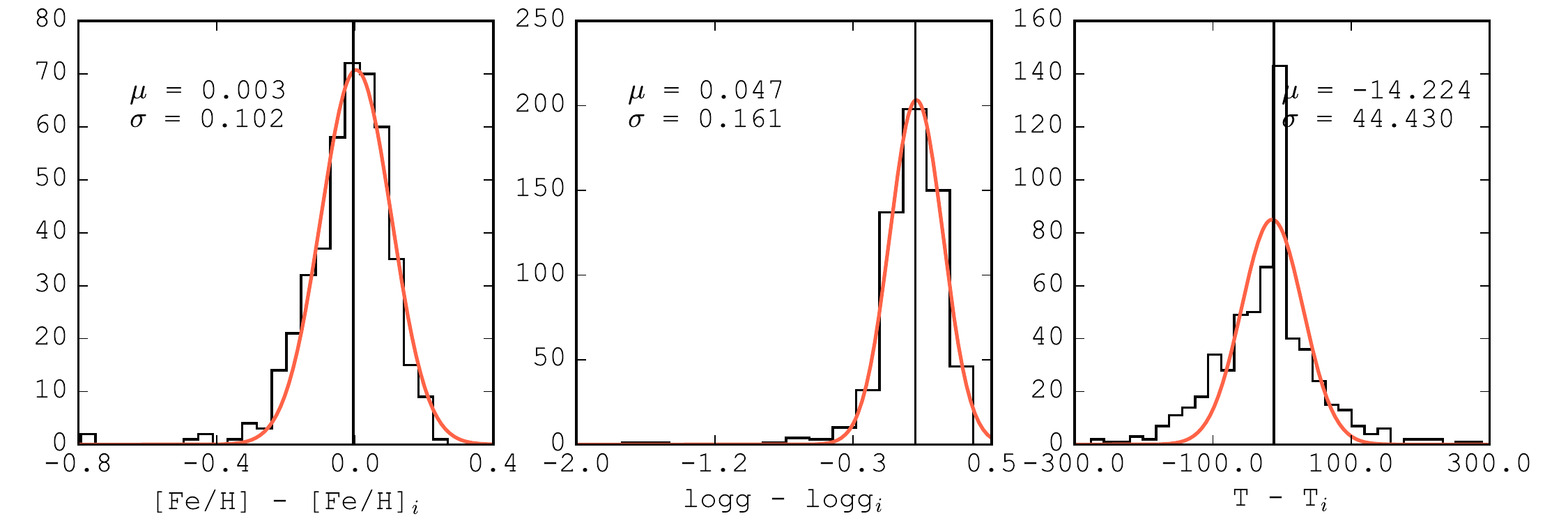}
      \caption{Histograms of the difference between the metallicity (left panel), surface gravity (middle panel) obtained with SPECIES, and from using the photometric relations from sections \ref{sec:initial_met} and \ref{sec:initial_logg} (right panel). The vertical lines correspond (from left to right) to the 16, 50 and 84 percentiles.}
         \label{fig:comparison_met_logg_photometry}
   \end{figure}

Finally, we compared the final metallicity, surface gravity and temperatures obtained with SPECIES, and the initial values derived from photometry in section \ref{sec:initial_met}, \ref{sec:initial_logg} and \ref{sec:ini_temperature}. The distributions we obtain for the difference between both quantities are shown in Figure \ref{fig:comparison_met_logg_photometry}. For all three parameters we find them to be distributed around zero (median of the distributions around -0.0001 and 0.02 for the metallicity and surface gravity, respectively), meaning excellent agreement. We find only a few cases that the values from SPECIES are smaller than from the photometric relations.

\subsection{Comparison with other catalogues}\label{sec:comparison}

In order to test the accuracy of SPECIES, we compared the spectral parameters for a set of stars obtained with our code, with ones listed in the literature. We chose five different catalogues for this comparison, since each had analysed a large sample of stars and they all used differing methods to compute the stellar parameters, providing a robust test of the SPECIES automatic calculations. The samples are briefly described as follows:

   \begin{figure*}
   \centering
            {\includegraphics[width=18cm]{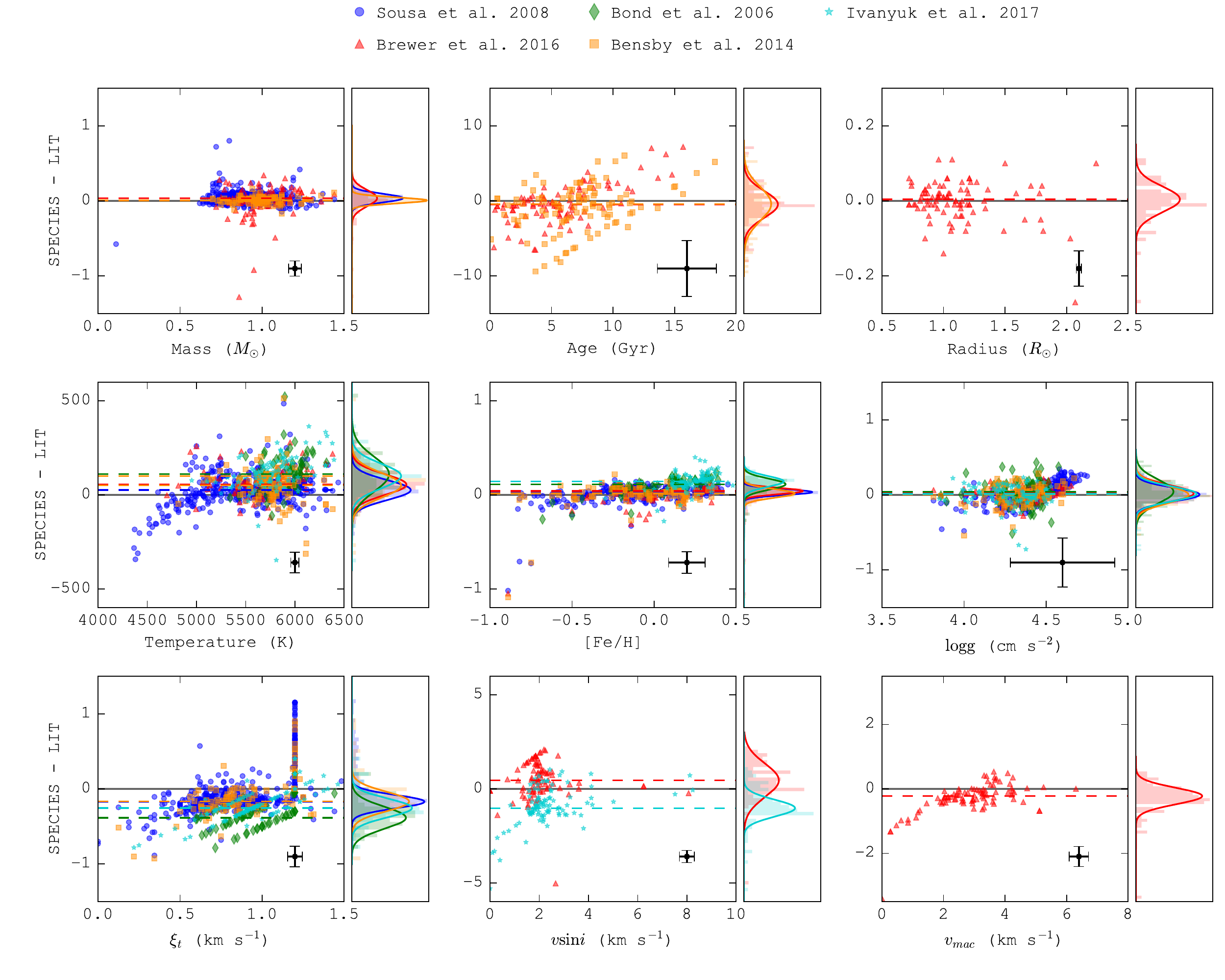}}
      \caption{Comparison between the values for different parameters obtained with our code and 
      from literature. 
      The y-axis in the plots correspond to the difference between both measurements,
      represented by different symbols and colors:
      blue circles to \cite{sousa2008}, red triangles to \cite{brewer2016}, green diamonds to \cite{bond2006}, orange squares to \citet{bensby2014}, and cyan stars to \citet{ivanyuk2017}. The black points in the bottom right of each plot represents the average uncertainty in the points.
      The histograms in the right panels of each of the plots show the distribution of the results, 
      fitted by Gaussian functions, with parameters given in Table \ref{tab:offsets_tables}.}
         \label{fig:compare_values}
   \end{figure*}

\begin{table*}
\caption{Parameters of the Gaussian distributions adjusted to the difference of stellar parameters from SPECIES and from the literature. $\mu$ and $\sigma$ correspond to the mean and standard deviation of the distributions, respectively. The offset for each catalogue is taken to be $\mu$ from this table.}
\label{tab:offsets_tables}
\centering
\begin{tabular}{l|cc|cc|cc|cc|cc}
& \multicolumn{2}{c}{\citeauthor{sousa2008}} & \multicolumn{2}{c}{\citeauthor{brewer2016}} & \multicolumn{2}{c}{\citeauthor{bond2006}} & \multicolumn{2}{c}{\citeauthor{bensby2014}} & \multicolumn{2}{c}{\citeauthor{ivanyuk2017}}\\
& \multicolumn{2}{c}{\citeyearpar{sousa2008}} & \multicolumn{2}{c}{\citeyearpar{brewer2016}} & \multicolumn{2}{c}{\citeyearpar{bond2006}} & \multicolumn{2}{c}{\citeyearpar{bensby2014}} & \multicolumn{2}{c}{\citeyearpar{ivanyuk2017}}\\
& $\mu$ & $\sigma$ & $\mu$ & $\sigma$ & $\mu$ & $\sigma$ & $\mu$ & $\sigma$  & $\mu$ & $\sigma$ \\
\hline
$M$ ($M_{\odot}$) & 0.04 & 0.06 & 0.04 & 0.11 &  &  & 0.01 & 0.04 &  &  \\
Age (Gyr) &  &  & -0.46 & 1.77 &  &  & -0.41 & 2.24 &  &  \\
$R$ ($R_{\odot}$) &  &  & 0.00 & 0.03 &  &  &  &  &  &  \\
$T$ (K) & 25.75 & 52.78 & 54.77 & 55.43 & 110.49 & 91.87 & 51.19 & 65.31 & 100.38 & 64.67 \\
$[$Fe$/$H$]$ & 0.03 & 0.03 & 0.04 & 0.03 & 0.11 & 0.06 & 0.02 & 0.05 & 0.14 & 0.06 \\
$\log$ g & 0.00 & 0.09 & 0.02 & 0.11 & 0.04 & 0.16 & 0.00 & 0.12 & 0.01 & 0.10 \\
$\xi_t$ (km s$^{-1}$) & -0.17 & 0.09 &  &  & -0.38 & 0.15 & -0.17 & 0.13 & -0.25 & 0.12 \\
$v\sin i$ (km s$^{-1}$) &  &  & 0.46 & 0.89 &  &  &  &  & -1.02 & 0.45 \\
$v_{mac}$ (km s$^{-1}$) &  &  & -0.22 & 0.25 &  &  &  &  &  &  \\
\end{tabular}
\end{table*}

\begin{figure}
   \centering
   \includegraphics[width=9cm]{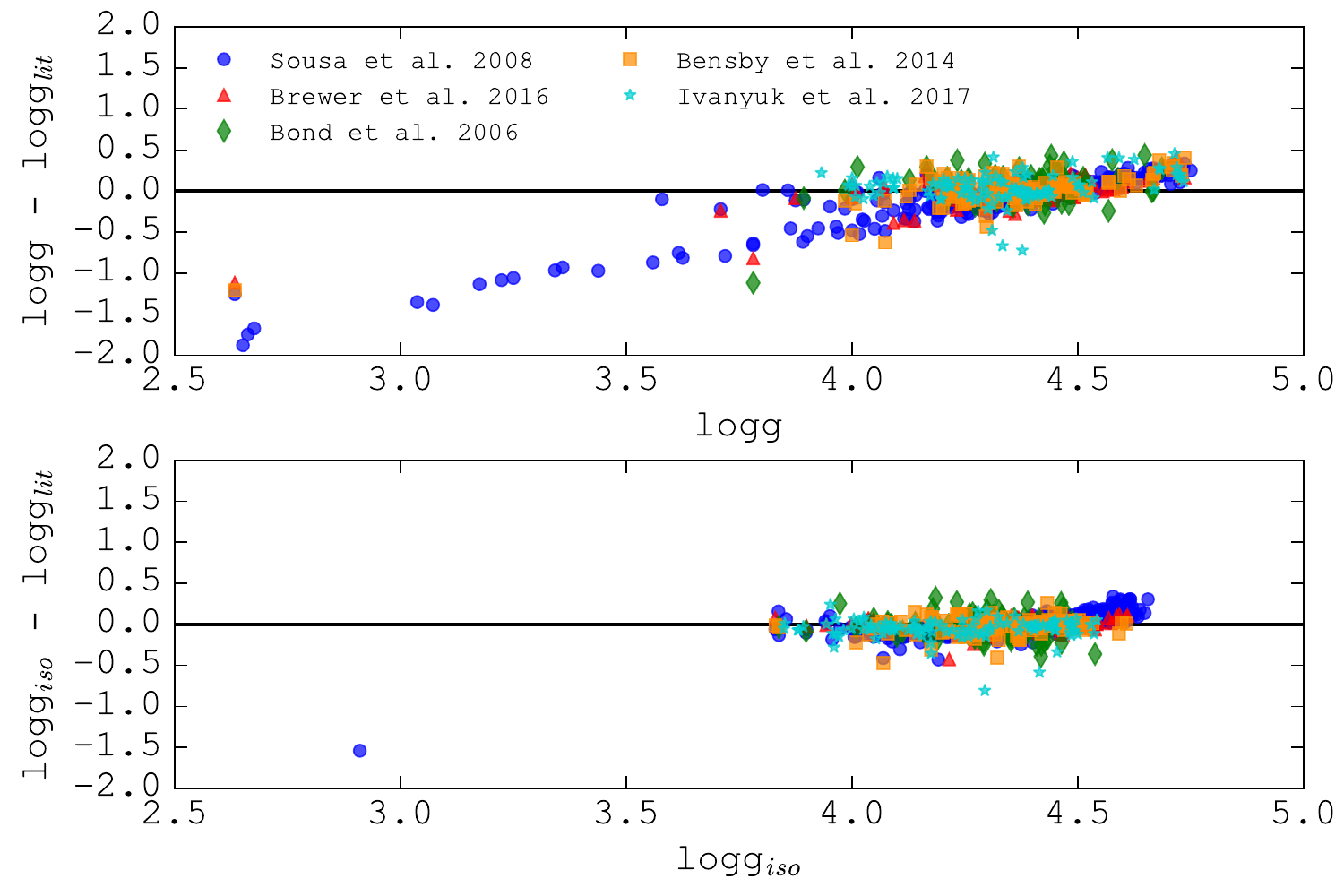}
      \caption{Comparison between the surface gravity from SPECIES, and from the literature. In the top panel, the surface gravity from SPECIES corresponds to the results obtained from the convergence of the atmospheric parameters (section \ref{sec:atmospheric_parameters}), without the option of recomputing using \isologg. In the bottom panel, the surface gravity from SPECIES corresponds to the \isologg obtained for the same stars (section \ref{sec:mass_age_plogg}).}
         \label{fig:compare_logg_and_literature}
   \end{figure}
   
      \begin{figure*}
   \centering
            {\includegraphics[width=18cm]{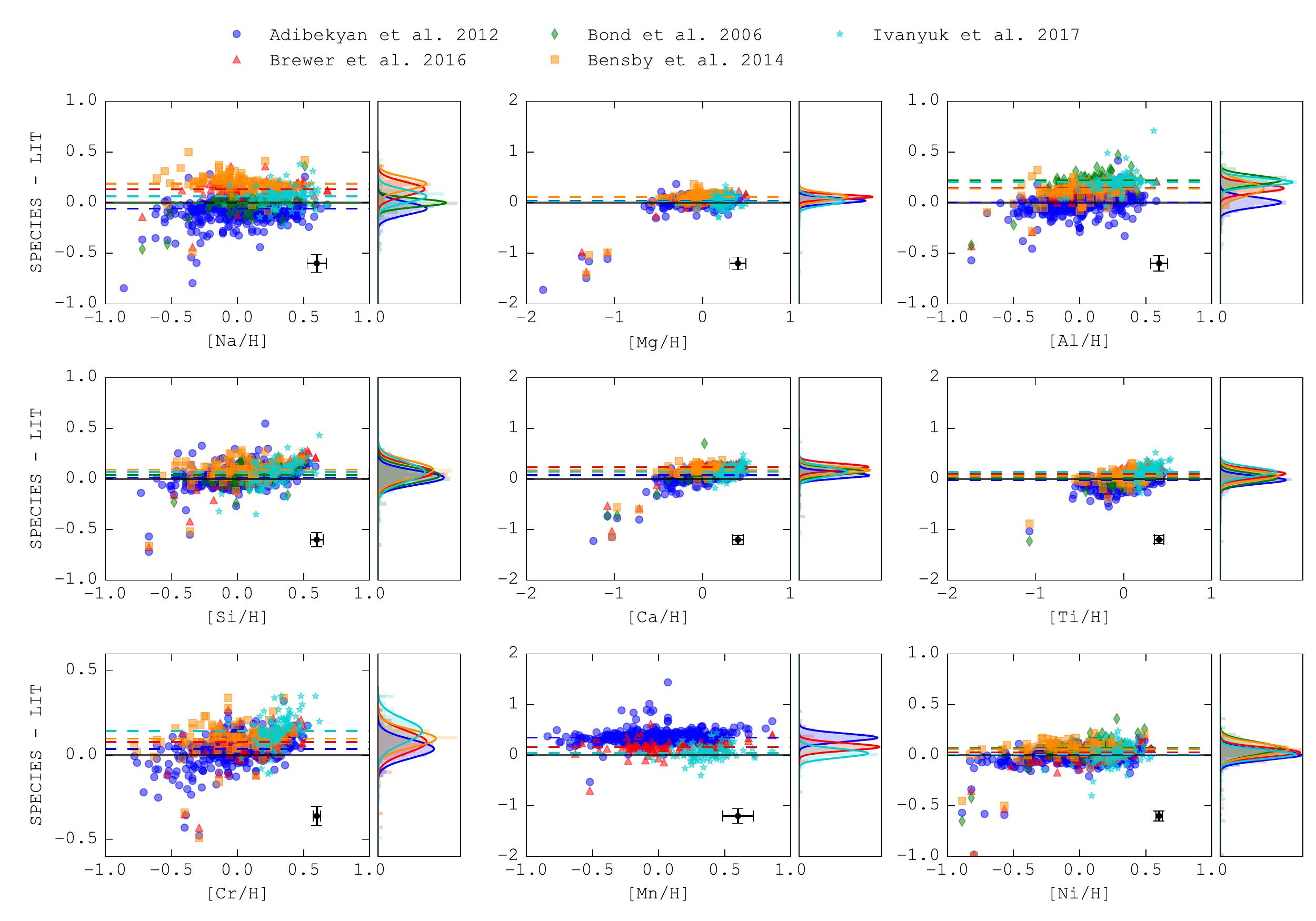}}
      \caption{Comparison between the abundances for different elements obtained with our code and 
      from literature. 
      The y-axis in the plots correspond to the difference between both measurements,
      represented by different symbols:
      blue circles to \cite{sousa2008}, red triangles to \cite{brewer2016}, green squares to \cite{bond2006}, orange squares to \cite{bensby2014}, and cyan stars to \cite{ivanyuk2017}. The black points in the bottom right of each plot represents the average uncertainty in the points.
      The histograms at the right panels of each plots show the distribution of the results, 
      fitted by Gaussian functions.}
         \label{fig:compare_abundances}
   \end{figure*}

\begin{table*}
\caption{Parameters of the Gaussian distributions adjusted to the difference of atomic abundance from SPECIES and from the literature. $\mu$ and $\sigma$ correspond to the mean and standard deviation of the distributions, respectively. The offset for each element and catalogue is taken to be $\mu$ from this table.}
\label{tab:offsets_abund}
\centering
\begin{tabular}{l|cc|cc|cc|cc|cc}
& \multicolumn{2}{c}{\citeauthor{adibekyan2012}} & \multicolumn{2}{c}{\citeauthor{brewer2016}} & \multicolumn{2}{c}{\citeauthor{bond2006}} & \multicolumn{2}{c}{\citeauthor{bensby2014}} & \multicolumn{2}{c}{\citeauthor{ivanyuk2017}}\\
& \multicolumn{2}{c}{\citeyearpar{adibekyan2012}} & \multicolumn{2}{c}{\citeyearpar{brewer2016}} & \multicolumn{2}{c}{\citeyearpar{bond2006}} & \multicolumn{2}{c}{\citeyearpar{bensby2014}} & \multicolumn{2}{c}{\citeyearpar{ivanyuk2017}}\\
& $\mu$& $\sigma$ & $\mu$ & $\sigma$ & $\mu$ & $\sigma$ & $\mu$ & $\sigma$  & $\mu$ & $\sigma$\\
\hline
$[$Na$/$H$]$ & -0.06 & 0.07 & 0.13 & 0.07 & -0.00 & 0.05 & 0.14 & 0.07 & 0.06 & 0.06 \\
$[$Mg$/$H$]$ & 0.04 & 0.06 & 0.12 & 0.05 &  &  & 0.06 & 0.07 & 0.03 & 0.09 \\
$[$Al$/$H$]$ & 0.00 & 0.05 & 0.14 & 0.04 & 0.22 & 0.05 & 0.09 & 0.07 & 0.20 & 0.04 \\
$[$Si$/$H$]$ & 0.01 & 0.07 & 0.07 & 0.08 & 0.04 & 0.07 & 0.04 & 0.08 & 0.07 & 0.10 \\
$[$Ca$/$H$]$ & 0.07 & 0.06 & 0.24 & 0.06 & 0.14 & 0.07 & 0.12 & 0.06 & 0.14 & 0.09 \\
$[$Ti$/$H$]$ & -0.02 & 0.08 & 0.10 & 0.08 & 0.01 & 0.09 & 0.06 & 0.09 & 0.14 & 0.09 \\
$[$Cr$/$H$]$ & 0.04 & 0.05 & 0.08 & 0.07 &  &  & 0.05 & 0.05 & 0.14 & 0.07 \\
$[$Mn$/$H$]$ & 0.35 & 0.08 & 0.16 & 0.07 &  &  &  &  & 0.04 & 0.08 \\
$[$Ni$/$H$]$ & 0.00 & 0.06 & 0.02 & 0.05 & 0.06 & 0.06 & 0.01 & 0.06 & 0.01 & 0.07 \\

\end{tabular}
\end{table*}

\begin{itemize}
\item \citet[hereafter SPOCS2]{brewer2016}, a continuation of \cite[SPOCS]{spocs2005}, in which stellar parameters were presented for $\sim$ 1000 stars. They used the spectral synthesis method to derive the atmospheric parameters, and interpolation using Yonsei-Yale (Y2) isochrones \citep{Y2} to obtain mass and age measurements. They set the microturbulence velocity to 4 km/s through their calculation, and derive a formula for the macroturbulence velocity very similar to the one used in this work (Eq \ref{eq:vmac}). In SPOCS2, the abundances list was increased, as well as the number of stars in their sample ($\sim$ 1600 stars).\newline

\item \citet[hereafter S08]{sousa2008}, in which they used the same method as we did to compute the atmospheric parameters (\teff, \logg, [Fe/H] and $\xi_t$), that is by computing the EWs using ARES for a set of iron lines and then using MOOG to derive their stellar parameters. In the case of the abundances for other chemical elements, we used the values from \citet[hereafter A12]{adibekyan2012}, which uses the atmospheric parameters derived in S08.\newline

\item \citet[hereafter B06]{bond2006}, where the procedure used to derive their parameters also relied on the measurement of EWs, assuming LTE to derive the atmospheric parameters. There are considerable differences between their method and ours. First, they measured the EWs of their lines by direct integration, instead of Gaussian fitting, as is done in this work. Second, the temperatures were derived using the star colors, following the relation from \citet{smith1995}. Finally, they derived the metallicity with two different methods, one using Str\"{o}mgren \textit{uvby} colors \citep{stromgren1966}, and the other using the measured EW. For this comparison, we are using the metallicity values derived through spectroscopy.\newline

\item \citet[hereafter B14]{bensby2014}, in which they also used EW measurements, along with LTE model stellar atmospheres, in order to determine the parameters. The differences between their method and ours are that in B14 they used the MARCS code \citep{MARCS} to solve the radiative transfer equations, computed the EW for each line using the IRAF task SPLOT, and used Y2 isochrones  to derive the mass and age of each star.\newline

\item \citet[hereafter I17]{ivanyuk2017}, where they used the Infrared Flux Method \citep[IRFM,][]{IRFM} calibration to derive effective temperatures, and the modified numerical scheme developed by \citep{Pavlenko2017} in order to compute the iron abundance, surface gravity, microturbulent and rotational velocity, from high S/N HARPS spectra observed as part of the Calan-Hertfordshire Extrasolar Planet Search (CHEPS) program \citep{jenkins2009}. These values were then used to derive the atomic abundances for several elements.
\end{itemize}

We selected the highest S/N spectra taken with HARPS (given that it is the highest resolution instrument currently accepted by SPECIES) for each star we wanted to analyse, which left us with 95 stars for SPOCS2, 435 stars for S08, 67 stars for B06, 99 stars for B14, and 103 stars for I17

The comparison between the atmospheric parameters (plus mass and age) from each catalogue and ours are shown in Figure \ref{fig:compare_values}.
In Figure \ref{fig:compare_abundances} the same comparison is shown for the chemical abundances (only the elements we had in common with each catalogue). 

\subsubsection{Fundamental physical parameters}\label{sec:comparison_atmospheric_parameters}

In order to study the agreement between our results for the fundamental parameters (\teff, $[$Fe/H$]$, log~g, $\xi_t$, $v\sin i$, $v_{mac}$, mass, age and radius), with that of the literature, we computed the difference between both measurements, obtaining for each parameter and catalogue a distribution of differences around zero. Then, for each distribution we adjusted a Gaussian function obtaining the mean ($\mu$) and standard deviation ($\sigma$) of the difference of results. For the analysis performed in the following sections, we consider the mean of the distribution as the offset between our results and the literature, and the significance of that offset will be given by the width of the distribution, and how far away from zero it is located. The difference in parameters, as well as the adjusted Gaussian distributions, are shown in Figure \ref{fig:compare_values}, and the Gaussian parameters ($\mu$, $\sigma$) are shown in Table \ref{tab:offsets_tables}.   

We find that, overall, the measurements are in good agreement ($\mu \leq 1.5\sigma$) among the different catalogues, albeit with a few exceptions.
These are found for the following quantities: for the temperature, $\mu = 1.55\sigma$ against I17. For the metallicity, $\mu = 1.80\sigma$ and $2.30\sigma$ against the values from B06 and I17, respectively. For the microturbulence, we find $\mu = 1.89\sigma$, $2.50\sigma$ and $2.1\sigma$ against S08, B06 and I17, respectively. Finally, for the rotational velocity, we find that $\mu = 2.30\sigma$ for the distribution of our results against the ones from I17.

%[ 0.  0.  0.]
%[  1.14440917e-09   3.99999991e-02   6.99999989e-02]
%[-0.17567999 -0.10000012  0.06439997]
%[-0.55984 -0.4005  -0.22476]
%[ 0.  0.  0.]
%[ 0.04  0.08  0.1 ]
%[-0.22   -0.13   -0.0602]
%[-0.3436 -0.229  -0.0558]
%[-4.12 -3.   -2.08]

The largest discrepancies are found against the values from B06 and I17. Those catalogues are the only ones that only use photometric calibrations to derive the stellar temperatures (as explained above). In order to check if the temperature is the source of the discrepancies, we recomputed the parameters using the values listed in B06 and I17, for the stars we had in common with those catalogues. We find that, while the offsets with metallicity are significantly improved (0.0 and 0.04 with respect to B06 and I17, respectively), the other parameters do not improve. We conclude that the differences in temperature against what was obtained in B06 and I17 produce the offsets in metallicity, but are not responsible for the discrepancies with the rest of the parameters.
The results for the rotational velocity are also very different between SPECIES and I17. The SPECIES results are smaller than for I07, except for very few exceptions. We believe this is caused by considering the line broadening as the contribution from rotational and macroturbulence velocities, instead of taking into account only the rotational contribution (as was done in I07). This manifests into lower rotational velocities than in I07.

As for the offsets seen with the other catalogues, these can be due to the different method and calibrations used to derive the parameters, and can be corrected with respect to the ones obtained with SPECIES by applying the values in Table \ref{tab:offsets_tables}.

We looked again at the differences with \logg and \isologg. In section~\ref{sec:mass_age_plogg}, we stated that \isologg (the surface gravity a star would have for the mass, age and radius derived from isochrones) is a better indication of the true \logg than the spectroscopic value. We now compare the surface gravities from the literature against those that SPECIES would obtain without the option to recompute the stellar parameters with \logg = \isologg, and against \isologg. This is shown in Figure~\ref{fig:compare_logg_and_literature}. We obtain large discrepancies between \logg and the literature for \logg $< 4.0$, but this difference disappears when using the \isologg value. This supports the statement we made in section~\ref{sec:mass_age_plogg}, that \isologg is a better representation of the true surface gravity of a star in a lot of cases. SPECIES will use \isologg as the correct results for the cases when \logg-\isologg$>0.22$ dex.

\subsubsection{Atomic abundances}\label{sec:compare_ab}

For the analysis of the atomic abundances from SPECIES, we followed the same procedure as in the previous section. We obtained the differences between the measurements from SPECIES and from the same catalogues already described and adjusted Gaussian distributions to the results. The results of this are shown in Figure \ref{fig:compare_abundances}, and the parameters of the Gaussian distributions ($\mu$, $\sigma$) are shown in Table \ref{tab:offsets_abund}.
Different abundances for the sun were used as references in each of the works, so we first need to correct the results in the literature for the differences between their reference solar abundances and the scale used in this work (solar chemical composition from \citealt{asplund2009}). We also remind the reader that the abundances for the S08 stars are listed in \citet{adibekyan2012}.

We find that the largest discrepancies are seen against the results of SPOCS2, and I17. 
For SPOCS2, $\mu = 1.86\sigma$, $2.4\sigma$, $3.5\sigma$, $4\sigma$ and $2.29\sigma$ for Na, Mg, Al, Ca and Mn, respectively. 
For I17, $\mu = 5\sigma$, $1.56\sigma$, $1.56\sigma$ and $2\sigma$ for Al, Ca, Ti and Cr, respectively.
We checked if the differences with I17 are again a consequence of the method they used to derive the temperature, by recomputing the abundances using the temperature from I17. We find a decrease in the offsets with respect to Ca and Ti ($\mu < 1.5\sigma$), but almost no change in the results for Al and Cr. The improvement in the differences for some of the elements was expected, given that in the previous section we found that the discrepancy with the metallicity is significantly decreased when using the temperature from I17, which thus will affect the final chemical abundance. By seeing no improvement in Al and Cr, we can conclude that those elements are less affected by temperature and metallicity than the rest of the species analysed. We performed the same analysis but using the temperature from SPOCS2, to see if there are changes with the chemical abundance. We find that the discrepancies with Na and Mn decreases, falling below the $1.5\,\sigma$ level, but for Mg, Al and Ca we do not see such improvements, with the offsets still above the $1.5\,\sigma$ level. This shows that differences in the temperature obtained between this work and SPOCS2 are not the source of the large abundance differences for Mg, Al and Ca. 
I17 also compared their abundances against results from other catalogues (some of them included in this work) and even though they found similar trends in abundance versus metallicity, they do see offsets between them. One of the explanations they find includes selection effects and differences in atomic line data.

Other large discrepancies we find are: $\mu_{B14} = 2\sigma_{B14}$ for Na, $\mu_{B06} = 4.4\sigma_{B06}$ for Al, $\mu_{B06} = 2\sigma_{B06}$ and $\mu_{B14} = 2\sigma_{B14}$ for Ca, and $\mu_{A12} = 4.4\sigma_{A12}$ for Mn. 
These can be explained by the differences in method used to derive the abundances, and differences in the line list used.

%\textbf{The source of the discrepancies seen above could be the calibrations applied within SPECIES to the atomic abundances, in order to match the Solar values ([M/H] $\sim$ 0). These correspond to 0.01, -0.07, 0.17, 0.04, 0.09, 0.01, 0.05, -0.02 and -0.04 for Na, Mg, Al, Si, Ca, Ti, Cr, Mn and Ni, respectively. If we remove these calibrations to our results, we are left with some cases where the offsets improve ($\mu \sim 0.0$ for Ca, compared with all the catalogues), others where the results do not vary (for example with Ti), and others where the offsets either increase or reduce. For Mg, if we remove the calibration, we are left with offsets larger than $1.5\,\sigma$ for all the catalogues, but if we do the same Al and Mn, we are left with $\mu < 1.5\,\sigma$ for I17 and B06, but with discrepancy against the rest of the catalogues. This shows us that, for some elements, the calibrations applied in SPECIES produce offsets between our results and the literature, and can explain the values listed in Table \ref{tab:offsets_abund}. We also find that, even when removing the calibrations, there is still a large discrepancy between our results and from B06 and I17. This is seen also when comparing the abundances of B06 and I17 against the rest of the catalogues. I17 addressed these offsets in their work in great detail.}

\subsubsection{Results for the Gaia Benchmark Stars sample}\label{sec:Gaia}

\begin{figure}
   \centering
   \includegraphics[width=9cm]{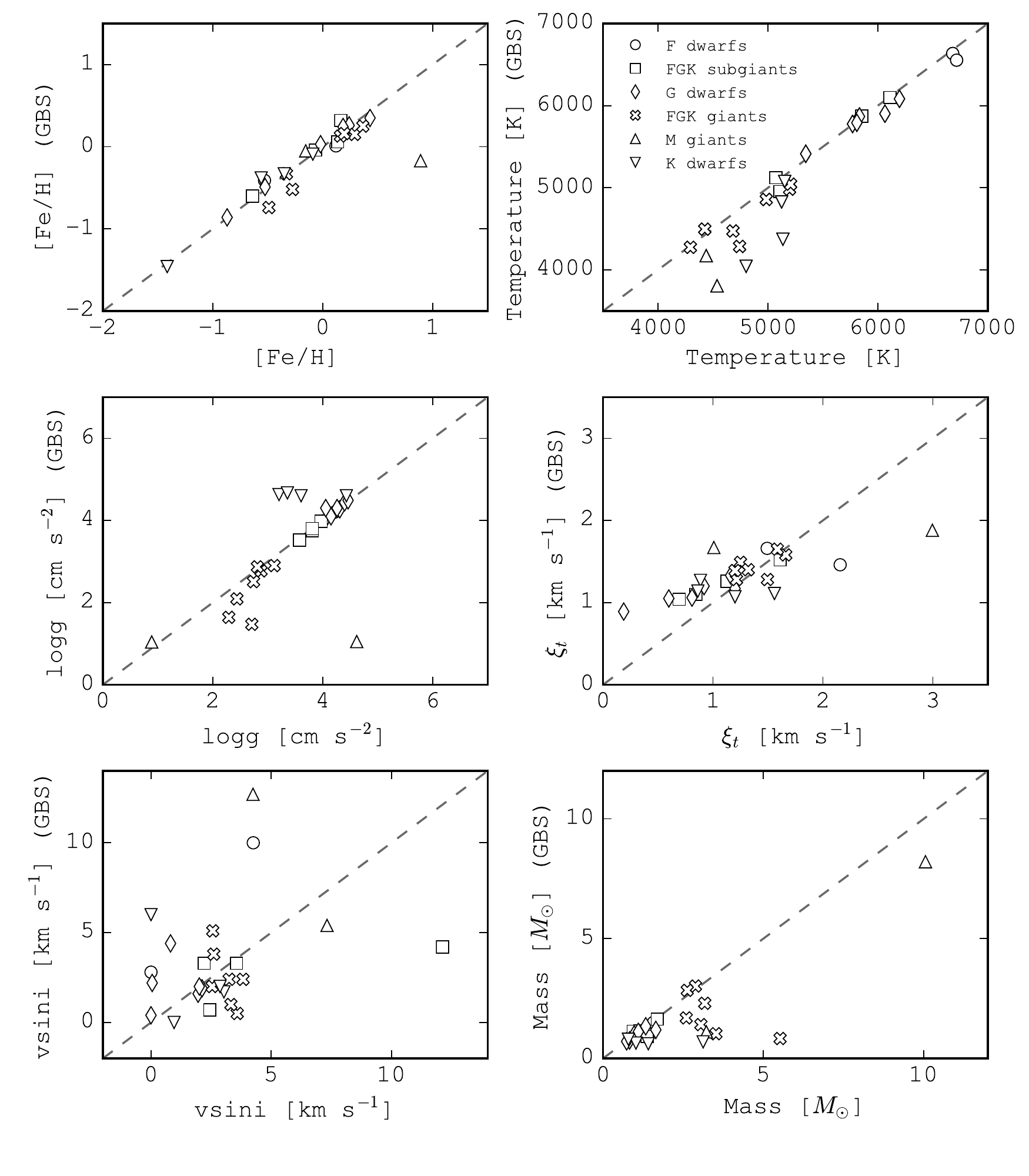}
      \caption{Results for the stellar parameters, obtained with SPECIES (x-axis), compared with what was found in the literature (y-axis), for the GBS sample. The different symbols denote different spectral types. The dashed line represents the 1:1 relation. The inset in the mass plot is a zoom of the populated region for $M < 5 M_{\odot}$.}
         \label{fig:FGK_Gaia_params}
   \end{figure}
   
\begin{figure}
   \centering
   \includegraphics[width=9cm]{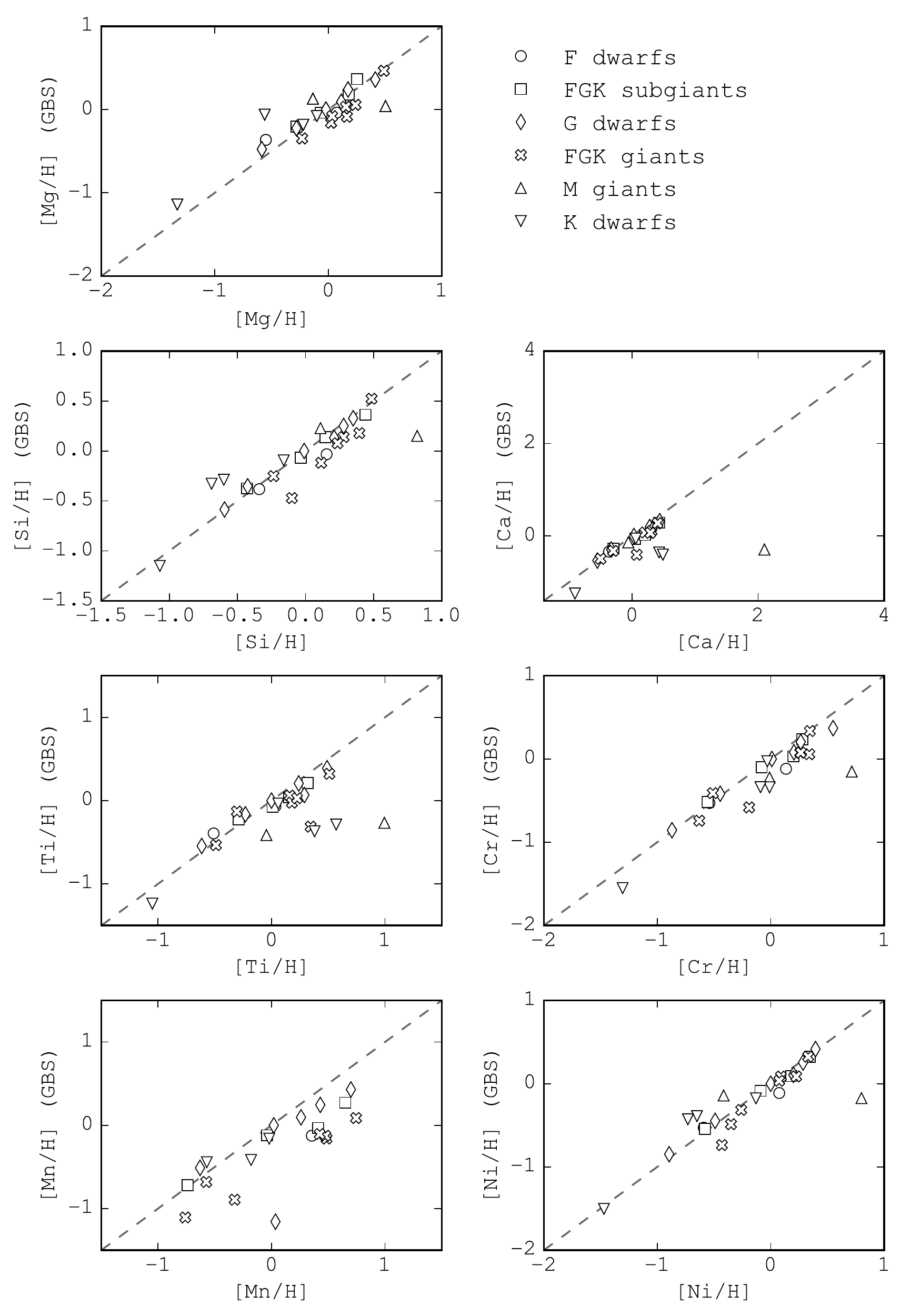}
      \caption{Results for the chemical abundance of $\alpha$ and iron peak elements, obtained with SPECIES (x-axis), compared with what was found in the literature (y-axis), for the GBS sample. The different symbols denote different spectral types. The dashed line represents the 1:1 relation.}
         \label{fig:FGK_Gaia_params_ab}
   \end{figure}
   
Finally, we compared the results obtained with SPECIES for the Gaia Benchmark Stars (GBS) sample. This sample consists of 34 FGK stars, presented in \citet{jofre2014}, spanning a wide range of metallicities and gravities, which translates into  different evolutionary stages.

The GBS sample was presented and studied in several works: \citet{jofre2014} for the determination of metallicity, \citet{heiter2015} for the effective temperature and surface gravity, and \citet{jofre2015} for chemical abundance of $\alpha$ and iron peak elements. In those papers, the parameters for each star were computed using different methods (except for the rotational velocity, for which they extract values from the literature). The input spectra were obtained from \citet{blanco-cuaresma2014}, and correspond to HARPS data. The results obtained with SPECIES for the GBS sample are shown in Figure~\ref{fig:FGK_Gaia_params}, for the atmospheric parameters, as well as rotational velocity and mass, and in Figure~\ref{fig:FGK_Gaia_params_ab} for chemical abundance. It is important to note that SPECIES could not converge to correct solutions for every star. Those corresponded, in most of the cases, to stars with very few spectral lines (mostly giant stars), or stars that are part of a spectroscopic binary system, where line blending was present in the spectra. The results obtained for each star are listed on Table~\ref{tab:FGK_gaia} and~\ref{tab:FGK_gaia_ab}.

We find that the results from SPECIES are systematically larger than the ones from the literature, for all the parameters analysed. In term of the spectral types, we find good agreement with the FGK subgiant, giant, and G dwarf samples (mean of difference for each parameter between SPECIES and the literature is less than $1\sigma$ the mean uncertainty from SPECIES), with the exception of the mass and Mn abundance of FGK giants, where SPECIES obtained values larger than $1\sigma$ from the mean uncertainty. It is not possible to draw more conclusions for the other spectral types (F dwarfs, M giants and K dwarfs), due to the low number of stars for each type ($<3$ stars for each case). We also looked at the \isologg obatined for the Gaia stars, and found them to be very similar to the surface gravity from \citet{heiter2015}. This favors the statement we made in sections~\ref{sec:mass_age_plogg} and \ref{sec:comparison_atmospheric_parameters}, in which \isologg is a good representation of the true \logg for a lot of cases, now including stars in different evolutionary stages, with \logg$<4.0$ dex.

\subsection{Correlation between parameters}\label{sec:correlations}

\begin{figure*}
   \centering
   \includegraphics[width=18cm]{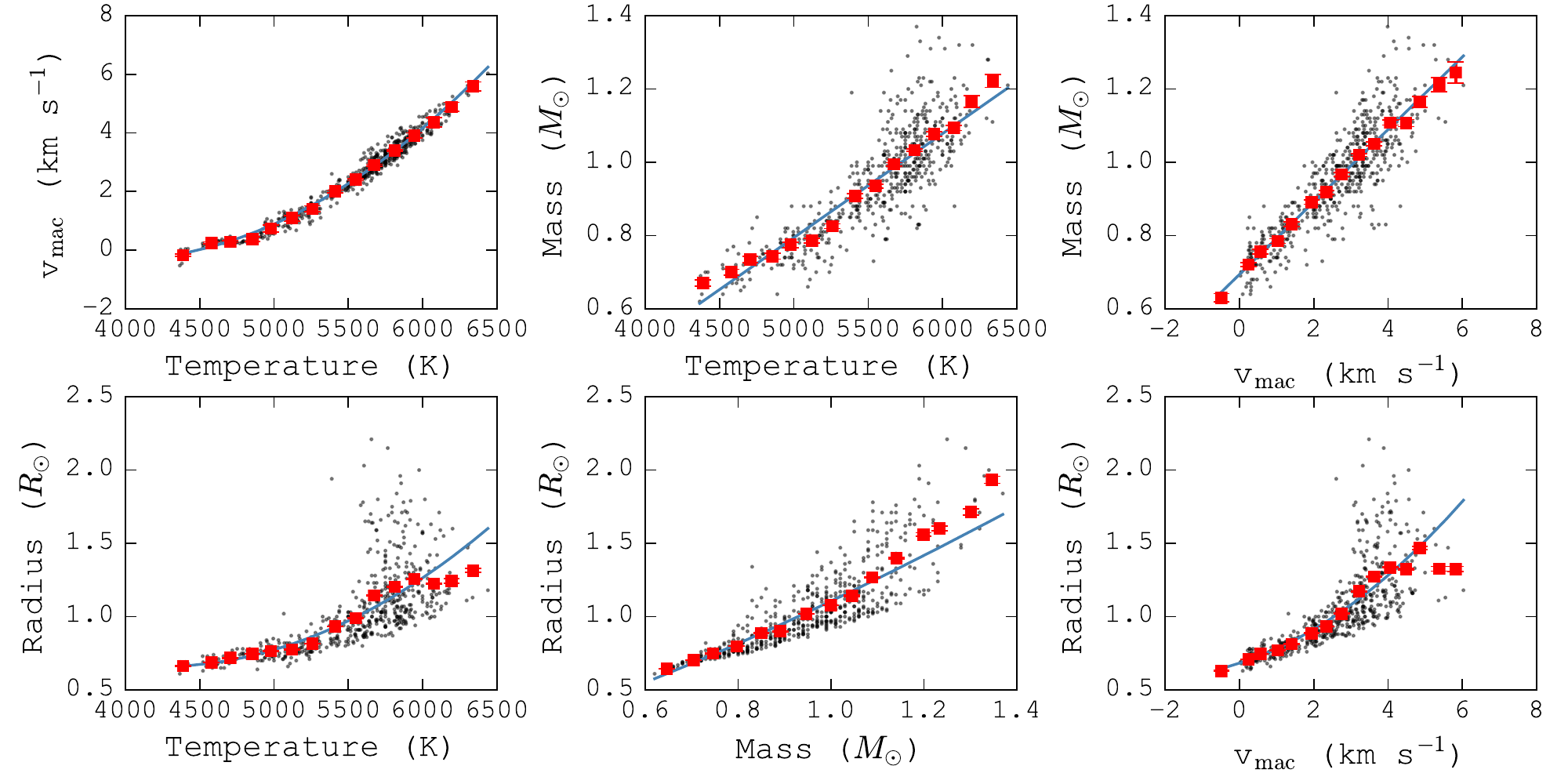}
      \caption{Correlations between the stellar parameters computed by SPECIES. The red squares are the binned data points, and the blue lines correspond to the fits described in Equation \ref{eq:correlations}. The data used correspond to points within 3$\sigma$ of their corresponding distribution.}
         \label{fig:correlations}
   \end{figure*}

\begin{figure*}
   \centering
   \includegraphics[width=18cm]{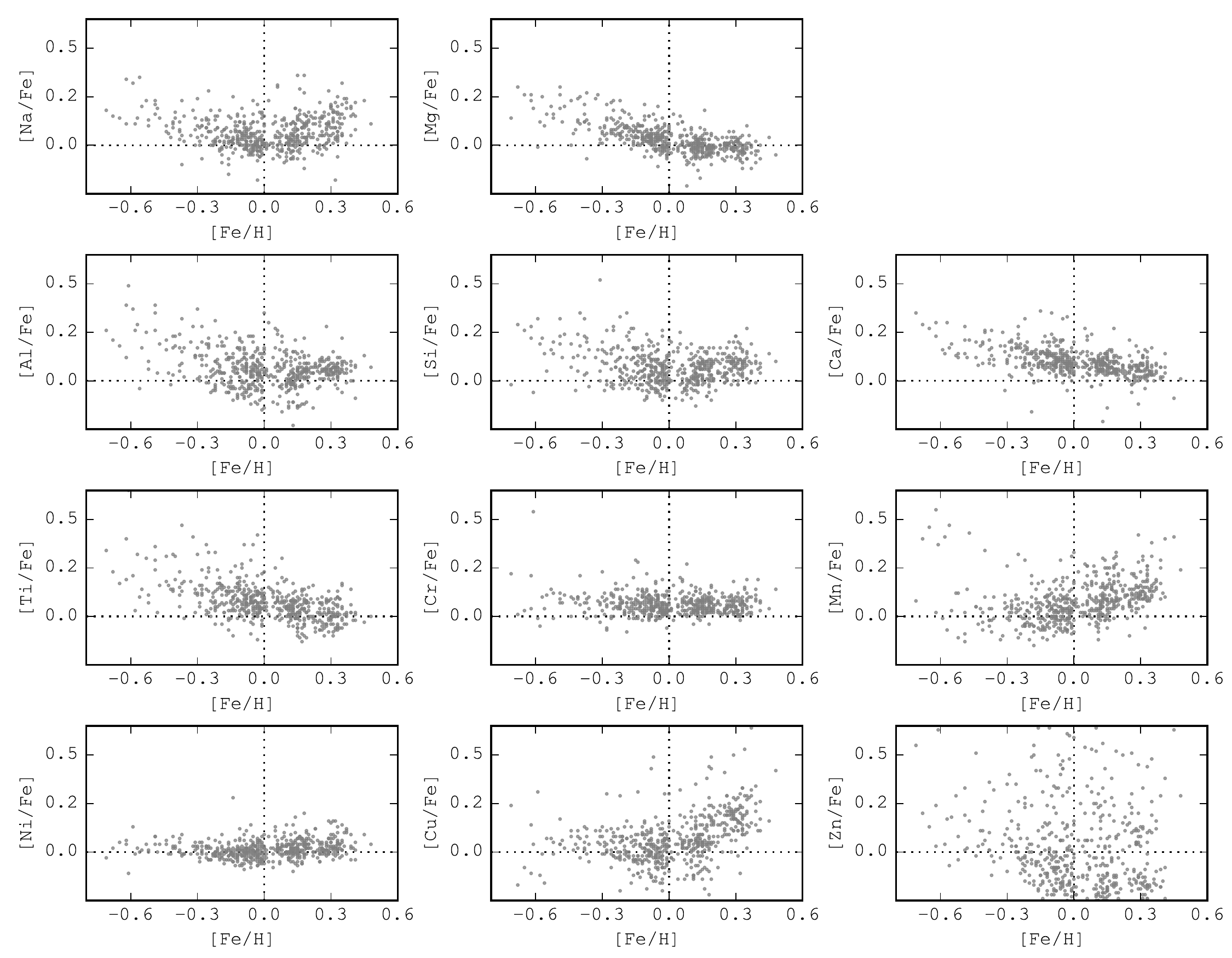}
      \caption{Abundances for the 11 elements analysed by SPECIES with respect to iron, versus metallicity. This includes all the stars studied in section~\ref{sec:compare_ab}, as well as from the Gaia benchmark sample (section~\ref{sec:Gaia}).}
         \label{fig:ab_vs_met}
   \end{figure*}

\begin{figure*}
   \centering
   \includegraphics[width=19cm]{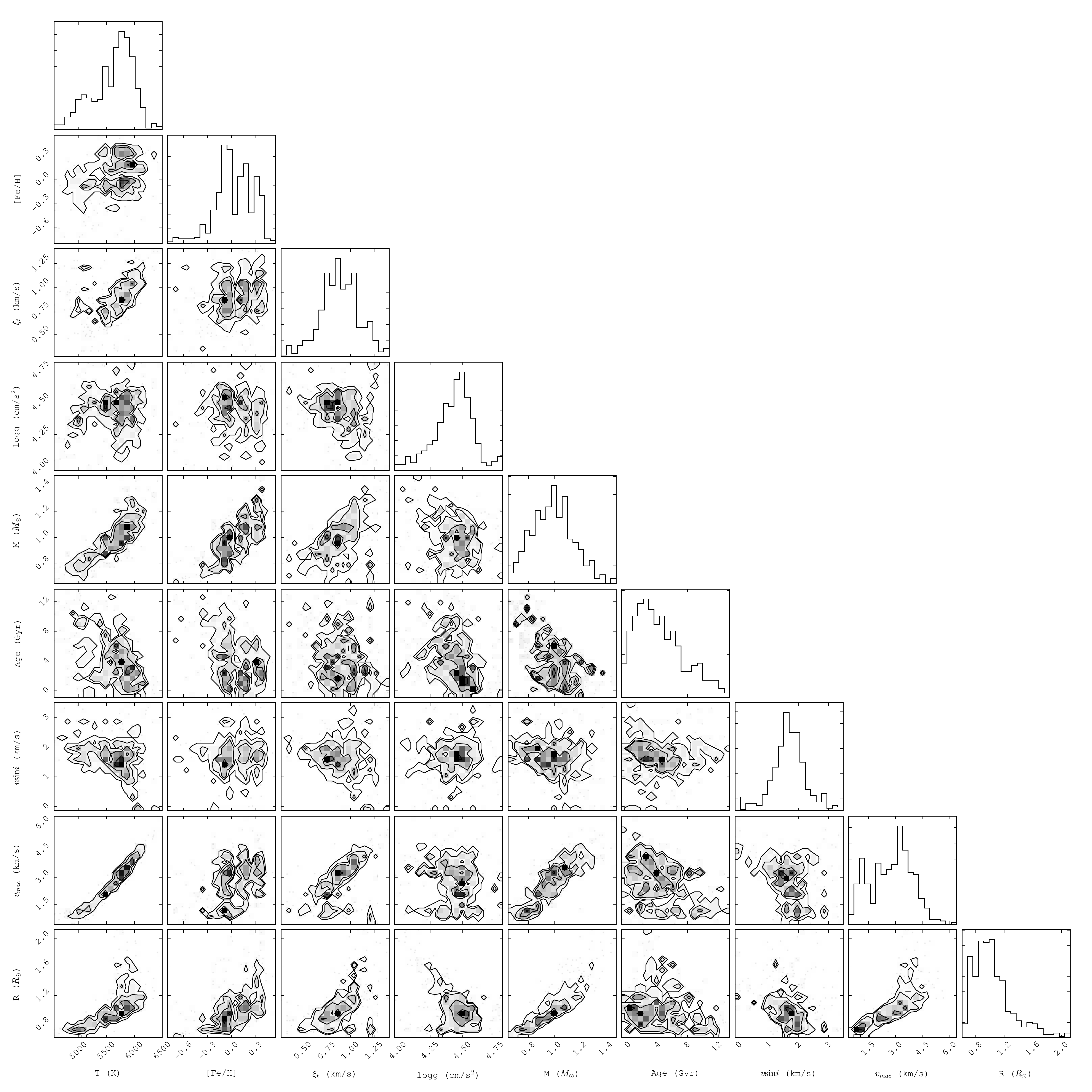}
   \caption{Correlation between the atmospheric parameters, as well as the mass and age for each stars, derived by this code for HARPS spectra. The histograms the the top of each column show the distribution of every single quantity. For \logg, $\xi_t$ and age, the points farther than 3$\sigma$ from the mean of the distribution were not included.}
       \label{fig:correlations_corner}
   \end{figure*}

We studied whether there were strong correlations between the stellar parameters, by plotting each quantity against the rest,  considering only points within 3$\sigma$ of the mean.
We find that the majority of the parameters derived using this code show no strong correlations between each other, as can be seen in Figure \ref{fig:correlations_corner}, even though we do find some exceptions. We find that the mass, radius, temperature and macroturbulent velocities show correlations among each other, as is shown in Figure \ref{fig:correlations}. We adjusted the following relations to those correlations: 

\begin{align}
\label{eq:correlations}
v_{mac}\: =&\: \frac{(T-5777)^2}{9.3\times 10^5} + \frac{(T-5777)}{251} + 3.54 \nonumber\\
M\: =&\: \frac{(T-5777)}{3371} + 1.03 \nonumber\\
M\: =&\: 0.10\, v_{mac} +0.66 \\
R\: =&\: \frac{(T-5777)^2}{5.84\times 10^6} + \frac{(T-5777)}{1802} + 1.1 \nonumber\\
R\: =&\: 1.08\,M^{1.28} \nonumber\\
R\: =&\: 0.01\, v_{mac}^2 + 0.08\, v_{mac} + 0.66 \nonumber
\end{align}

All these relations are shown as the blue lines in Figure \ref{fig:correlations}.

The mass correlation with temperature reflects the known mass-luminosity relationship for stars \citep{Kuiper1938}, for which $L \propto M^{\alpha}$. Dwarf stars increase in luminosity for higher temperatures, therefore the relation can be interpreted as larger mass for higher surface temperature.
The correlation between macroturbulence velocity and temperature is produced by the method we used to derive $v_{mac}$, following equation \ref{eq:vmac} (Section \ref{sec:broadening}), and the increased depth of the convective envelope with decreasing temperature. In the equation for instance, the metallicity dependence is not as strong as the temperature dependence, which explains why we do not see such a strong correlation between macroturbulence velocity and metallicity (Figure \ref{fig:correlations_corner}). 
The relation between stellar mass and radius has been well studied over the years, and for main sequence stars, \citet{demircan1991} found that where $R = 1.06\,M^{0.945}$, for $M < 1.66 M_{\odot}$. The fit we performed to the SPECIES results is in agreement with the previous relation.

The rest of the strong correlations seen in Figure \ref{fig:correlations} are a consequence of the relations mentioned previously. The relation between mass and microturbulence is due to the relation between microturbulence and temperature, and of mass with temperature. The macroturbulence with microturbulence is also due to the relation between both quantities and temperature. Finally, the relation between macroturbulence and mass is produced by the effect temperature has on both parameters.

When looking at the chemical compositions, we studied its abundance with respect to iron, versus the star's overall metallicity. This is shown in Figure~\ref{fig:ab_vs_met}. We find that, for most of the elements, their abundance is greater than iron in metal-poor stars, and resembles (Mg, Al, Ca, Ti, Cr, Ni) or is greater (Na, Mn, Cu) than iron for metal-rich ([Fe/H] $\geq$ 0.0) stars. This behaviour is very similar to that of \citet{ivanyuk2017}, where they also compare their results with catalogues in the literature (some of them are also included in this work). We cannot make any conclusions about the behaviour of Zn, given that the spread in the results is too large.

%When looking at the chemical abundances, we find that they are very correlated with the metallicity, as is shown in Figure \ref{fig:ab_vs_met}. We adjusted linear fits to each relation, of the form [M/H]$=a\cdot$[Fe/H]$+ b$, with $a$ the slope, $b$ the offset, and M the corresponding element. These fits, along with their parameters, are shown in Figure \ref{fig:ab_vs_met} as the red lines. 

%For most of the elements, we also find that their abundance is greater than for iron, increasing in metal-poor stars, whilst their abundance is almost the same as for iron in metal rich stars (slope of the linear fits $<1$). This behaviour is followed by all the elements, except for Mn and Cu, for which the slope of the linear fits is $>1$.

%In general, the relations we find are very close to being 1:1 correlated ($a\sim 1$ and $b\sim 0$), which solidifies the notion that the metallicity (measure of the amount of iron compared to hydrogen) is a good indicator of the chemical abundance in stellar atmospheres.

\subsection{Offsets between different instruments}\label{sec:compare_inst}

\begin{figure*}
   \centering
   \includegraphics[width=18cm]{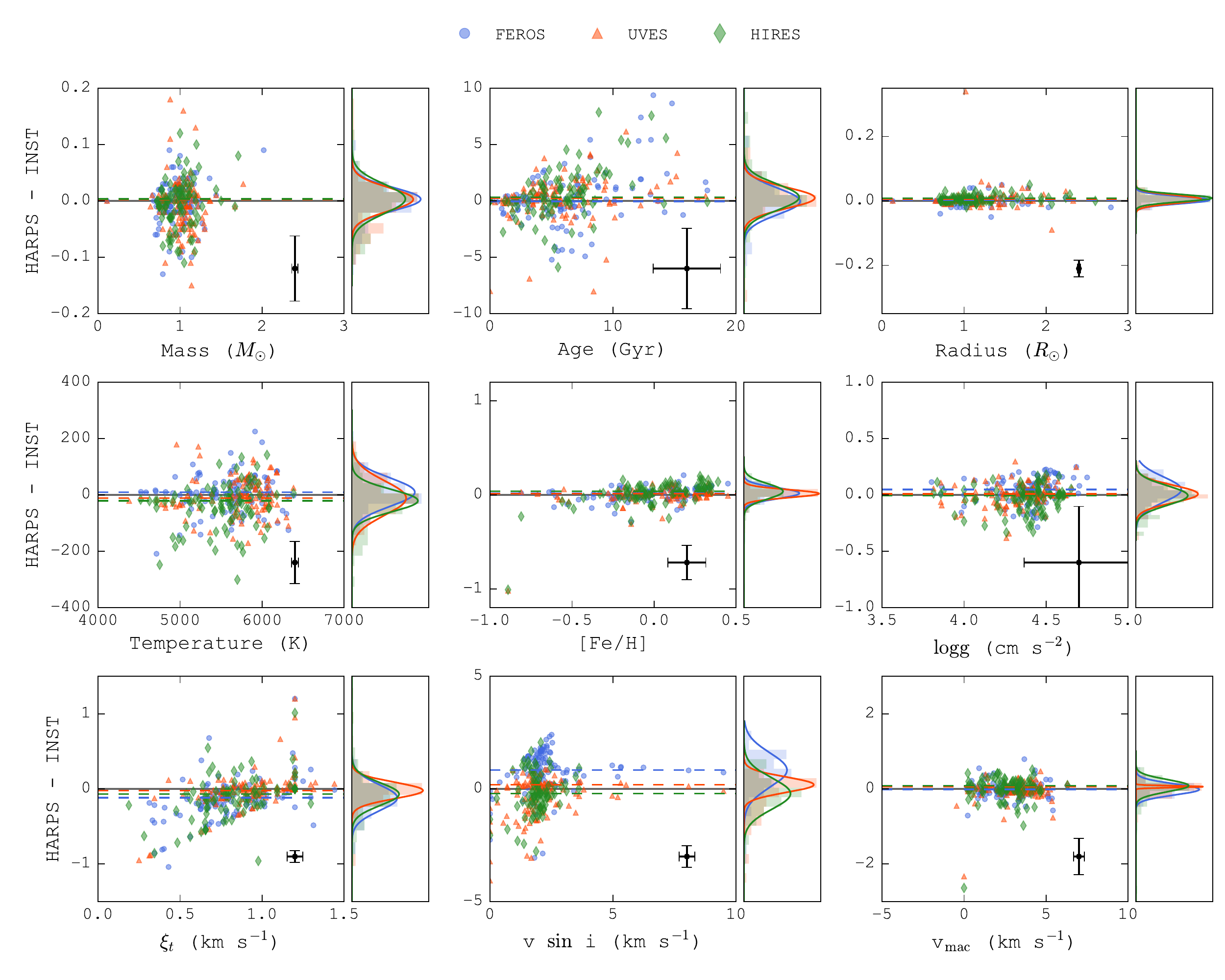}
      \caption{Comparison between the stellar parameters obtained using spectra from different instruments, with respect to the ones obtained with HARPS. The y-axis for each plot corresponds to the difference between the parameters. Blue dots correspond to FEROS data, red triangles to UVES data, and green diamonds to HIRES data. For each panel, the right-hand plot corresponds to the distribution of the values, and solid line to the Gaussian distribution adjusted to each histogram. The black points in the bottom right of each plot represents the average uncertainty in the points.}
         \label{fig:comparison_inst}
   \end{figure*}
   
\begin{figure*}
   \centering
   \includegraphics[width=18cm]{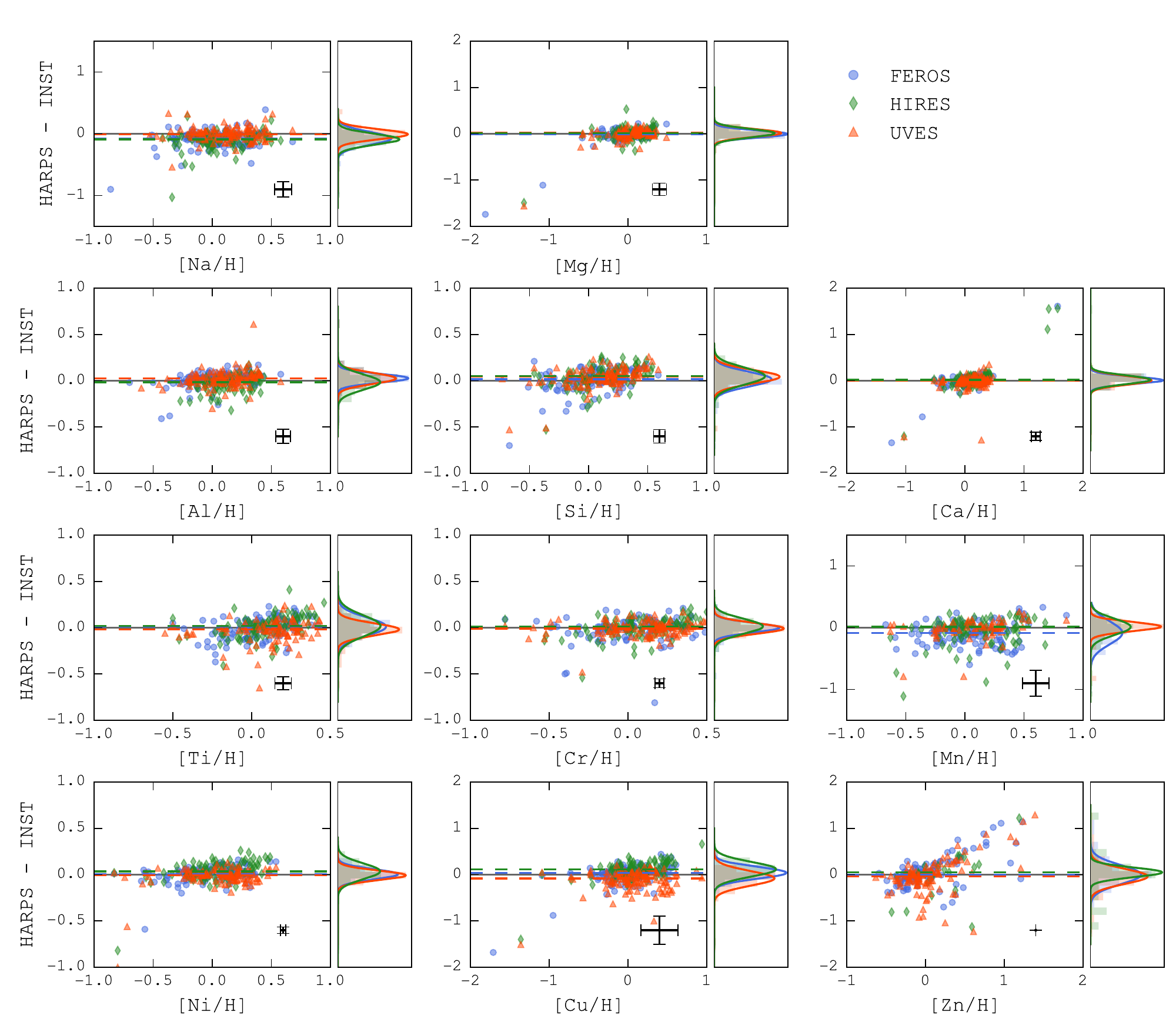}
      \caption{Comparison between the abundances obtained with spectra from different instruments, with respect to the values obtained with HARPS spectra. The y-axis for each panel corresponds to the difference between the values from HARPS, and from FEROS (blue squares), UVES (red triangles) or HIRES (green diamonds). The right-hand plots show the distribution of the values, along with their Gaussian distribution fit. The black points in the bottom right of each plot represents the average uncertainty in the points.}
         \label{fig:comparison_abundances_inst}
   \end{figure*}
   
\begin{table}
\caption{Gaussian distribution parameters ($\mu$, $\sigma$) obtained for the difference between stellar parameters from HARPS, and from other instruments. The distributions are shown in on top of the histograms in Figure \ref{fig:comparison_inst}}
\label{tab:offsets_inst}
\centering
\begin{tabular}{c|cc|cc|cc}
& \multicolumn{2}{c}{FEROS} & \multicolumn{2}{c}{UVES} & \multicolumn{2}{c}{HIRES} \\
& $\mu$ & $\sigma$ & $\mu$ & $\sigma$ & $\mu$ & $\sigma$ \\
\hline
$M$ ($M_{\odot}$) & 0.00 & 0.02 & 0.00 & -0.02 & 0.00 & 0.03 \\
Age (Gyr) & -0.04 & 1.16 & 0.27 & 0.93 & 0.33 & 1.15 \\
$R$ ($R_{\odot}$) & 0.00 & 0.01 & 0.01 & 0.01 & 0.01 & 0.01 \\
$T$ (K) & 9.66 & 50.47 & -11.24 & 61.00 & -20.63 & 39.48 \\
$[$Fe$/$H$]$ & 0.02 & 0.06 & 0.01 & 0.04 & 0.04 & 0.08 \\
$\log$ g & 0.05 & 0.11 & 0.01 & 0.08 & -0.01 & 0.09 \\
$\xi_t$ & \multirow{2}{*}{-0.12} & \multirow{2}{*}{0.17} & \multirow{2}{*}{-0.02} & \multirow{2}{*}{0.11} & \multirow{2}{*}{-0.07} & \multirow{2}{*}{0.15} \\
(\kms) &&&&&&\\
$v\sin\,i$ & \multirow{2}{*}{0.84} & \multirow{2}{*}{0.76} & \multirow{2}{*}{0.18} & \multirow{2}{*}{0.34} & \multirow{2}{*}{-0.21} & \multirow{2}{*}{0.69} \\
(\kms) &&&&&&\\
$v_{mac}$ & \multirow{2}{*}{-0.01} & \multirow{2}{*}{0.15} & \multirow{2}{*}{0.06} & \multirow{2}{*}{0.02} & \multirow{2}{*}{0.09} & \multirow{2}{*}{0.16} \\
(\kms) &&&&&&\\

\end{tabular}
\end{table}

\begin{table}
\caption{Gaussian distribution parameters ($\mu$, $\sigma$) obtained for the difference between atomic abundances from HARPS, and from other instruments. The distributions are shown in on top of the histograms in Figure \ref{fig:comparison_abundances_inst}}
\label{tab:offsets_inst_ab}
\centering
\begin{tabular}{c|cc|cc|cc}
& \multicolumn{2}{c}{FEROS} & \multicolumn{2}{c}{UVES} & \multicolumn{2}{c}{HIRES} \\
& $\mu$ & $\sigma$ & $\mu$ & $\sigma$ & $\mu$ & $\sigma$ \\
\hline
$[$Na$/$H$]$ & -0.08 & 0.11 & -0.01 & 0.08 & -0.09 & 0.09 \\
$[$Mg$/$H$]$ & -0.01 & 0.08 & 0.02 & 0.08 & 0.01 & 0.09 \\
$[$Al$/$H$]$ & 0.03 & 0.05 & 0.02 & 0.06 & -0.02 & 0.08 \\
$[$Si$/$H$]$ & 0.02 & 0.08 & 0.04 & 0.07 & 0.05 & 0.09 \\
$[$Ca$/$H$]$ & 0.01 & 0.08 & 0.01 & 0.09 & 0.02 & 0.09 \\
$[$Ti$/$H$]$ & 0.01 & 0.10 & -0.02 & 0.07 & 0.02 & 0.11 \\
$[$Cr$/$H$]$ & 0.00 & 0.05 & -0.01 & 0.06 & 0.01 & 0.08 \\
$[$Mn$/$H$]$ & -0.08 & 0.19 & 0.02 & 0.07 & 0.01 & 0.13 \\
$[$Ni$/$H$]$ & 0.00 & 0.05 & -0.01 & 0.05 & 0.04 & 0.09 \\
$[$Cu$/$H$]$ & 0.04 & 0.13 & -0.08 & 0.16 & 0.12 & 0.16 \\
$[$Zn$/$H$]$ & -0.00 & 0.20 & -0.04 & 0.16 & 0.05 & 0.10 \\
\end{tabular}
\end{table}

In order to use this code with spectra from different instruments, it is necessary to understand any offsets that are present between the parameters computed with spectra taken from different spectrographs. We compared the results we obtained for the stars used in the previous section, using four of the available instruments accepted by our code (HARPS, FEROS, UVES and HIRES). The results of this comparison are shown in  Figure \ref{fig:comparison_inst} for the atmospheric parameters and in Figure \ref{fig:comparison_abundances_inst} for the abundances. It is important to point out that not all the stars were observed with all instruments, therefore the number of stars compared per instrument varies, with 118 for FEROS, 115 for UVES, and 89 for HIRES. 

We followed the same procedure used in the previous sections to analyse the significance of the differences in results. In this case, the comparison was done with respect to the HARPS results. The parameters for the Gaussian distributions adjusted to the difference in results are listed in Table \ref{tab:offsets_inst} and Table \ref{tab:offsets_inst_ab}, for the atmospheric parameters and atomic abundance, respectively.

From both tables, it can be seen that all the quantities are in good agreement among all the instruments, with $\mu < 1.5\sigma$ for each of them. This shows that SPECIES delivers consistent results with spectra from different high resolution spectrographs. 
We do want to mention that the only quantity with an offset larger than $1\sigma$ is seen in $v\sin i$, with respect to FEROS, with $\mu = 1.15 \sigma$. We believe this is caused mainly because of the lower spectral resolution ($R=48000$) of FEROS, with respect to the rest of the instruments ($R = 115000$, $110000$ and $67000$ for HARPS, UVES and HIRES, respectively, as it appears in their documentation). This produces larger instrumental broadening, which, together with the macroturbulence contribution, dominate the absorption line profiles. In those cases, the rotational broadening has to be larger than $\sim$2 \kms in order to make a contribution to the line profile that is measurable with FEROS (\citealp{murgas13}). This is also the reason behind the almost constant increase in difference for larger HARPS $v\sin i$, for low rotational velocities (up to 2 \kms). For those low values, the line profiles measured with FEROS are still dominated by macroturbulence and instrumental broadening, resulting in a constant $v\sin i$ for FEROS. This leads us to then set a minimum limit for the rotational velocity measured using FEROS spectra of 2 \kms. For stars with slower rotation, the FEROS spectral resolution makes it difficult to obtain accurate results.
In the future, we will perform the same comparison, but using spectra from the other instruments accepted by SPECIES that are not included in this analysis (i.e. CORALIE, AAT).

\section{Summary and conclusions}\label{sec:conclusions}

In this paper we have presented a new code to derive stellar parameters in an automated way, using high resolution stellar spectra and minimal photometric inputs. 
The parameters calculated by SPECIES agree with previously published values at the 1$\sigma$ level, for works using the same method as the one used in this work (EW measurements), as well as others (synthetic spectra).
The code presented here computes all the stellar parameters in a self-consistent way, and we include in our values the rotational and macroturbulence velocity for each star, which is not present in most of the major catalogues that employ the EW method. 

We also show the methods we used to derive the uncertainties for the atmospheric parameters, by providing analytic formulas that can be later used by others in the study of correlations between each parameter.
We have listed the correlations present in our values, which can be linked to the physics that govern stars, or to the methods we use to derive them. We recommend the use of SPECIES for FGK dwarf and subgiant stars, for which we had tested it against a large sample of stars, and to use it with caution for giant stars, which will be tested more in future works. 
SPECIES has been used in \citet{Bluhm2016}, \citet{Jones2017}, \citet{Diaz2018}, and Pantoja et al. (2018, submitted to MNRAS).

In future works we will apply the SPECIES code to accept spectra from more instruments, we will study a wide range of stars across a large evolutionary range to probe in detail the underlying nature of element production, and also we aim to include a module in SPECIES that will allow the calculation of precise parameters for M dwarf stars, where dust and molecules play a significant role.

Finally we note that SPECIES takes on the order of five minutes on a standard iMac desktop with a 3.2Gb processor to obtain all the parameters for a single stellar spectrum. It can be run in single spectra mode or in parallel to simultaneously analyse large data sets.

\begin{table*}
\centering
\caption{Line data used in the computation of the atmospheric parameters.}
\label{tab:linelist}
\begin{tabular}{cccc|cccc|cccc}
Wavelength & $\chi_I$ & $\log$gf & Name & Wavelength & $\chi_I$ & $\log$gf & Name & Wavelength & $\chi_I$ & $\log$gf & Name \\
\hline
5494.47 & 4.07 & -1.96 & FeI & 5905.68 & 4.65 & -0.78 & FeI & 6393.61 & 2.43 & -1.43 & FeI \\
5522.45 & 4.21 & -1.47 & FeI & 5927.8 & 4.65 & -1.07 & FeI & 6421.36 & 2.28 & -1.98 & FeI \\
5524.24 & 4.15 & -2.84 & FeI & 5929.68 & 4.55 & -1.16 & FeI & 6436.41 & 4.19 & -2.4 & FeI \\
5539.29 & 3.64 & -2.59 & FeI & 5930.19 & 4.65 & -0.34 & FeI & 6481.88 & 2.28 & -2.94 & FeI \\
5552.69 & 4.95 & -1.78 & FeI & 5933.81 & 4.64 & -2.14 & FeI & 6498.95 & 0.96 & -4.66 & FeI \\
5560.22 & 4.43 & -1.1 & FeI & 5934.67 & 3.93 & -1.08 & FeI & 6518.37 & 2.83 & -2.56 & FeI \\
5568.86 & 3.63 & -2.91 & FeI & 5947.53 & 4.61 & -2.04 & FeI & 6533.94 & 4.56 & -1.28 & FeI \\
5577.03 & 5.03 & -1.49 & FeI & 5956.71 & 0.86 & -4.56 & FeI & 6574.25 & 0.99 & -4.96 & FeI \\
5586.77 & 3.37 & -0.1 & FeI & 5976.79 & 3.94 & -1.3 & FeI & 6581.22 & 1.48 & -4.68 & FeI \\
5587.58 & 4.14 & -1.7 & FeI & 5984.83 & 4.73 & -0.29 & FeI & 6591.31 & 4.59 & -2.04 & FeI \\
5595.05 & 5.06 & -1.78 & FeI & 6003.02 & 3.88 & -1.02 & FeI & 6593.88 & 2.43 & -2.3 & FeI \\
5608.98 & 4.21 & -2.31 & FeI & 6007.97 & 4.65 & -0.76 & FeI & 6608.04 & 2.28 & -3.96 & FeI \\
5609.97 & 3.64 & -3.18 & FeI & 6008.57 & 3.88 & -0.92 & FeI & 6609.12 & 2.56 & -2.65 & FeI \\
5611.36 & 3.63 & -2.93 & FeI & 6015.24 & 2.22 & -4.66 & FeI & 6627.56 & 4.55 & -1.5 & FeI \\
5618.64 & 4.21 & -1.34 & FeI & 6019.37 & 3.57 & -3.23 & FeI & 6633.76 & 4.56 & -0.81 & FeI \\
5619.61 & 4.39 & -1.49 & FeI & 6027.06 & 4.07 & -1.2 & FeI & 6667.43 & 2.45 & -4.37 & FeI \\
5635.83 & 4.26 & -1.59 & FeI & 6056.01 & 4.73 & -0.46 & FeI & 6667.72 & 4.58 & -2.1 & FeI \\
5636.71 & 3.64 & -2.53 & FeI & 6065.49 & 2.61 & -1.49 & FeI & 6699.14 & 4.59 & -2.11 & FeI \\
5650.0 & 5.1 & -0.8 & FeI & 6078.5 & 4.79 & -0.38 & FeI & 6703.58 & 2.76 & -3.0 & FeI \\
5651.48 & 4.47 & -1.79 & FeI & 6079.02 & 4.65 & -0.97 & FeI & 6704.49 & 4.22 & -2.64 & FeI \\
5652.33 & 4.26 & -1.77 & FeI & 6082.72 & 2.22 & -3.53 & FeI & 6713.75 & 4.79 & -1.41 & FeI \\
5661.02 & 4.58 & -2.42 & FeI & 6089.57 & 5.02 & -0.87 & FeI & 6725.36 & 4.1 & -2.21 & FeI \\
5661.35 & 4.28 & -1.83 & FeI & 6093.65 & 4.61 & -1.32 & FeI & 6726.67 & 4.61 & -1.05 & FeI \\
5677.69 & 4.1 & -2.64 & FeI & 6094.38 & 4.65 & -1.56 & FeI & 6733.15 & 4.64 & -1.44 & FeI \\
5678.39 & 3.88 & -2.97 & FeI & 6096.67 & 3.98 & -1.76 & FeI & 6739.52 & 1.56 & -4.85 & FeI \\
5680.24 & 4.19 & -2.29 & FeI & 6098.25 & 4.56 & -1.81 & FeI & 6745.97 & 4.07 & -2.71 & FeI \\
5701.56 & 2.56 & -2.16 & FeI & 6120.26 & 0.91 & -5.86 & FeI & 6750.16 & 2.42 & -2.58 & FeI \\
5717.84 & 4.28 & -0.98 & FeI & 6137.0 & 2.2 & -2.91 & FeI & 6753.47 & 4.56 & -2.35 & FeI \\
5731.77 & 4.26 & -1.1 & FeI & 6151.62 & 2.18 & -3.26 & FeI & 6756.55 & 4.29 & -2.78 & FeI \\
5738.24 & 4.22 & -2.24 & FeI & 6157.73 & 4.07 & -1.26 & FeI & 6786.86 & 4.19 & -1.9 & FeI \\
5741.86 & 4.26 & -1.69 & FeI & 6165.36 & 4.14 & -1.48 & FeI & 6793.26 & 4.07 & -2.43 & FeI \\
5742.96 & 4.18 & -2.35 & FeI & 6173.34 & 2.22 & -2.84 & FeI & 6796.12 & 4.14 & -2.4 & FeI \\
5752.04 & 4.55 & -0.92 & FeI & 6187.4 & 2.83 & -4.13 & FeI & 6804.3 & 4.58 & -1.85 & FeI \\
5754.41 & 3.64 & -2.85 & FeI & 6188.0 & 3.94 & -1.6 & FeI & 6806.86 & 2.73 & -3.14 & FeI \\
5759.26 & 4.65 & -2.07 & FeI & 6199.51 & 2.56 & -4.35 & FeI & 6810.27 & 4.61 & -1.0 & FeI \\
5760.36 & 3.64 & -2.46 & FeI & 6200.32 & 2.61 & -2.39 & FeI & 5100.66 & 2.81 & -4.16 & FeII \\
5775.09 & 4.22 & -1.11 & FeI & 6213.44 & 2.22 & -2.54 & FeI & 5132.67 & 2.81 & -3.95 & FeII \\
5778.46 & 2.59 & -3.44 & FeI & 6219.29 & 2.2 & -2.39 & FeI & 5136.8 & 2.84 & -4.32 & FeII \\
5784.67 & 3.4 & -2.53 & FeI & 6220.79 & 3.88 & -2.36 & FeI & 5197.58 & 3.23 & -2.23 & FeII \\
5793.92 & 4.22 & -1.62 & FeI & 6226.74 & 3.88 & -2.08 & FeI & 5234.63 & 3.22 & -2.22 & FeII \\
5806.73 & 4.61 & -0.93 & FeI & 6232.65 & 3.65 & -1.21 & FeI & 5264.81 & 3.34 & -3.21 & FeII \\
5811.91 & 4.14 & -2.36 & FeI & 6240.65 & 2.22 & -3.23 & FeI & 5284.11 & 2.89 & -3.01 & FeII \\
5814.82 & 4.28 & -1.81 & FeI & 6246.33 & 3.6 & -0.73 & FeI & 5414.08 & 3.22 & -3.61 & FeII \\
5835.11 & 4.26 & -2.18 & FeI & 6252.57 & 2.4 & -1.64 & FeI & 5425.26 & 3.2 & -3.27 & FeII \\
5837.7 & 4.29 & -2.3 & FeI & 6265.14 & 2.18 & -2.51 & FeI & 5627.5 & 3.39 & -4.14 & FeII \\
5849.69 & 3.69 & -2.95 & FeI & 6270.23 & 2.86 & -2.55 & FeI & 5991.38 & 3.15 & -3.55 & FeII \\
5853.15 & 1.48 & -5.09 & FeI & 6280.62 & 0.86 & -4.34 & FeI & 6084.11 & 3.2 & -3.8 & FeII \\
5855.09 & 4.61 & -1.56 & FeI & 6297.8 & 2.22 & -2.7 & FeI & 6113.33 & 3.21 & -4.12 & FeII \\
5856.1 & 4.29 & -1.57 & FeI & 6301.51 & 3.65 & -0.72 & FeI & 6149.25 & 3.89 & -2.72 & FeII \\
5858.79 & 4.22 & -2.19 & FeI & 6303.47 & 4.32 & -2.62 & FeI & 6239.95 & 3.89 & -3.44 & FeII \\
5859.6 & 4.55 & -0.63 & FeI & 6311.5 & 2.83 & -3.16 & FeI & 6247.56 & 3.87 & -2.32 & FeII \\
5861.11 & 4.28 & -2.35 & FeI & 6315.81 & 4.07 & -1.67 & FeI & 6369.46 & 2.89 & -4.21 & FeII \\
5862.37 & 4.55 & -0.42 & FeI & 6322.69 & 2.59 & -2.38 & FeI & 6416.93 & 3.89 & -2.7 & FeII \\
5879.49 & 4.61 & -1.99 & FeI & 6330.85 & 4.73 & -1.22 & FeI & 6432.68 & 2.89 & -3.58 & FeII \\
5880.03 & 4.56 & -1.94 & FeI & 6335.34 & 2.2 & -2.28 & FeI & 6456.39 & 3.9 & -2.1 & FeII \\
5881.28 & 4.61 & -1.76 & FeI & 6380.75 & 4.19 & -1.34 & FeI & 6516.08 & 2.89 & -3.38 & FeII \\
5902.48 & 4.59 & -1.86 & FeI & 6392.54 & 2.28 & -3.97 & FeI &  & & & \\
\end{tabular}
\end{table*}

\begin{longtab}
\centering
\begin{longtable}{cccc|cccc|cccc}
\caption{Line data used in the computation of the chemical abundances.}\\
\label{tab:linelist_ab}
Wavelength & $\chi_I$ & $\log$gf & Name & Wavelength & $\chi_I$ & $\log$gf & Name & Wavelength & $\chi_I$ & $\log$gf & Name \\
\hline
5148.83 & 2.102 & -2.044 & NaI & 5185.9 & 1.893 & -1.41 & TiI & 5662.93 & 3.695 & -1.975 & FeI \\
5682.63 & 2.102 & -0.706 & NaI & 5226.53 & 1.566 & -1.26 & TiI & 5667.45 & 5.064 & -1.875 & FeI \\
5688.19 & 2.104 & -1.406 & NaI & 5206.02 & 0.941 & 0.02 & CrI & 5667.51 & 4.178 & -1.576 & FeI \\
5688.2 & 2.104 & -0.452 & NaI & 5208.4 & 0.941 & 0.17 & CrI & 5679.02 & 4.652 & -0.92 & FeI \\
6154.22 & 2.102 & -1.547 & NaI & 5214.13 & 3.369 & -0.74 & CrI & 5679.11 & 5.033 & -2.021 & FeI \\
6160.74 & 2.104 & -1.246 & NaI & 5272.0 & 3.449 & -0.42 & CrI & 5691.49 & 4.301 & -1.52 & FeI \\
5528.4 & 4.346 & -0.498 & MgI & 5275.27 & 2.889 & -0.244 & CrI & 5717.83 & 4.284 & -1.13 & FeI \\
5711.08 & 4.346 & -1.724 & MgI & 5275.31 & 4.106 & -3.342 & CrI & 5731.76 & 4.256 & -1.3 & FeI \\
5785.31 & 5.108 & -2.11 & MgI & 5275.74 & 2.889 & -0.023 & CrI & 5775.08 & 4.22 & -1.298 & FeI \\
6318.71 & 5.108 & -2.103 & MgI & 5275.75 & 4.613 & -2.612 & CrI & 6085.25 & 2.759 & -3.095 & FeI \\
5557.06 & 3.143 & -2.11 & AlI & 5275.77 & 3.556 & -4.56 & CrI & 6271.27 & 3.332 & -2.703 & FeI \\
6696.02 & 3.143 & -1.347 & AlI & 5287.17 & 3.438 & -0.87 & CrI & 6591.31 & 4.593 & -2.07 & FeI \\
6698.67 & 3.143 & -1.647 & AlI & 5296.69 & 0.983 & -1.36 & CrI & 6625.02 & 1.011 & -5.35 & FeI \\
5645.61 & 4.93 & -2.14 & SiI & 5297.37 & 2.9 & 0.167 & CrI & 6625.06 & 5.32 & -6.617 & FeI \\
5666.67 & 5.616 & -1.797 & SiI & 5298.01 & 2.9 & -0.06 & CrI & 6627.54 & 4.549 & -1.68 & FeI \\
5675.41 & 5.619 & -1.234 & SiI & 5298.27 & 0.983 & -1.14 & CrI & 6633.41 & 4.835 & -1.49 & FeI \\
5690.42 & 4.93 & -1.87 & SiI & 5329.13 & 2.914 & -0.008 & CrI & 6646.91 & 4.435 & -5.242 & FeI \\
5701.1 & 4.93 & -2.05 & SiI & 5329.78 & 2.914 & -0.795 & CrI & 6646.93 & 2.609 & -3.99 & FeI \\
5747.66 & 5.614 & -1.544 & SiI & 5329.8 & 3.857 & -4.31 & CrI & 6745.1 & 4.58 & -2.16 & FeI \\
5772.14 & 5.082 & -1.75 & SiI & 5345.79 & 1.004 & -0.896 & CrI & 6745.95 & 4.076 & -2.77 & FeI \\
5780.38 & 4.92 & -2.35 & SiI & 5345.85 & 4.618 & -3.048 & CrI & 6746.95 & 2.609 & -4.35 & FeI \\
6125.02 & 5.614 & -1.465 & SiI & 5348.31 & 1.004 & -1.21 & CrI & 6783.7 & 2.588 & -3.98 & FeI \\
6131.57 & 5.616 & -1.557 & SiI & 5386.96 & 3.369 & -0.743 & CrI & 6786.85 & 4.191 & -2.07 & FeI \\
6131.85 & 5.616 & -1.617 & SiI & 5409.78 & 1.03 & -0.67 & CrI & 6793.25 & 4.584 & -4.275 & FeI \\
6142.48 & 5.619 & -1.296 & SiI & 5442.4 & 3.422 & -1.06 & CrI & 6793.25 & 4.076 & -2.326 & FeI \\
6145.01 & 5.616 & -1.311 & SiI & 5694.74 & 3.857 & -0.241 & CrI & 6796.04 & 4.186 & -2.563 & FeI \\
6155.13 & 5.619 & -0.755 & SiI & 5694.78 & 4.535 & -4.146 & CrI & 6796.12 & 4.143 & -2.53 & FeI \\
6237.31 & 5.614 & -0.975 & SiI & 5694.78 & 4.618 & -1.573 & CrI & 5234.61 & 10.448 & -2.697 & FeI \\
6243.81 & 5.616 & -1.244 & SiI & 5702.3 & 3.449 & -0.67 & CrI & 5234.62 & 3.221 & -2.23 & FeI \\
6407.29 & 5.871 & -1.393 & SiI & 5712.73 & 5.522 & -7.605 & CrI & 5264.8 & 3.23 & -3.12 & FeI \\
6414.97 & 5.871 & -1.036 & SiI & 5712.77 & 3.011 & -1.049 & CrI & 5284.07 & 10.531 & -1.381 & FeI \\
6526.63 & 5.871 & -1.607 & SiI & 5781.16 & 3.011 & -1.0 & CrI & 5284.1 & 2.891 & -2.99 & FeI \\
6527.2 & 5.871 & -1.072 & SiI & 5781.17 & 3.322 & -0.854 & CrI & 5325.55 & 3.221 & -3.12 & FeI \\
6721.84 & 5.863 & -1.527 & SiI & 5781.24 & 4.618 & -2.893 & CrI & 5414.07 & 3.221 & -3.54 & FeI \\
6741.62 & 5.984 & -1.75 & SiI & 5783.06 & 3.323 & -0.5 & CrI & 5425.24 & 3.199 & -3.16 & FeI \\
5260.38 & 2.521 & -1.719 & CaI & 5784.96 & 3.321 & -0.38 & CrI & 5534.81 & 11.035 & -3.108 & FeI \\
5261.7 & 2.521 & -0.579 & CaI & 5787.91 & 3.322 & -0.083 & CrI & 5534.83 & 3.245 & -2.73 & FeI \\
5349.46 & 2.709 & -0.31 & CaI & 6330.09 & 0.941 & -2.92 & CrI & 5534.89 & 10.545 & -0.44 & FeI \\
5512.98 & 2.933 & -0.464 & CaI & 6882.51 & 3.438 & -0.375 & CrI & 5991.37 & 3.153 & -3.54 & FeI \\
5590.11 & 2.521 & -0.571 & CaI & 5004.89 & 2.92 & -1.63 & MnI & 6247.55 & 3.892 & -2.31 & FeI \\
5867.13 & 5.55 & -4.655 & CaI & 5117.93 & 3.134 & -1.2 & MnI & 6247.57 & 5.956 & -4.827 & FeI \\
5867.56 & 2.933 & -1.57 & CaI & 5255.33 & 3.133 & -0.851 & MnI & 6456.37 & 3.903 & -2.1 & FeI \\
6102.09 & 5.229 & -4.839 & CaI & 5255.38 & 5.52 & -8.779 & MnI & 5115.39 & 3.834 & -0.11 & NiI \\
6156.02 & 2.521 & -2.506 & CaI & 5377.6 & 3.844 & -0.166 & MnI & 5129.37 & 3.679 & -0.63 & NiI \\
6166.43 & 2.521 & -1.142 & CaI & 5413.66 & 3.859 & -0.647 & MnI & 5155.12 & 3.898 & -0.65 & NiI \\
6169.04 & 2.523 & -0.797 & CaI & 5505.86 & 2.178 & -2.527 & MnI & 5155.76 & 3.898 & 0.074 & NiI \\
6169.56 & 2.526 & -0.478 & CaI & 5537.75 & 2.187 & -2.328 & MnI & 5176.56 & 3.898 & -0.44 & NiI \\
6471.66 & 2.526 & -0.686 & CaI & 5151.88 & 5.033 & -8.85 & FeI & 5220.29 & 3.74 & -1.31 & NiI \\
6717.68 & 2.709 & -0.524 & CaI & 5151.91 & 1.011 & -3.322 & FeI & 5392.33 & 4.154 & -1.32 & NiI \\
5173.74 & 0.0 & -1.06 & TiI & 5228.37 & 4.22 & -1.29 & FeI & 5589.35 & 3.898 & -1.14 & NiI \\
5186.33 & 2.117 & -0.77 & TiI & 5250.64 & 2.198 & -2.181 & FeI & 5593.73 & 3.898 & -0.84 & NiI \\
5192.96 & 0.021 & -0.95 & TiI & 5307.36 & 1.608 & -2.987 & FeI & 5625.31 & 4.089 & -0.7 & NiI \\
5194.04 & 2.103 & -0.65 & TiI & 5522.44 & 4.209 & -1.55 & FeI & 5637.11 & 4.089 & -0.82 & NiI \\
5201.05 & 3.57 & -4.08 & TiI & 5528.89 & 4.473 & -2.02 & FeI & 5682.19 & 4.105 & -0.47 & NiI \\
5201.08 & 2.092 & -0.66 & TiI & 5529.16 & 3.642 & -2.73 & FeI & 5748.35 & 1.676 & -3.26 & NiI \\
5210.38 & 0.048 & -0.82 & TiI & 5539.28 & 3.642 & -2.66 & FeI & 5760.83 & 4.105 & -0.8 & NiI \\
5384.63 & 0.826 & -2.77 & TiI & 5543.93 & 4.218 & -1.14 & FeI & 5805.21 & 4.167 & -0.64 & NiI \\
5389.98 & 1.873 & -1.1 & TiI & 5543.97 & 4.154 & -6.058 & FeI & 5831.39 & 5.004 & -5.748 & NiI \\
5449.15 & 1.443 & -1.87 & TiI & 5546.99 & 4.218 & -1.91 & FeI & 6086.28 & 4.266 & -0.53 & NiI \\
5453.64 & 1.443 & -1.6 & TiI & 5549.94 & 3.695 & -2.91 & FeI & 6111.07 & 4.088 & -0.87 & NiI \\
5648.56 & 2.495 & -0.26 & TiI & 5633.94 & 4.991 & -0.27 & FeI & 6130.13 & 4.266 & -0.96 & NiI \\
5662.15 & 2.318 & 0.01 & TiI & 5634.01 & 5.086 & -2.633 & FeI & 6175.36 & 4.089 & -0.53 & NiI \\
5673.42 & 3.148 & -0.244 & TiI & 5636.69 & 3.64 & -2.61 & FeI & 6186.71 & 4.105 & -0.96 & NiI \\
5673.43 & 3.112 & -1.533 & TiI & 5638.26 & 4.22 & -0.87 & FeI & 6259.59 & 4.089 & -1.237 & NiI \\
5679.91 & 2.472 & -0.57 & TiI & 5638.33 & 4.584 & -2.929 & FeI & 6414.58 & 4.154 & -1.18 & NiI \\
5689.46 & 2.297 & -0.36 & TiI & 5641.43 & 4.256 & -1.18 & FeI & 6424.85 & 4.167 & -1.355 & NiI \\
5880.27 & 1.053 & -2.0 & TiI & 5641.48 & 3.642 & -3.079 & FeI & 5105.53 & 1.389 & -1.542 & CuI \\
6002.63 & 2.16 & -1.49 & TiI & 5649.98 & 5.1 & -0.92 & FeI & 5218.19 & 3.817 & 0.364 & CuI \\
6303.75 & 1.443 & -1.58 & TiI & 5650.7 & 5.086 & -0.96 & FeI & 5220.06 & 3.817 & -0.59 & CuI \\
6312.23 & 1.46 & -1.55 & TiI & 5651.46 & 4.473 & -2.0 & FeI & 5700.23 & 1.642 & -2.583 & CuI \\
6599.1 & 0.9 & -2.085 & TiI & 5652.31 & 4.26 & -1.95 & FeI & 6362.33 & 5.796 & 0.15 & ZnI \\
5154.06 & 1.566 & -1.75 & TiI & 5662.51 & 4.178 & -0.573 & FeI & && & \\
\end{longtable}
\end{longtab}

\begin{sidewaystable*}
\caption{Sample of the catalogue produced by our code. Only a few columns and rows are shown.}\label{tab:sample_table}
\centering   
\begin{tabular}{ccccccccccccccccc}
\hline\hline
Starname & Instrument & Temperature & $[$Fe$/$H$]$ & \logg & $\xi_t$ & v$\sin$i & $\sigma_{vsini}$ & v$_{mac}$ & $\sigma_{v_{mac}}$ & $[$Na$/$H$]$ & $\sigma_{[Na/H]}$ & $\sigma_T$ & $\sigma_{[Fe/H]}$ & Mass & $\sigma_{mass}$ & use\_T$_c$\\
\hline\hline
HD870 & HARPS & 5409.3 & -0.09 & 4.418 & 0.75 & 2.189 & 0.288 & 2.359 & 0.288 & -0.1 & 0.008 & 44.629 & 0.099 & 0.8 & 0.022 & no \\
HD1461 & HARPS & 5750.0 & 0.19 & 4.371 & 0.7 & 0.62 & 0.243 & 3.526 & 0.24 & 0.31 & 0.023 & 39.318 & 0.084 & 1.07 & 0.025 & no \\
HD21411 & HARPS & 5500.0 & -0.2 & 4.451 & 0.61 & 2.482 & 0.255 & 2.555 & 0.254 & -0.22 & 0.031 & 41.635 & 0.086 & 0.89 & 0.024 & no \\
HD21411 & HARPS & 5500.0 & -0.2 & 4.451 & 0.61 & 2.482 & 0.255 & 2.555 & 0.254 & -0.22 & 0.031 & 41.635 & 0.086 & 0.89 & 0.024 & no \\
HD28471 & HARPS & 5797.0 & -0.03 & 4.462 & 0.78 & 1.076 & 0.21 & 3.533 & 0.209 & -0.02 & 0.087 & 29.294 & 0.059 & 1.0 & 0.023 & no \\
HD28701 & HARPS & 5782.0 & -0.26 & 4.453 & 0.79 & 1.686 & 0.244 & 3.534 & 0.244 & -0.12 & 0.06 & 51.461 & 0.097 & 0.94 & 0.037 & no \\
HD44447 & HARPS & 6044.0 & -0.19 & 4.338 & 1.1 & 0.62 & 0.432 & 4.663 & 0.432 & -0.11 & 0.045 & 35.545 & 0.061 & 1.05 & 0.03 & no \\
HD44447 & HARPS & 6044.0 & -0.19 & 4.338 & 1.1 & 0.62 & 0.432 & 4.663 & 0.432 & -0.11 & 0.045 & 35.545 & 0.061 & 1.05 & 0.03 & no \\
HD76151 & HARPS & 5816.0 & 0.13 & 4.516 & 0.76 & 0.62 & 0.244 & 3.571 & 0.243 & 0.19 & 0.055 & 29.937 & 0.062 & 1.07 & 0.009 & no \\
HD88656 & HARPS & 5193.1 & -0.04 & 4.555 & 0.5 & 2.392 & 0.451 & 1.586 & 0.451 & -0.06 & 0.018 & 31.749 & 0.075 & 0.84 & 0.029 & no \\
HD101581 & HARPS & 5016.0 & -0.53 & 4.211 & 1.33 & 1.791 & 0.043 & 1.676 & 0.043 & 0.09 & 0.25 & 151.346 & 0.325 & 1.59 & 0.036 & no \\
HD104263 & HARPS & 5439.0 & -0.0 & 4.25 & 0.65 & 1.372 & 0.24 & 2.782 & 0.239 & 0.07 & 0.026 & 30.535 & 0.068 & 0.84 & 0.022 & no \\
HD116920 & HARPS & 4954.5 & -0.23 & 4.283 & 0.57 & 1.98 & 0.869 & 1.568 & 0.869 & -0.12 & 0.012 & 48.33 & 0.112 & 0.8 & 0.013 & no \\
HD117618 & HARPS & 6060.0 & 0.1 & 4.448 & 0.96 & 1.528 & 0.282 & 4.581 & 0.281 & 0.07 & 0.044 & 48.018 & 0.089 & 1.14 & 0.028 & no \\
HD119638 & HARPS & 6148.0 & -0.1 & 4.43 & 1.11 & 0.0 & 0.0 & 4.996 & 0.195 & -0.03 & 0.047 & 29.4 & 0.049 & 1.1 & 0.024 & no \\
HD171990 & HARPS & 5985.0 & -0.01 & 3.952 & 1.32 & 1.237 & 0.344 & 5.32 & 0.344 & 0.16 & 0.07 & 43.878 & 0.081 & 1.24 & 0.026 & no \\
HD192310 & HARPS & 5188.0 & 0.04 & 4.578 & 0.65 & 1.628 & 0.572 & 1.497 & 0.571 & 0.16 & 0.02 & 46.888 & 0.107 & 2.04 & 0.148 & no \\
HD193193 & HARPS & 5985.0 & -0.05 & 4.319 & 1.06 & 0.911 & 0.269 & 4.406 & 0.268 & 0.07 & 0.054 & 52.485 & 0.096 & 1.11 & 0.019 & no \\
HD220339 & HARPS & 5075.0 & -0.31 & 4.579 & 0.71 & 2.067 & 0.823 & 1.251 & 0.822 & -0.16 & 0.059 & 56.498 & 0.127 & 0.74 & 0.061 & no \\
HD222422 & HARPS & 5478.0 & -0.09 & 4.415 & 0.48 & 1.879 & 0.215 & 2.538 & 0.215 & -0.14 & 0.037 & 32.178 & 0.071 & 0.92 & 0.026 & no \\
HIP22059 & HARPS & 4850.0 & -0.29 & 4.208 & 0.85 & 2.76 & 0.217 & 1.401 & 0.216 & -0.06 & 0.055 & 60.444 & 0.132 & 0.71 & 0.029 & no \\
HIP22059 & HARPS & 4850.0 & -0.29 & 4.208 & 0.85 & 2.76 & 0.217 & 1.401 & 0.216 & -0.06 & 0.055 & 60.444 & 0.132 & 0.71 & 0.029 & no \\
HIP98764 & HARPS & 4934.3 & -0.3 & 4.307 & 0.6 & 2.42 & 0.98 & 1.377 & 0.979 & -0.17 & 0.046 & 72.338 & 0.165 & 0.71 & 0.033 & no \\
HD23079 & FEROS & 6051.2 & -0.06 & 4.54 & 0.995 & 0.62 & 0.298 & 4.398 & 0.298 & -0.01 & 0.077 & 58.391 & 0.103 & 1.1 & 0.038 & no \\
HD39091 & FEROS & 5987.2 & 0.1 & 4.284 & 1.031 & 0.62 & 0.375 & 4.603 & 0.375 & 0.24 & 0.098 & 59.603 & 0.115 & 1.1 & 0.028 & yes \\
HD72673 & FEROS & 5232.2 & -0.41 & 4.367 & 0.764 & 2.693 & 0.391 & 1.96 & 0.391 & -0.24 & 0.034 & 49.732 & 0.106 & 0.77 & 0.037 & no \\
HD94151 & FEROS & 5618.2 & 0.07 & 4.383 & 0.968 & 0.674 & 0.343 & 3.074 & 0.342 & 0.23 & 0.066 & 59.058 & 0.128 & 0.97 & 0.027 & no \\
HD126525 & FEROS & 5696.2 & -0.03 & 4.461 & 0.91 & 0.871 & 0.293 & 3.197 & 0.293 & 0.06 & 0.018 & 62.972 & 0.129 & 1.0 & 0.024 & no \\
HD128674 & FEROS & 5629.2 & -0.32 & 4.495 & 0.974 & 1.796 & 0.467 & 2.908 & 0.467 & -0.24 & 0.035 & 60.819 & 0.118 & 0.94 & 0.022 & no \\
HD159868 & FEROS & 5594.2 & -0.03 & 3.931 & 1.086 & 0.927 & 0.372 & 3.813 & 0.372 & 0.1 & 0.103 & 32.967 & 0.07 & 1.33 & 0.018 & no \\
HD177565 & FEROS & 5657.2 & 0.09 & 4.395 & 1.025 & 0.62 & 0.227 & 3.182 & 0.226 & 0.06 & 0.203 & 36.204 & 0.077 & 0.99 & 0.02 & no \\
HD192310 & FEROS & 4983.9 & -0.13 & 4.04 & 0.93 & 0.62 & 0.407 & 1.964 & 0.406 & 0.17 & 0.199 & 55.973 & 0.136 & 2.88 & 0.484 & yes \\
HD196050 & FEROS & 5856.2 & 0.19 & 4.102 & 1.332 & 0.62 & 0.447 & 4.423 & 0.446 & 0.4 & 0.131 & 66.806 & 0.136 & 1.16 & 0.034 & no \\
HD202206 & FEROS & 5769.2 & 0.3 & 4.4 & 1.117 & 0.62 & 0.293 & 3.565 & 0.288 & 0.55 & 0.1 & 40.089 & 0.083 & 1.07 & 0.034 & no \\
HD16141 & UVES & 5785.2 & 0.15 & 4.155 & 1.062 & 0.62 & 0.186 & 4.065 & 0.156 & 0.15 & 0.008 & 40.874 & 0.082 & 1.22 & 0.016 & no \\
HD44447 & UVES & 6090.2 & -0.03 & 4.465 & 0.527 & 0.62 & 0.432 & 4.586 & 0.429 & -0.13 & 0.011 & 87.011 & 0.15 & 1.12 & 0.046 & no \\
HD45289 & UVES & 5738.2 & -0.01 & 4.314 & 0.871 & 0.62 & 0.184 & 3.61 & 0.158 & 0.04 & 0.011 & 30.37 & 0.061 & 0.99 & 0.012 & no \\
HD114613 & UVES & 5726.2 & 0.19 & 3.975 & 1.09 & 0.62 & 0.284 & 4.07 & 0.279 & 0.26 & 0.077 & 37.504 & 0.074 & 1.35 & 0.004 & no \\
HD140901 & UVES & 5660.2 & 0.16 & 4.509 & 0.795 & 0.62 & 0.42 & 2.826 & 0.42 & 0.15 & 0.142 & 46.902 & 0.103 & 0.99 & 0.032 & no \\
HD104263 & HIRES & 5589.2 & 0.11 & 4.533 & 0.855 & 1.659 & 0.437 & 2.707 & 0.437 & 0.14 & 0.096 & 100.985 & 0.208 & 0.82 & 0.045 & no \\
HD157338 & HIRES & 6238.2 & -0.0 & 4.764 & 1.432 & 0.62 & 1.809 & 4.735 & 1.809 & -0.0 & 0.038 & 263.0 & 0.315 & 1.16 & 0.022 & no \\
HD157347 & HIRES & 5738.2 & 0.02 & 4.503 & 0.363 & 2.116 & 0.442 & 3.207 & 0.44 & 0.02 & 0.229 & 77.35 & 0.103 & 0.99 & 0.026 & no \\
HD159868 & HIRES & 5638.2 & -0.01 & 4.008 & 1.108 & 2.033 & 0.299 & 3.819 & 0.298 & 0.1 & 0.081 & 101.287 & 0.205 & 1.42 & 0.05 & no \\
HD195564 & HIRES & 5677.5 & 0.01 & 4.385 & 1.701 & 2.005 & 0.863 & 3.21 & 0.863 & -0.05 & 0.06 & 283.308 & 0.397 & 1.34 & 0.035 & yes \\
\end{tabular}
\end{sidewaystable*}

\begin{table*}
\caption{Results from SPECIES for the GBS samples, for the stellar parameters. The spectral types were drawn from \citet{heiter2015}, and denote the temperature classification, plus the luminosity class (V: dwarf, IV: subgiant, III: giant).}
\label{tab:FGK_gaia}
\centering
\begin{tabular}{llllllll}
\hline\hline
Starname & Sp Type & $[$Fe$/$H$]$ & $T$ (K) & $\log$ g & $\xi_t$ (km s$^{-1}$) & $v\sin i$ (km s$^{-1}$) & $M$ ($M_{\odot}$) \\
\hline\hline
18Sco & G V & 0.11 $\pm$ 0.04 & 5872 $\pm$ 20 & 4.53 $\pm$ 0.15 & 0.87 $\pm$ 0.02 & 1.35 $\pm$ 0.16 & 1.1 $\pm$ 0.0 \\
61CygB & K V & -0.48 $\pm$ 0.37 & 4989 $\pm$ 171 & 3.28 $\pm$ 2.67 & 0.95 $\pm$ 0.18 & 2.61 $\pm$ 2.43 & 2.7 $\pm$ 1.3 \\
Arcturus & FGK III & -0.20 $\pm$ 0.37 & 4752 $\pm$ 354 & 2.90 $\pm$ 1.90 & 1.73 $\pm$ 0.04 & 3.29 $\pm$ 1.66 & 3.5 $\pm$ 0.2 \\
Gmb1830 & K V & -1.17 $\pm$ 0.10 & 5368 $\pm$ 61 & 4.66 $\pm$ 1.08 & 1.41 $\pm$ 0.08 & 0.00 $\pm$ 0.00 & 0.1 $\pm$ 0.0 \\
HD107328 & FGK III & -0.41 $\pm$ 0.09 & 4526 $\pm$ 47 & 2.02 $\pm$ 0.46 & 1.70 $\pm$ 0.03 & 2.30 $\pm$ 0.47 & 3.2 $\pm$ 0.0 \\
HD220009 & FGK III & -0.86 $\pm$ 0.41 & 4120 $\pm$ 354 & 1.06 $\pm$ 2.78 & 1.34 $\pm$ 0.04 & -0.07 $\pm$ 3.60 & 3.7 $\pm$ 0.1 \\
HD22879 & G V & -0.85 $\pm$ 0.13 & 5858 $\pm$ 80 & 4.37 $\pm$ 0.28 & 0.68 $\pm$ 0.06 & 0.00 $\pm$ 0.00 & 0.8 $\pm$ 0.0 \\
HD49933 & F V & -0.48 $\pm$ 0.25 & 6732 $\pm$ 137 & 4.33 $\pm$ 1.44 & 2.24 $\pm$ 0.26 & 4.04 $\pm$ 1.90 & 1.3 $\pm$ 0.0 \\
Procyon & F V & 0.14 $\pm$ 0.26 & 6744 $\pm$ 137 & 4.06 $\pm$ 1.71 & 1.59 $\pm$ 0.18 & 0.00 $\pm$ 0.00 & 1.6 $\pm$ 0.1 \\
Sun & G V & 0.00 $\pm$ 0.05 & 5799 $\pm$ 25 & 4.46 $\pm$ 0.19 & 0.90 $\pm$ 0.02 & 1.56 $\pm$ 0.21 & 1.0 $\pm$ 0.0 \\
alfCenA & G V & 0.26 $\pm$ 0.09 & 5831 $\pm$ 42 & 4.34 $\pm$ 0.16 & 0.98 $\pm$ 0.03 & 1.78 $\pm$ 0.34 & 1.1 $\pm$ 0.0 \\
betAra & M III & -0.35 $\pm$ 0.69 & 4116 $\pm$ 500 & 0.59 $\pm$ 5.05 & 3.10 $\pm$ 0.10 & 6.88 $\pm$ 4.08 & 14.3 $\pm$ 0.6 \\
betGem & FGK III & 0.18 $\pm$ 0.08 & 5010 $\pm$ 38 & 3.17 $\pm$ 0.24 & 1.30 $\pm$ 0.03 & 2.16 $\pm$ 0.24 & 3.2 $\pm$ 0.2 \\
betHyi & FGK IV & -0.05 $\pm$ 0.08 & 5874 $\pm$ 38 & 4.04 $\pm$ 0.19 & 1.21 $\pm$ 0.04 & 3.05 $\pm$ 0.28 & 1.2 $\pm$ 0.0 \\
betVir & G V & 0.21 $\pm$ 0.06 & 6223 $\pm$ 28 & 4.21 $\pm$ 0.21 & 1.24 $\pm$ 0.03 & 1.66 $\pm$ 0.34 & 1.4 $\pm$ 0.0 \\
delEri & FGK IV & 0.16 $\pm$ 0.07 & 5133 $\pm$ 35 & 3.87 $\pm$ 0.23 & 0.93 $\pm$ 0.03 & 1.87 $\pm$ 0.24 & 0.9 $\pm$ 0.0 \\
epsEri & K V & -0.07 $\pm$ 0.13 & 5276 $\pm$ 35 & 4.46 $\pm$ 1.74 & 1.13 $\pm$ 0.04 & 2.53 $\pm$ 2.23 & 0.8 $\pm$ 0.0 \\
epsFor & FGK IV & -0.39 $\pm$ 0.07 & 5370 $\pm$ 4 & 4.26 $\pm$ 0.42 & 0.69 $\pm$ 0.03 & 2.15 $\pm$ 0.38 & 1.9 $\pm$ 0.1 \\
epsVir & FGK III & -0.06 $\pm$ 0.12 & 4830 $\pm$ 1 & 2.15 $\pm$ 0.24 & 1.48 $\pm$ 0.05 & 2.16 $\pm$ 0.32 & 4.5 $\pm$ 0.1 \\
etaBoo & FGK IV & -0.07 $\pm$ 0.26 & 5670 $\pm$ 1 & 3.03 $\pm$ 0.50 & 1.71 $\pm$ 0.13 & 11.96 $\pm$ 0.43 & 2.1 $\pm$ 0.0 \\
gamSge & M III & 0.14 $\pm$ 0.26 & 4481 $\pm$ 158 & 2.63 $\pm$ 0.92 & 1.98 $\pm$ 0.12 & 4.10 $\pm$ 0.79 & 4.5 $\pm$ 0.0 \\
ksiHya & FGK III & 0.21 $\pm$ 0.08 & 5186 $\pm$ 36 & 3.16 $\pm$ 0.21 & 1.35 $\pm$ 0.03 & 4.04 $\pm$ 0.21 & 0.1 $\pm$ 0.0 \\
muAra & G V & 0.38 $\pm$ 0.21 & 5967 $\pm$ 106 & 4.62 $\pm$ 0.48 & 1.10 $\pm$ 0.08 & 1.99 $\pm$ 0.44 & 1.3 $\pm$ 0.2 \\
muLeo & FGK III & 0.21 $\pm$ 0.15 & 4458 $\pm$ 1 & 2.23 $\pm$ 0.39 & 1.60 $\pm$ 0.07 & 1.21 $\pm$ 0.37 & 2.3 $\pm$ 0.0 \\
\end{tabular}
\end{table*}

\begin{table*}
\caption{Results from SPECIES for the GBS samples, for the chemical abundance of the elements in common with \citet{jofre2015}.}
\label{tab:FGK_gaia_ab}
\centering
\begin{tabular}{lccccccc}
\hline\hline
Starname & $[$Mg$/$H$]$ & $[$Si$/$H$]$ & $[$Ca$/$H$]$ & $[$Ti$/$H$]$ & $[$Cr$/$H$]$ & $[$Mn$/$H$]$ & $[$Ni$/$H$]$ \\
\hline\hline
18Sco & 0.08 $\pm$ 0.07 & 0.08 $\pm$ 0.02 & 0.14 $\pm$ 0.02 & 0.16 $\pm$ 0.10 & 0.10 $\pm$ 0.01 & 0.12 $\pm$ 0.07 & 0.08 $\pm$ 0.03 \\
61CygB & -0.46 $\pm$ 0.14 & -0.87 $\pm$ 0.02 & 0.63 $\pm$ 0.04 & 0.51 $\pm$ 0.11 & -0.02 $\pm$ 0.05 & -0.49 $\pm$ 0.10 & -0.91 $\pm$ 0.18 \\
Arcturus & -0.00 $\pm$ 0.08 & -0.13 $\pm$ 0.04 & -0.03 $\pm$ 0.01 & 0.29 $\pm$ 0.10 & -0.23 $\pm$ 0.02 & -0.32 $\pm$ 0.01 & -0.22 $\pm$ 0.02 \\
Gmb1830 & -1.27 $\pm$ 0.23 & -0.90 $\pm$ 0.20 & -0.82 $\pm$ 0.02 & -0.84 $\pm$ 0.13 & -1.10 $\pm$ 0.05 & nan $\pm$ nan & -1.25 $\pm$ 0.16 \\
HD107328 & 0.01 $\pm$ 0.06 & -0.08 $\pm$ 0.04 & -0.21 $\pm$ 0.01 & -0.16 $\pm$ 0.04 & -0.43 $\pm$ 0.04 & -0.61 $\pm$ 0.06 & -0.43 $\pm$ 0.01 \\
HD220009 & -0.35 $\pm$ 0.05 & -0.40 $\pm$ 0.03 & -0.56 $\pm$ 0.05 & -0.72 $\pm$ 0.01 & -0.81 $\pm$ 0.02 & -1.05 $\pm$ 0.10 & -0.87 $\pm$ 0.01 \\
HD22879 & -0.58 $\pm$ 0.04 & -0.56 $\pm$ 0.05 & -0.55 $\pm$ 0.00 & -0.61 $\pm$ 0.07 & -0.86 $\pm$ 0.01 & 0.04 $\pm$ 0.00 & -0.88 $\pm$ 0.02 \\
HD49933 & -0.53 $\pm$ 0.17 & -0.32 $\pm$ 0.03 & -0.33 $\pm$ 0.02 & -0.45 $\pm$ 0.02 & -0.51 $\pm$ 0.03 & nan $\pm$ nan & -0.54 $\pm$ 0.01 \\
Procyon & 0.08 $\pm$ 0.07 & 0.17 $\pm$ 0.07 & 0.23 $\pm$ 0.01 & 0.04 $\pm$ 0.06 & 0.12 $\pm$ 0.03 & 0.34 $\pm$ 0.42 & 0.08 $\pm$ 0.03 \\
Sun & -0.03 $\pm$ 0.07 & -0.01 $\pm$ 0.02 & 0.02 $\pm$ 0.04 & 0.02 $\pm$ 0.06 & 0.01 $\pm$ 0.01 & 0.01 $\pm$ 0.04 & 0.00 $\pm$ 0.02 \\
alfCenA & 0.18 $\pm$ 0.06 & 0.29 $\pm$ 0.02 & 0.29 $\pm$ 0.02 & 0.27 $\pm$ 0.02 & 0.26 $\pm$ 0.01 & 0.42 $\pm$ 0.03 & 0.28 $\pm$ 0.03 \\
betAra & -0.29 $\pm$ 0.23 & 0.17 $\pm$ 0.28 & -0.46 $\pm$ 0.07 & -0.59 $\pm$ 0.06 & -0.41 $\pm$ 0.16 & -0.62 $\pm$ 0.28 & -0.53 $\pm$ 0.06 \\
betGem & 0.15 $\pm$ 0.13 & 0.26 $\pm$ 0.02 & 0.18 $\pm$ 0.01 & 0.19 $\pm$ 0.01 & 0.25 $\pm$ 0.01 & 0.40 $\pm$ 0.29 & 0.08 $\pm$ 0.01 \\
betHyi & -0.08 $\pm$ 0.07 & -0.04 $\pm$ 0.02 & 0.01 $\pm$ 0.02 & 0.01 $\pm$ 0.04 & -0.06 $\pm$ 0.03 & -0.06 $\pm$ 0.04 & -0.08 $\pm$ 0.01 \\
betVir & 0.11 $\pm$ 0.08 & 0.23 $\pm$ 0.05 & 0.29 $\pm$ 0.01 & 0.27 $\pm$ 0.10 & 0.21 $\pm$ 0.03 & 0.19 $\pm$ 0.08 & 0.20 $\pm$ 0.01 \\
delEri & 0.18 $\pm$ 0.05 & 0.13 $\pm$ 0.01 & 0.21 $\pm$ 0.02 & 0.19 $\pm$ 0.03 & 0.20 $\pm$ 0.01 & 0.40 $\pm$ 0.15 & 0.18 $\pm$ 0.03 \\
epsEri & -0.06 $\pm$ 0.00 & -0.21 $\pm$ 0.01 & 0.10 $\pm$ 0.07 & 0.06 $\pm$ 0.03 & 0.04 $\pm$ 0.03 & -0.00 $\pm$ 0.09 & -0.12 $\pm$ 0.03 \\
epsFor & -0.18 $\pm$ 0.07 & -0.27 $\pm$ 0.05 & -0.12 $\pm$ 0.05 & 0.07 $\pm$ 0.03 & -0.42 $\pm$ 0.00 & -0.51 $\pm$ 0.01 & -0.36 $\pm$ 0.00 \\
epsVir & 0.05 $\pm$ 0.19 & 0.25 $\pm$ 0.01 & -0.02 $\pm$ 0.02 & -0.23 $\pm$ 0.00 & -0.06 $\pm$ 0.03 & 0.09 $\pm$ 0.24 & -0.12 $\pm$ 0.00 \\
etaBoo & 0.13 $\pm$ 0.27 & 0.38 $\pm$ 0.08 & 0.19 $\pm$ 0.02 & -0.12 $\pm$ 0.14 & -0.07 $\pm$ 0.09 & 0.38 $\pm$ 0.09 & 0.08 $\pm$ 0.12 \\
gamSge & 0.15 $\pm$ 0.04 & 0.17 $\pm$ 0.04 & 1.93 $\pm$ 1.85 & 0.51 $\pm$ 0.04 & 0.34 $\pm$ 0.05 & 0.09 $\pm$ 0.04 & -0.06 $\pm$ 0.02 \\
ksiHya & 0.12 $\pm$ 0.10 & 0.22 $\pm$ 0.06 & 0.21 $\pm$ 0.01 & 0.23 $\pm$ 0.00 & 0.24 $\pm$ 0.02 & 0.39 $\pm$ 0.38 & 0.10 $\pm$ 0.01 \\
muAra & 0.22 $\pm$ 0.07 & 0.28 $\pm$ 0.02 & 0.30 $\pm$ 0.13 & 0.38 $\pm$ 0.06 & 0.42 $\pm$ 0.04 & 0.58 $\pm$ 0.11 & 0.35 $\pm$ 0.02 \\
muLeo & 0.38 $\pm$ 0.08 & 0.38 $\pm$ 0.01 & 0.17 $\pm$ 0.03 & 0.20 $\pm$ 0.05 & 0.20 $\pm$ 0.02 & 0.49 $\pm$ 0.36 & 0.13 $\pm$ 0.03 \\
\end{tabular}
\end{table*}

\begin{acknowledgements}
MGS acknowledges support from CONICYT-PCHA/Doctorado Nacional/2014-21141037. JSJ acknowledges support by Fondecyt grant 1161218 and partial support by CATA-Basal (PB06, CONICYT).
We acknowledge the very helpful comments from Y. V. Pavlenko. 
This research has made use of the VizieR catalogue access tool, CDS,
 Strasbourg, France. The original description of the VizieR service was
 published in A\&AS 143, 23.
\end{acknowledgements}

%-------------------------------------------------------------------

\bibliographystyle{aa}
\bibliography{refs_paper}

\appendix
\section{Second method for the uncertainty in microturbulent velocity}\label{sec:err_vt2}

As mentioned in section \ref{sec:error_microturbulence}, we include two different estimations for the uncertainty in the microturbulent velocity. The first one was already described (shown in the final catalogue as \texttt{err\_vt}). Here we described the second method used (\texttt{err\_vt2}).

We used Equation 12 from \citet{Magain1984}, 
\begin{equation}
\sigma_{\xi_t} \simeq c(\sigma_{\delta}^2/\sigma_{W}^2)\left.\frac{\partial \xi_t}{\partial S_{RW}}\right|_{S_{RW}=0}
\end{equation}

where $\xi_t$ is the microturbulent velocity, $c = \partial A/\partial W$ change of abundance with equivalent width,  $\sigma_{\delta}^2$ is the variance of the uncertainty of the equivalent widths (EWs), $\sigma_{W}^2$ is the variance in the EWs, and $S_{RW}$ is the slope of the linear fit performed to EWs vs \FeI abundance.

We assumed the same approximations than in \citet{Magain1984}, with one of them being that $c = c_i = \partial A_i/\partial W_i$, the same for all the lines. In order to compute that value, we plotted $c_i$ vs $W_i$ and found the ranges for which $c$ is constant. We adopted $c$ within that range to be the final value.

The dependency of $\xi_t$ with $S_{RW}$ can be adjusted by the cubic spline from Eq. \ref{eq:vt_cubic_spline}, with coefficients given by Eq. \ref{eq:coefs_vt}. The final value for the microturbulence velocity obtained by SPECIES is reached when $S_{RW} = 0$ (Section \ref{sec:atmospheric_parameters}), therefore in order to obtain ($\partial \xi_t/\partial S_{RW})|_{0}$ it is necessary to derive Eq \ref{eq:vt_cubic_spline}, and replace $\xi_t$ by the SPECIES value in Eq. \ref{eq:coefs_vt}.

Finally, $\sigma_{\delta}^2$ and $\sigma_{W}^2$ are obtained from the ARES files for each star. 

\section{Second method for the uncertainty in temperature}\label{sec:err_T2}

The method described here is very similar to the one used in Section \ref{sec:error_temperature}, meaning that the uncertainty in the temperature is composed by two parts: 

\begin{equation}
\sigma_T^2\: =\: \left(\left.\frac{\partial T}{\partial \xi_t}\right|_{\xi_t}\right)^2\, \sigma_{\xi_t}^2 \: +\: \left(\left.\frac{\partial T}{\partial \chi_I}\right|_{\chi_I = 0}\right)^2\,\sigma_{\chi_I}^2,
\end{equation}

where the first term corresponds to the contribution from the uncertainty in the microturbulence, and the second term is the contribution from the uncertainty in the slope of the dependence between the individual \FeI abundances and the excitation potential, $\chi_I$. The first term is the same as the one derived in section \ref{sec:error_temperature}, but the second term is different and described here.

From equation 16.4 of \cite{gray2005}, we can derive that:

\begin{equation}
\log\left(\frac{w}{\lambda}\right)\, \propto \, \log A -\, \theta_{\text{ex}}\,\chi_I,
\end{equation}

where $\log (w/\lambda)$ is the reduced equivalent width of the line, $\log A = \log(N_E/N_H)$ the abundance of the element $E$ to hydrogen, $\theta_{\text{ex}} = 5040/T$, and $\chi_I$ is the excitation potential. If we assume that the equivalent width of the line will not change with respect to $\chi_I$, but $\log A$ will, then when we differentiate with respect to $\chi_I$, $\partial/\partial \chi_I$, we obtain

\begin{equation}
\frac{\partial T}{\partial\chi_I}\, = \, s_{\chi_I}\,\frac{T^2}{5040}
\end{equation}

where $\partial \log A/\partial\chi_I = s_{\chi_I}$ is the slope of the correlation between individual line abundances and excitation potential, and is one of the results obtained from the MOOG output file. The contribution from the uncertainty in the excitation potential will then be the above expression multiplied by the error in the slope.

\section{Column description}\label{sec:column_description}

The columns returned by SPECIES are as follow:

%['Starname', 'Instrument', 'RV', '[Fe/H]', 'err_[Fe/H]', 'Temperature', 'err_T', 'logg', 'err_logg', 'vt', 'err_vt', 'nFeI', 'nFeII', 'exception', 'vsini', 'err_vsini', 'vmac', 'err_vmac', '[Na/H]', 'e_[Na/H]', 'nNaI', '[Mg/H]', 'e_[Mg/H]', 'nMgI', '[Al/H]', 'e_[Al/H]', 'nAlI', '[Si/H]', 'e_[Si/H]', 'nSiI', '[Ca/H]', 'e_[Ca/H]', 'nCaI', '[TiI/H]', 'e_[TiI/H]', 'nTiI', '[TiII/H]', 'e_[TiII/H]', 'nTiII', '[Cr/H]', 'e_[Cr/H]', 'nCrI', '[Mn/H]', 'e_[Mn/H]', 'nMnI', '[Ni/H]', 'e_[Ni/H]', 'nNiI', '[Cu/H]', 'e_[Cu/H]', 'nCuI', '[Zn/H]', 'e_[Zn/H]', 'nZnI', 'exception_Fe', 'exception_Ti', '[FeI/H]', '[FeII/H]', 'Mass', 'err_mass', 'Age', 'err_age', 'iso_logg', 'err_iso_logg', 'Radius', 'err_radius', 'use_Tc', 'use_vt', 'use_iso_logg', 'err_vt2', 'err_T2', 'T_photo', 'err_T_photo', 'T_photo_relation']

\begin{itemize}
\item[·] Col 1: Star name.
\item[·] Col 2: Instrument used to obtain the spectra (HARPS, FEROS, HIRES, UVES, CORALIE, AAT).
\item[·] Col 3: Velocity in \kms, obtained from the CCF (section \ref{sec:EW}), used to correct the spectrum to the restframe.
\item[·] Cols 4-11: Atmospheric stellar parameters and their corresponding uncertainty (metallicity, temperature, surface gravity and microturbulent velocity, respectively)
\item[·] Cols 12-13: Number of \FeI and \FeII lines used for the computation of the atmospheric parameters, respectively.
\item[·] Col 14: Exception to the atmospheric parameters. A value of 1 means the parameters were computed correctly. A value of 2 means that there were problems in the computation (parameters were repeated more than 200 times, or all of them were outside the permitted ranges), or that the code could not converge to a final value after performing over 1 million iterations.
\item[·] Cols 15-18: $v\sin i$ and $v_{mac}$ for each star, with their respective uncertainties.
\item[·] Cols 19-54: Abundances for Na, Mg, Al, Si, Ca, Ti I, Ti II, Cr, Mn, Ni, Cu and Zn, as well as their uncertainties (standard deviation from the mean) and the number of lines used.
\item[·] Cols 55-56: Individual abundances for \FeI and \FeII, respectively.
\item[·] Cols 57-64: Mass, age, \isologg (explained in Section ~\ref{sec:comparison_atmospheric_parameters}), and radius obtained, along with their uncertainties. 
\item[·] Col 65: Tells whether the temperature computed from the method of \citet{casagrande2010} was used as the final temperature or not, for the cases when no convergence was reached in the atmospheric parameters (Section ~\ref{sec:atmospheric_parameters}).
\item[·] Col 66: Tells whether the microturbulent velocity was set to 1.2 km~s$^{-1}$, for the cases when no convergence was reached in the atmospheric parameters (section \ref{sec:atmospheric_parameters}). 
\item[·] Col 67: Tells whether the surface gravity was set to be equal to \isologg (section \ref{sec:mass_age_plogg}).
\item[.] Cols 68-69: Error in the microturbulence and temperature, computed using the methods described in Sections ~\ref{sec:err_vt2} and \ref{sec:err_T2}.
\item[·] Cols 70-72: Value, uncertainty and relation used to obtain the temperature from photometry (section \ref{sec:ini_temperature}).

\end{itemize}

\end{document}